\documentclass{aastex61_v1}

\usepackage{amssymb,amsmath,fancyhdr,graphicx,epstopdf}

\newcommand{\angstrom}{\text{\normalfont\AA} }

\submitjournal{ApJS}

\shorttitle{The VLBA Extragalactic Proper Motion Catalog}
\shortauthors{Truebenbach \& Darling}

\begin{document}

\title{The VLBA Extragalactic Proper Motion Catalog and a Measurement of the Secular Aberration Drift}

\correspondingauthor{Alexandra E. Truebenbach}
\email{alexandra.truebenbach@colorado.edu}

\author{Alexandra E. Truebenbach}
\affil{Center for Astrophysics and Space Astronomy, \\
Department of Astrophysical and Planetary Sciences, \\
University of Colorado, 389 UCB, Boulder, CO 80309-0389, USA}

\author{Jeremy Darling}
\affiliation{Center for Astrophysics and Space Astronomy, \\
Department of Astrophysical and Planetary Sciences, \\
University of Colorado, 389 UCB, Boulder, CO 80309-0389, USA}



\begin{abstract}
We present a catalog of extragalactic proper motions created using archival VLBI data and our own VLBA astrometry. The catalog contains 713 proper motions, with average uncertainties of $\sim 24 \ \mu$as yr$^{-1}$, including 40 new or improved proper motion measurements using relative astrometry with the VLBA. The observations were conducted in X-band and yielded positions with uncertainties $\sim 70 \ \mu$as. We add 10 new redshifts using spectroscopic observations taken at Apache Point Observatory and Gemini North. With the VLBA Extragalactic Proper Motion Catalog, we detect the secular aberration drift -- the apparent motion of extragalactic objects caused by the solar system's acceleration around the Galactic center -- at a 6.3$\sigma$ significance. We model the aberration drift as a spheroidal dipole, with the square root of the power equal to $4.89 \pm 0.77$ $\mu$as yr$^{-1}$, an amplitude of $1.69 \pm 0.27$ $\mu$as yr$^{-1}$, and an apex at ($275.2 \pm 10.0^\circ$, $-29.4 \pm 8.8^\circ$). Our dipole model detects the aberration drift at a higher significance than some previous studies \citep[e.g.,][]{TitovLambert2013}, but at a lower amplitude than expected or previously measured. The full aberration drift may be partially removed by the no-net-rotation constraint used when measuring archival extragalactic radio source positions. Like the cosmic microwave background dipole, which is induced by the observer's motion, the aberration drift signal should be subtracted from extragalactic proper motions in order to detect cosmological proper motions, including the Hubble expansion, long-period stochastic gravitational waves, and the collapse of large-scale structure.
\end{abstract}




\section{Introduction}\label{intro}
Observations of extragalactic radio sources have been conducted with Very Long Baseline Interferometry (VLBI) since the 1970s with the purpose of measuring the Earth's orientation and monitoring the terrestrial and celestial reference frames. These observations have been conducted primarily by the United States Navy and the National Aeronautic and Space Administration (NASA), and, since 1998, have been coordinated by the International VLBI Service for Geodesy and Astrometry (IVS; \citealp{SchluterBehrend2007}). By measuring group delays -- the difference in arrival times of radio wave packets -- at widely separated radio antennae, VLBI experiments can produce radio positions typically with milliarcsecond precision or better. These positions are used to determine the International Celestial Reference System (ICRS), a barycentric reference system standardized by the International Astronomical Union (IAU). Currently, the ICRS is defined by the second realization of the International Celestial Reference Frame (ICRF2; \citealp{Maetal1998}; \citealp{Feyetal2009}). ICRF2 contains 3414 total sources, including 295 ``defining'' sources whose positions are used to fix the axes of the ICRS. 

When point-like extragalactic radio sources such as quasars were selected to anchor the ICRF, it was assumed that these sources are approximately fixed on the sky and exhibit no measurable proper motion. However, as the precision of VLBI measurements have improved, it has become clear that quasars are {\it not} fixed over human timescales and have proper motions on the order of microarcseconds to milliarcseconds \citep[e.g.,][]{Eubanks1997, Gwinnetal1997, FeisselGontier2000}. The intrinsic radio proper motion of a quasar is predominately due to the motion of plasma in relativistic jets produced by the quasars (e.g., \citealp{Feyetal1997}). These intrinsic proper motions are random in orientation on the sky. Quasars also show correlated proper motions from cosmological effects \citep[e.g.,][]{Gwinnetal1997, Quercellinietal2009, Nusseretal2012, Darling2013}, as well as from observer-induced signatures such as the secular aberration drift \citep[e.g.,][]{Fanselow1983,Bastian1995,Eubanksetal1995,Soversetal1998,Mignard2002,Kovalevsky2003,KopeikinMakarov2006,Titovetal2011, TitovLambert2013,Xuetal2012,Xuetal2013}. Both cosmological and observer-induced proper motions are often much smaller than the quasars' intrinsic proper motions, but the correlated nature of these effects allows a statistical detection if the sample size is large enough. Using a large sample of extragalactic proper motions, it is possible to obtain geometrical distances independent of canonical cosmological distance ladders \citep[e.g.,][]{DingCroft2009, Brodericketal2011}, to measure or constrain the deflection of quasar light by a primordial gravitational wave background spanning $10^{-18} - 10^{-8}$ Hz \citep[][]{Gwinnetal1997, BookFlanagan2011}, to test the isotropy of Hubble expansion \citep{Darling2014}, and to measure the collapse of large-scale structure \citep[e.g.,][]{Quercellinietal2009,Nusseretal2012,Darling2013}. 

The secular aberration drift is an observer-induced effect caused by the acceleration of the solar system around the Galactic center. The observed signal can be modeled as a curl-free dipole with an apex at the Galactic center (266.4$^\circ$, -28.9$^\circ$) and has been detected by \cite{Titovetal2011} and confirmed by \cite{Xuetal2012}, \cite{TitovLambert2013} (hereafter TL13), and \cite{Xuetal2013}. Using proper motions of masers associated with young massive stars, \cite{Reidetal2009} created a model of the Galactic plane to measure that the solar system is $8.4 \pm 0.6$ kpc from the Galactic center and has a barycentric circular rotation speed of $254 \pm 16$ km s$^{-1}$. This yields an acceleration of $0.79 \pm 0.11$ cm s$^{-1}$ yr$^{-1}$ and a dipole amplitude of $5.40 \pm 0.78 \ \mu$as yr$^{-1}$. In previous studies, extragalactic proper motions have been used to measure a solar acceleration of $0.93 \pm 0.11 $ cm s$^{-1}$ yr$^{-1}$ (dipole amplitude of $6.4 \pm 1.1 \ \mu$as yr$^{-1}$; TL13) and $0.85 \pm 0.05 $ cm s$^{-1}$ yr$^{-1}$ (dipole amplitude of $5.8 \pm 0.3$ $\mu$as yr$^{-1}$; \citealp{Xuetal2013}).

Although the secular aberration drift signal is small compared to typical extragalactic proper motions, it is important that it be well measured. The detection of the secular aberration drift will give an independent estimate of the solar acceleration without reliance on Galactic objects. If not corrected for in geodetic experiments, the drift can cause a deformation of the celestial reference frame axes, which can lead to inaccurate estimates of other geodetic parameters \citep{Titov2010}. Additionally, the secular aberration drift must be removed from extragalactic proper motions in order to detect and measure cosmological effects.

In this paper, we present the VLBA Extragalactic Proper Motion Catalog containing 713 proper motions created using $\sim 30$ years of archival VLBI data and our own NRAO Very Long Baseline Array (VLBA) observations. We then use the catalog to measure the secular aberration drift as a means of demonstrating one of the uses of a large, precise catalog of extragalactic proper motions. This catalog is 66\% larger than previous catalogs used to measured the aberration drift -- TL13 use 429 quasars -- due both to geodetic VLBI observations conducted in the intervening years and to the addition of our own astrometric observations. In addition, we use an analytic least-squares bootstrap technique to determine the proper motions, which provides more accurate estimation of the proper motions and associated uncertainties than previous techniques. Section \ref{creation} describes the catalog creation, Section \ref{redshifts} presents our optical redshift measurements, Section \ref{VLBAastro} presents our VLBA astrometric measurements and derived proper motions, Section \ref{catalog} presents the completed catalog, Section \ref{drift} measures the secular aberration drift, and Section \ref{Conclusions} summarizes and suggests future additions to this work. 

\section{Catalog Creation}\label{creation}
 We created our catalog of extragalactic radio proper motions using the 2017a Goddard VLBI global solution \footnote{https://gemini.gsfc.nasa.gov/solutions/2017\_astro/2017a\_ts.html} 
(e.g., \citealp{Feyetal2009}). The global solution uses group delays and radio wave arrival times of a series of distant compact radio sources (typically quasars) to simultaneously solve for both a terrestrial and celestial reference frame using the fitting program CALC/SOLVE\footnote{https://lupus.gsfc.nasa.gov/software\_calc\_solve.htm}. Earth orientation parameters are also solved for after each 24-hour session. 

The 2017a solution is computed from more than 30 years of dual-band VLBI observations -- 1979 August 3 to 2017 March 27 -- and uses a total of 5696 diurnal sessions and greater than $10^7$ measurements of group delays. All session with durations of 18 hours and longer and source elevations higher than $5^\circ$ were used in the fitting process. No net translation and rotation constraints were applied to the positions and velocities of all stations except for stations near Chile and Japan. The recent earthquakes in 2010 and 2011, respectively, caused non-linear motions for nearby stations, which are modeled using post-seismic deformation models \citep{Altamimietal2016}. Atmospheric gradient delays were modeled following \cite{MacMillanetal1995} and \cite{MacMillanetal1997} and the troposphere zenith delay was calculated using logged pressure and temperature (e.g., \citealp{Saastamoinen1972})\footnote{See https://gemini.gsfc.nasa.gov/solutions/2016a/gsf2016a.eops.txt for further details on initial parameters and assumptions}. 

In addition to the 2017a global solution, an astrometric time-series of 4618 extragalactic radio sources is also produced by Goddard\footnote{See https://gemini.gsfc.nasa.gov/solutions/2017\_astro/2017a\_ts.html}. Instead of treating the radio source positions as global parameters, which are assumed to be constant with time, the positions are incrementally treated as local parameters and are estimated once for each 24-hour session. Five separate solutions are combined to create the final 2017a time series. In the first solution, the positions of all 295 ICRF2 defining sources are global parameters and are tied to their ICRF2 positions with a a no-net-rotation (NNR) constraint, while the positions of all other sources are solved for at each session. In the second solution, the NNR constraint is removed for a quarter of the defining sources (every fourth source in RA) and the positions of this quarter are treated as local parameters and are solved for at each session. In the third, fourth, and fifth solutions, the next successive quarter of the defining sources are treated as local parameters. By incrementally treating all source positions as local parameters, the positions are allowed to vary between sessions and a coordinate time-series can be constructed. Then, we can fit a trend line to the coordinate vs. epoch for each source and solve for each source's proper motion.

During the creation of the solutions, a NNR constraint is applied to fix the ICRS axes \citep{Feyetal2009}. This constraint is needed to remove degeneracy when solving for local parameters of each 24-hour observing session. However, \cite{Titov2010} argues that a tight NNR constraint can remove systemic proper motions from the final catalog. To mitigate this effect, the NNR constraint is loosened during the creation of the Goddard 2017a coordinate time series. By creating five separate solutions where only a portion of the sources are required to satisfy the NNR constraint in each solution, all source positions are allowed to rotate with respect to the ICRS axes, while still enabling a non-degenerate local solution.

We fit a line to each coordinate time series (right ascension and declination) for each radio source using an analytical least-squares parameter estimation. If we assume each source has a constant proper motion (this is a reasonable approximation for most extragalactic sources, with a few exceptions discussed below), then we can solve for the proper motion, $\mu$, by minimizing the $\chi^2$ statistic,
\begin{equation}
S = \sum_{i=1}^{N} \frac{(r_i - \theta_0 - \mu t_i)^2}{\sigma_{r_i}^2}
\end{equation}       
where $r_i$ is the celestial position of the source at time $t_i$, $\sigma_{r_i}$ is the uncertainty of the source position, and $\theta_0$ is the y-intercept of the line -- a physically meaningless quantity in this application. Because the fitting model is linear, we can directly solve for $\mu$ and $\theta_0$ --
\begin{equation}
\mu = \frac{s \ s_{tr} - s_t \ s_r}{\Delta},
\end{equation}
and
\begin{equation}
\theta_0 = \frac{s_{t^2} \ s_r - s_t \ s_{tr}}{\Delta},
\end{equation}
where
\begin{equation}
s = \sum_{i=1}^N \frac{1}{\sigma_{r_i}^2},
\end{equation}
\begin{equation}
 s_t = \sum\limits_{i=1}^N \frac{t_i}{\sigma_{r_i}^2},
\end{equation}
\begin{equation}
 s_r = \sum_{i=1}^N \frac{r_i}{\sigma_{r_i}^2},
 \end{equation}
 \begin{equation}
s_{t^2} = \sum_{i=1}^N \frac{t_i^2}{\sigma_{r_i}^2},
\end{equation}
\begin{equation}
 s_{tr} = \sum_{i=1}^N \frac{t_i \ r_i}{\sigma_{r_i}^2},
\end{equation}
and 
\begin{equation}
\Delta = s \ s_{t^2} - s_t^2.
\end{equation}

We solved for proper motions separately in right ascension and declination -- $r$ in the above equations is thus $\alpha$ or $\delta$, respectively. Proper motions in right ascension, and their associated uncertainties, are corrected for declination. We excluded all sources that had been observed for less than 10 years or for fewer than 10 sessions. We also excluded sessions from before 1990 because fewer antennas and fewer monitored radio sources made VLBI data taken before this time less accurate (e.g., \citealp{Gontieretal2001}; \citealp{Malkin2004}; \citealp{Feissel-Vernieretal2004}; \citealp{LambertGontier2009}; \citealp{Titovetal2011}). We also removed the 39 ``special handling'' sources from ICRF2 \citep{Feyetal2009}. These sources show significant position instability in either right ascension and/or declination, indicating that their proper motion is largely from relativistic radio jets, rather than from the global effects we hope to measure. Thus, the inclusion of these sources in our catalog would impede our goal of detecting small, correlated proper motions. 

There is a large variation in the uncertainties of individual positions within many of the time series. Additionally, some of these positions with high uncertainties are separated from the other measurements by a large gap in time, thereby giving them a disproportionately large influence on the resulting proper motion fit. To assess the influence of individual measurements on each fit and to better estimate the uncertainty of the fits, we employed 500 iterations of a bootstrap re-sampling on each time series. The reported proper motions in our catalog are the median of the bootstrap distribution. The proper motion uncertainties are calculated from the variance of the distribution using
\begin{equation}
\sigma_\mu^2 = \frac{1}{N-1} \ \sum_i^N (\mu_{i} - \bar{\mu} )^2
\end{equation}
where $\bar{\mu}$ is the mean of the distribution, $\mu_i$ is the proper motion for an individual bootstrap iteration, and $N$ is the total number of bootstrap iterations.

\section{Optical Redshifts}\label{redshifts}
When available, we include redshifts in our catalog. Although not necessary for measurement of the secular aberration drift, redshifts are crucial for measurement of many of the cosmological effects described in Section \ref{intro}. The majority of the redshifts were obtained from the 2017 May 31 version of the Optical Characteristic of Astrometric Radio Sources (OCARS; \citealp{Malkin2016}) Catalog. This catalog contains known redshifts and optical or infrared magnitudes of radio sources observed in astrometric and geodetic VLBI observations. These source characteristics are primarily obtained from the NASA/IPAC Extragalactic Database\footnote{https://ned.ipac.caltech.edu} (NED) or the SIMBAD Astronomical Database\footnote{simbad.u-strasbg.fr/simbad/} \citep{Wengeretal2000}. About 8\% of our proper motion catalog is either missing redshifts or the redshifts are listed as ``questionable'' in OCARS.

We observed 28 catalog objects with either no redshift or a ``questionable'' OCARS redshift at the Apache Point Observatory (APO) 3.5m telescope and/or at Gemini North to increase the fraction of our proper motion catalog that is usable in cosmological studies. Table \ref{redshifttable} lists the observed objects and their measured redshifts. Ultimately, we measured 10 redshifts, ranging from $z= 0.21 - 2.86$. For cases where no redshift was determined, many of the spectra showed no significant emission or absorption lines. In a few cases, the only detectable lines were consistent with the local standard of rest ($z=0$), even though we expect all of our catalog objects to be extragalactic. With only foreground, Galactic absorption features present, we were unable to measure  optical redshifts for these extragalactic radio sources. These objects have a redshift of ``Galactic'' in Table \ref{redshifttable}. The spectroscopic observations are described below.
 
 \begin{deluxetable}{llllll}
 \tablecaption{Optical Redshifts Measured at Apache Point Observatory / Gemini North}
 \tablewidth{0pc}
 \tablehead{\colhead{IVS Name} & \colhead{Mag} & \colhead{Filter} & \colhead{$z$} & \colhead{Obs.}
 }
 \startdata
 0017+200 & 20.6 & V & \nodata & APO \\
 0019+058 & 18.8 & V & $2.86 \pm 0.02$ & APO \\
 0056-001 & 17.1 & V & $0.719 \pm 0.001$ & APO \\
 0106+138 & 19.0 & V & $1.697 \pm 0.005$ & APO \\
 0159+723 & 19.2 & V & Galactic & APO \\
 0253+033 & 18.0 & V & \nodata  & APO \\
 0300+470 & 16.6 & V & \nodata  & APO \\
 0302+625 & \nodata & \nodata & Galactic & APO \\
 0420+417 & 19.2 & R & \nodata  & Gemini \\
 0422+004 & 16.5 & V & 0.268$^a$ & NED \\
 0426+273 & 19.6 & R & \nodata  & APO \\
 0459+135 & 20.5 & V & $0.35 \pm 0.01$ & APO \\
 0529+483 & 19.9 & V & \nodata  & APO \\
 1013+127 & 18.6 & V & \nodata  & APO \\
 1147+245 & 15.7 & V & $0.209 \pm 0.001$ & APO \\
 1444+313 & 15.0 & r & Galactic & APO \\
 1506+591 & 18.9 & V & $0.310 \pm 0.004$ & APO \\
 1525+610 & 19.9 & G & $0.2456 \pm 0.0007$ & APO \\
 1717+178 & 19.9 & V & 0.14$^b$ & NED \\
 1754+155 & 17.2 & R & $2.06 \pm 0.04$ & APO \\
 1823+689 & 19.0 & R & $2.143 \pm 0.001$ & APO \\
 2013+163 & 17.3 & R & Galactic & APO \\
 2021+317 & 19.0 & R & \nodata  & Gemini \\
 2051+745 & 20.4 & V & $0.92 \pm 0.01$ & APO \\
 2225+033 & 17.5 & V & \nodata  & APO \\
 2315+032 & 20.4 & R & \nodata & APO \\
 2319+444 & 20.7 & R & \nodata  & APO \\
 \enddata
 \tablecomments{Columns from left to right: (a) The IVS name of the target, (b) the magnitude of the object from OCARS, (c) the optical filter in which the magnitude was measured, (d) the redshift of the object if a measurement was possible, and (e) the observatory where the object was observed (both objects observed at Gemini were first observed at APO). An ellipsis for the redshift indicates that no redshift was measured. A redshift of ``Galactic'' indicates that only Galactic ($z=0$) lines were detected, even though we expect all catalog objects to be extragalactic. We were unable to measure optical redshifts for these extragalactic radio sources. For all of our redshift observations, we also include the associated uncertainty based on the scatter in the line identifications. The two redshifts listed without uncertainties were obtained from the literature, where no uncertainties were provided.}
 \tablenotemark{a}{\cite{Shawetal2013}}
 \tablenotemark{b}{\cite{Sowards-Emmerdetal2005}}
 \label{redshifttable}
 \end{deluxetable}

\subsection{Apache Point Observatory}\label{APO}
We conducted observations on the 3.5m telescope at Apache Point Observatory from 2015 April 18 to 2016 June 30. We used the Dual Imaging Spectrograph (DIS) with a 1.5$''$ slit and two gratings centered at 4400 \angstrom and 7500 \angstrom with linear dispersions of 1.83 \angstrom pixel$^{-1}$ and 2.31 \angstrom pixel$^{-1}$, respectively. The final spectra have spatial resolution of $0.162''$ pixel$^{-1}$. We observed each target for a total of between $\sim 15$ and 75 minutes, depending on the target's optical magnitude. 

\begin{figure}
\label{APOspectra}
\centering
\centerline{\includegraphics[width=.5\textwidth, trim=.6cm 0cm 1.5cm 1.3cm,clip]{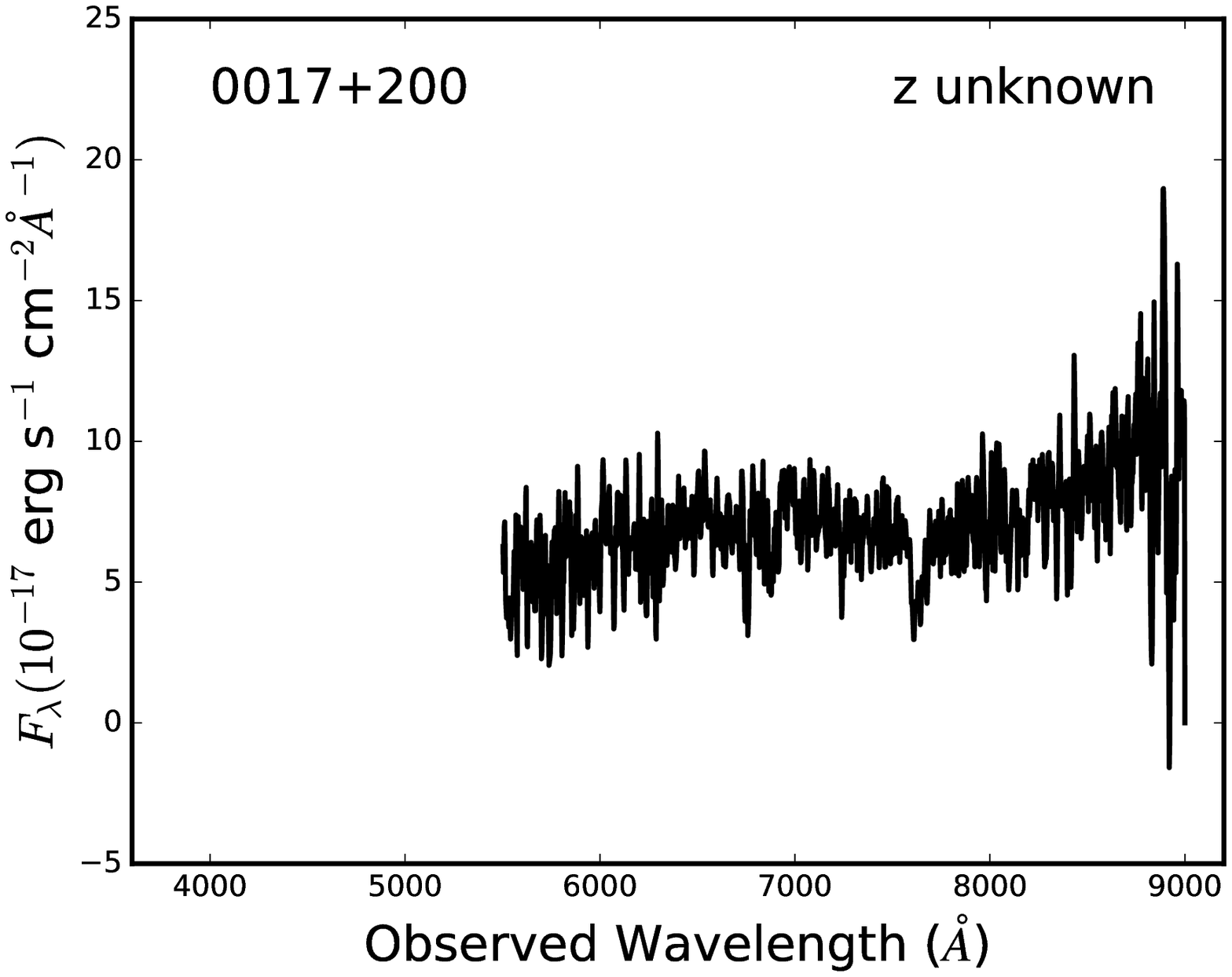}\includegraphics[width=.5\textwidth, trim=.6cm 0cm 1.5cm 1.3cm,clip]{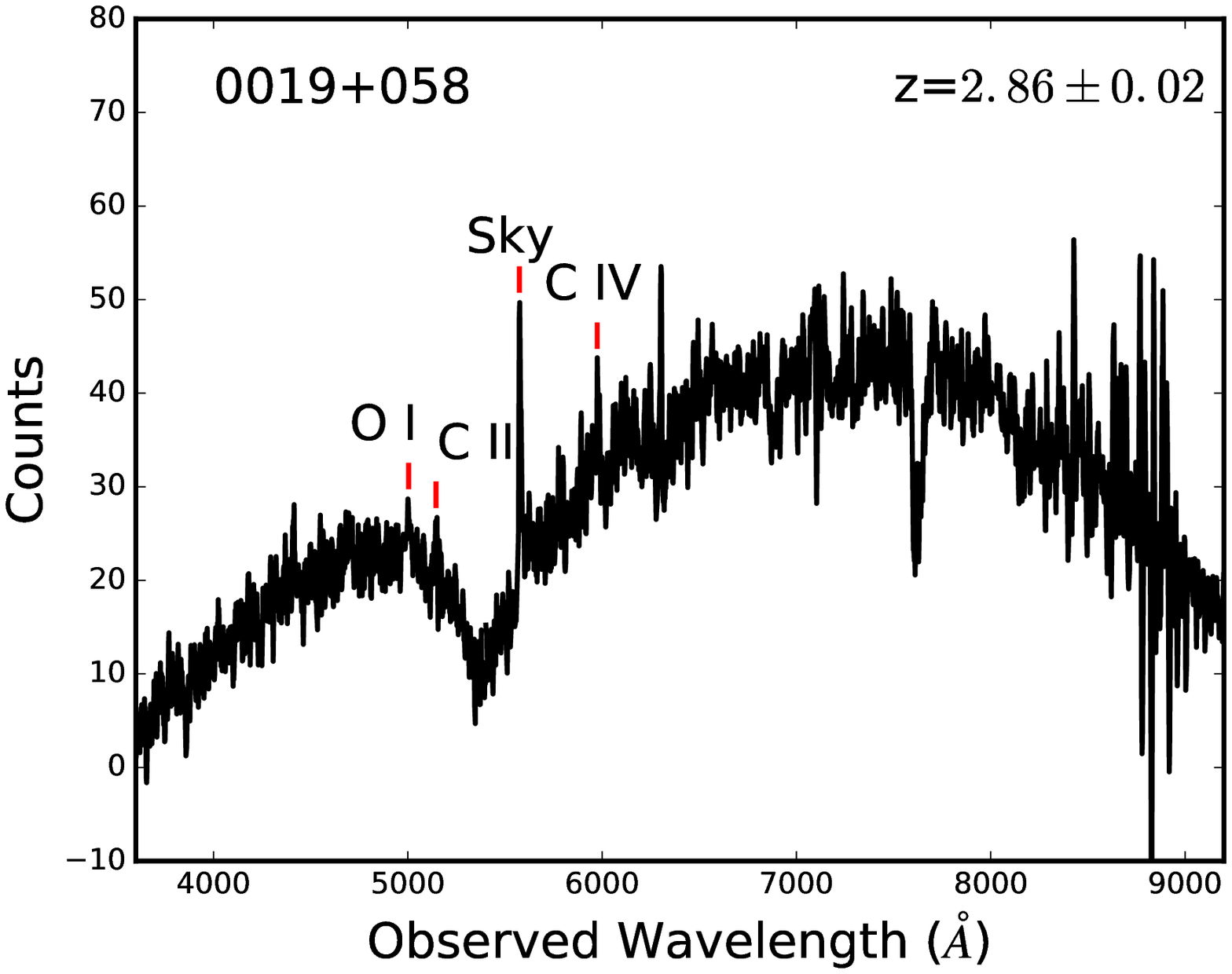}}
\vspace{0.01\textwidth}
\centerline{\includegraphics[width=.5\textwidth, trim=.6cm 0cm 1.5cm 1.3cm,clip]{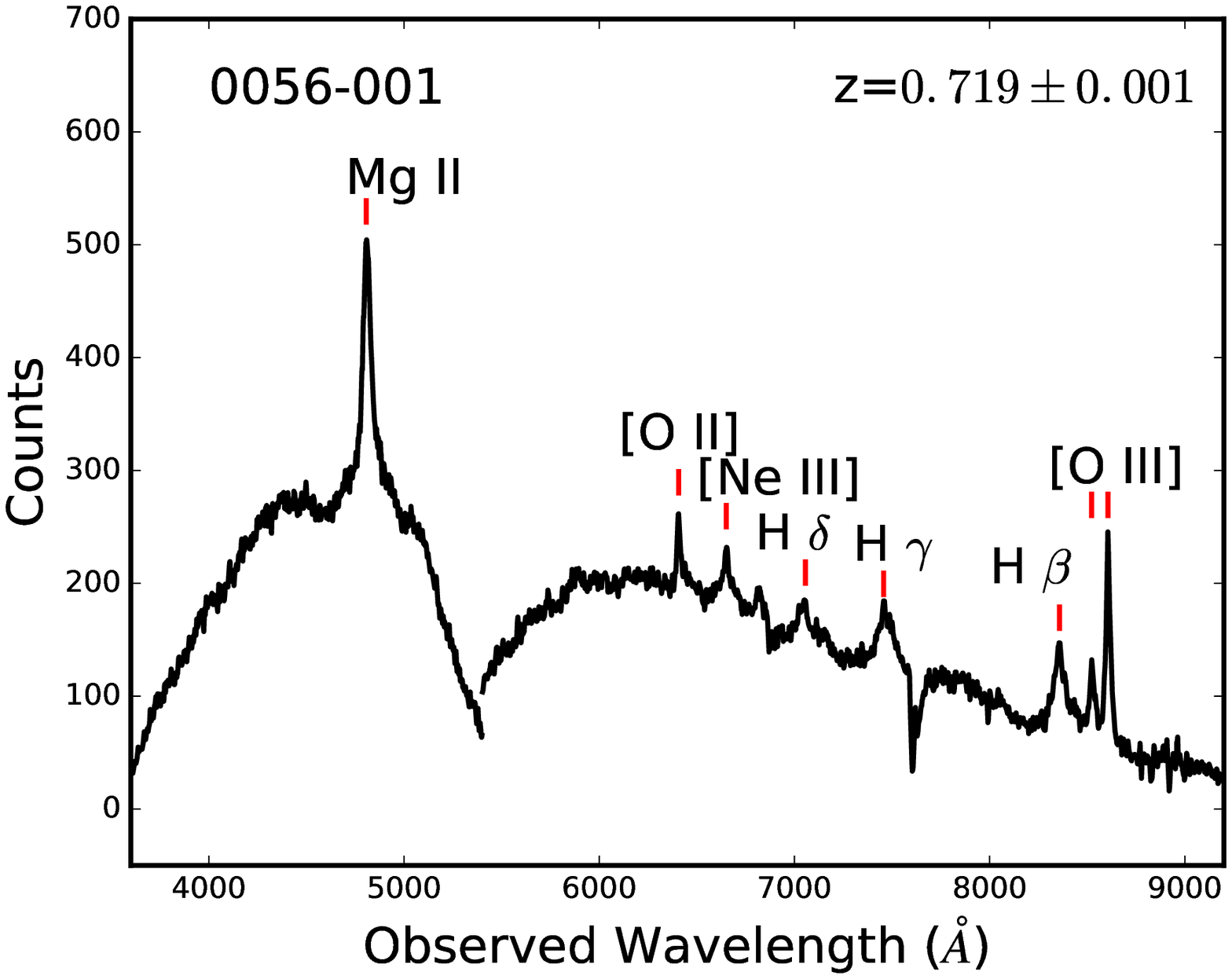}\includegraphics[width=.5\textwidth, trim=.4cm 0cm 1.5cm 1.3cm,clip]{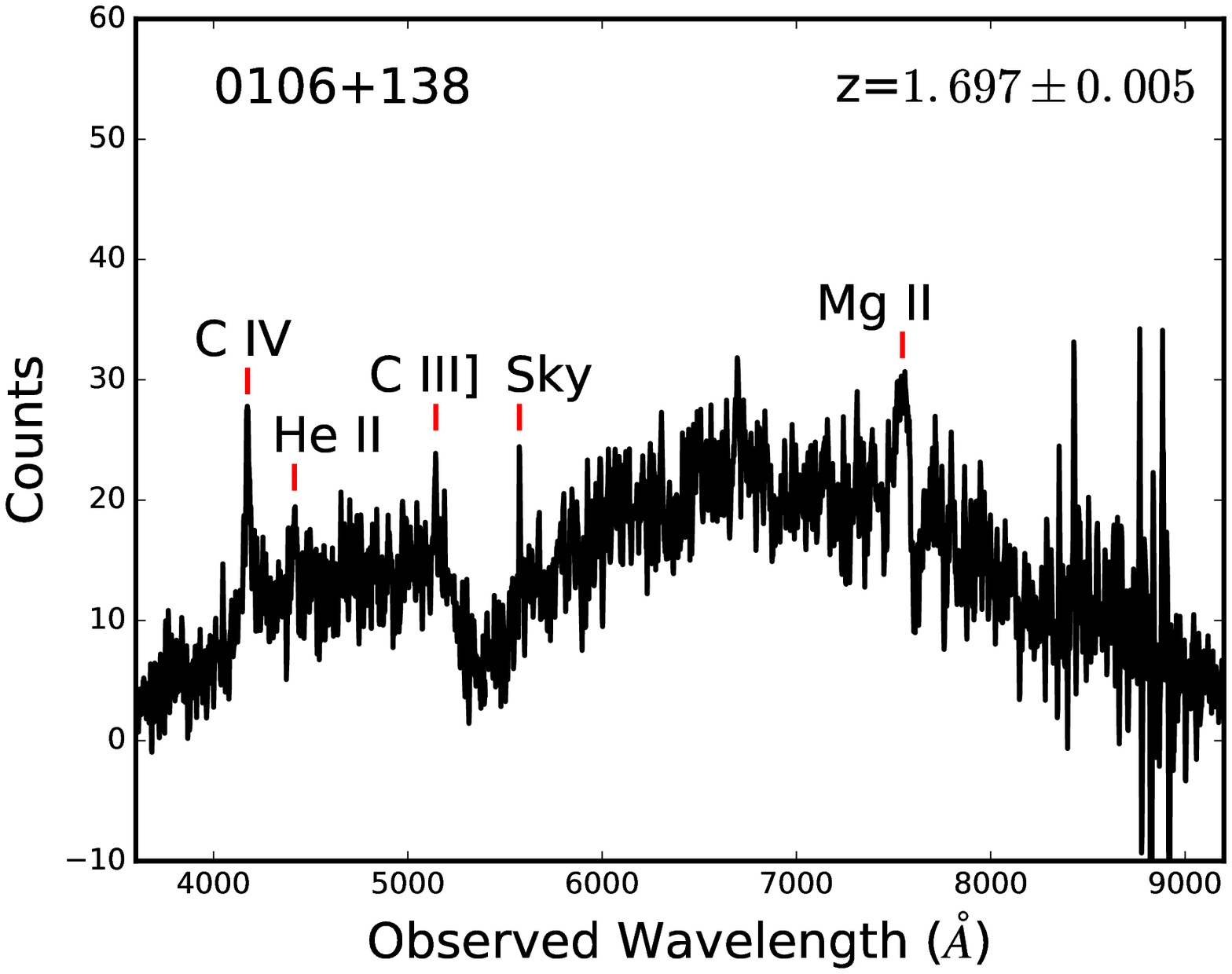}}
\vspace{0.01\textwidth}
\centerline{\includegraphics[width=.5\textwidth, trim=.4cm 0cm 1.5cm 1.3cm,clip]{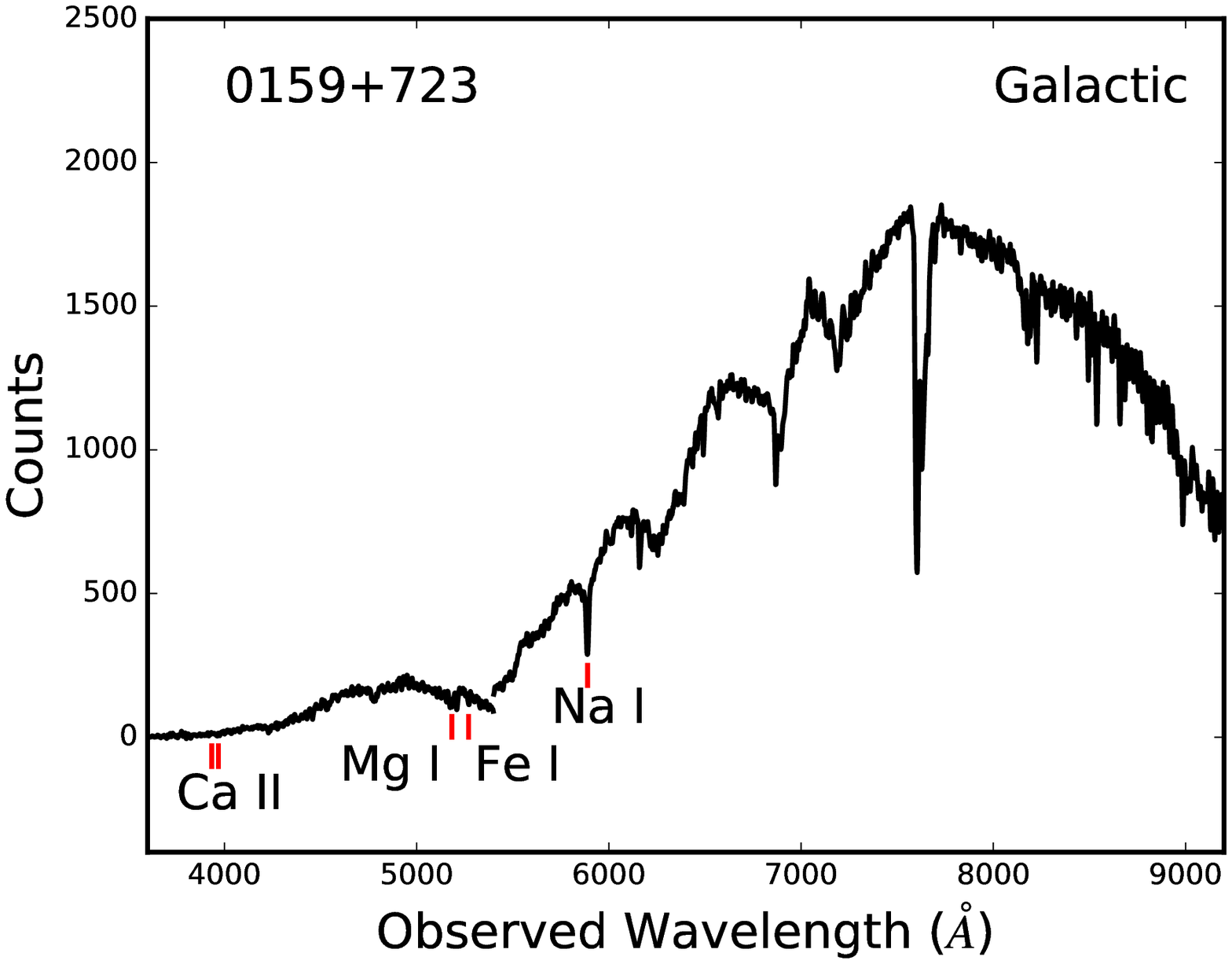}\includegraphics[width=.5\textwidth, trim=.2cm 0cm 1.5cm 1.3cm,clip]{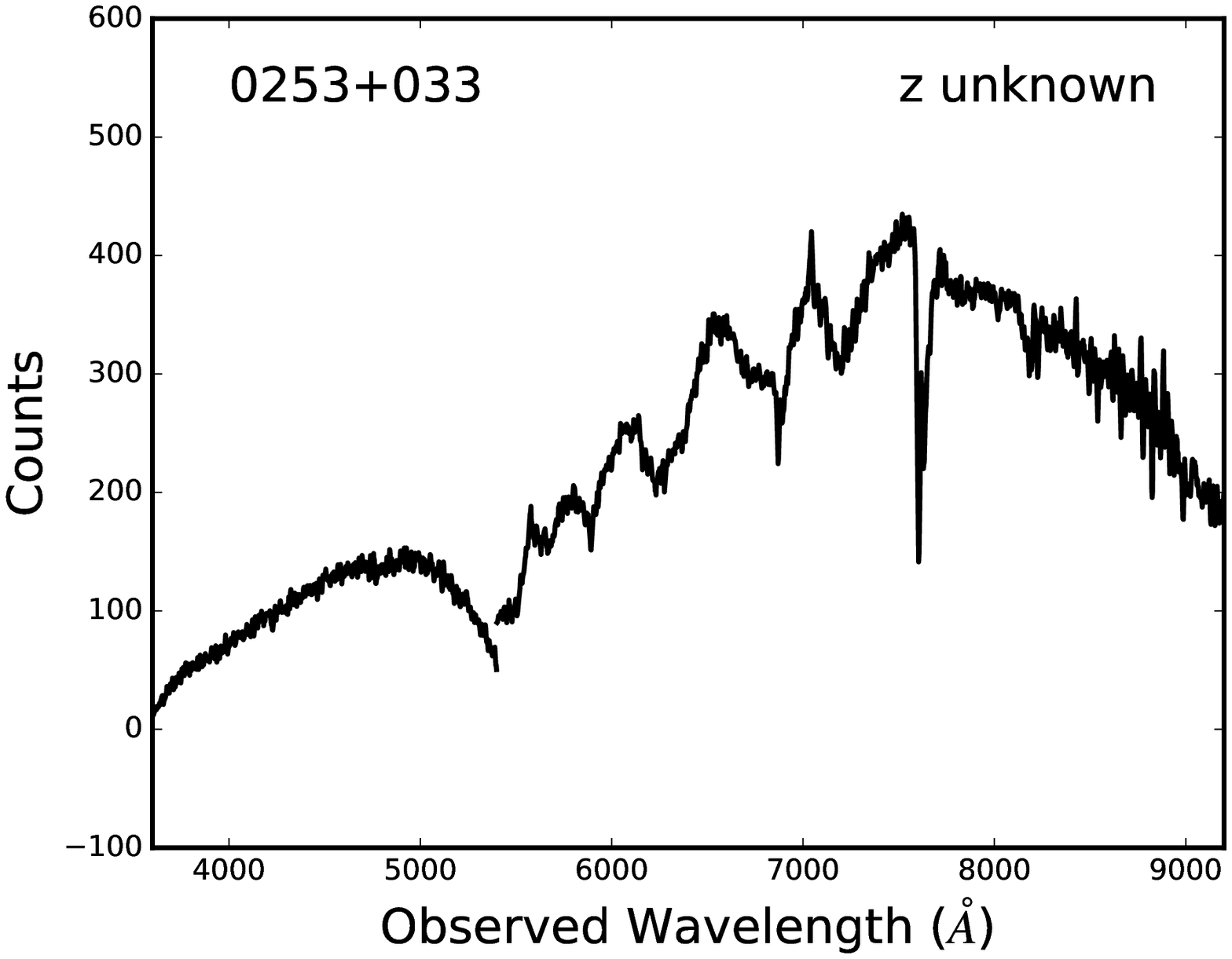}}
\end{figure}

We reduced the data using the Image Reduction and Analysis Facility (IRAF) package. The images were overscan corrected, trimmed, bias subtracted, flat fielded, wavelength calibrated, and background sky subtracted. Then we median stacked the reduced images for each source and extracted a one-dimensional final spectrum. For objects where a flux calibrator was observed on the same night, we flux calibrated the reduced, extracted spectra using the IRAF model of the calibrator's flux density and a mean extinction curve measured at Apache Point\footnote{http://www.apo.nmsu.edu/arc35m/Instruments/DIS/images/apoextinct.dat}. Figure \ref{APOspectra} shows the final, one-dimensional spectra taken at APO. The measured redshifts and key lines used to determine those redshifts are listed on the plots. In the rest of this subsection, we discuss each spectrum individually.

\paragraph{0017+200}
No emission or absorption lines were detected for this object. Additionally, the continuum on the blue CCD had too low of a signal-to-noise ratio (SNR) to extract. Together these factors prevent us from determining a redshift.

\paragraph{0019+058}
We find $z=2.86$ based on three AGN emission lines -- O I, C II, and C IV. There is also a prominent emission line at $\sim 5700 $ \angstrom that coincides with a sky emission line. It is likely that this line is an artifact of incomplete sky subtraction. There is a scatter of $10 - 30 $ \angstrom in the line identifications at $z=2.86$. This equates to a redshift uncertainty of $\sigma_z = 0.2$. OCARS gives this object a lower limit of $z> 0.64$, which is in good agreement with our measurement.

\paragraph{0056-001}
We find $z=0.719$ based on several emission lines, including H $\gamma$, H $\beta$ and [O III]. The scatter is $< 6 $ \angstrom, which equates to $\sigma_z=0.001$. After we observed this object at Apache Point, OCARS added a new redshift for this object -- $z=0.719$. Our measurements match this redshift exactly.

\paragraph{0106+138}
We find $z=1.697$ from several emission lines shown in Figure \ref{APOspectra}. The scatter is $3 - 7 $ \angstrom, which equates to $\sigma_z=0.005$. This is in good agreement with the redshift measured by SDSS: $z=1.7005 \pm 0.0003$ \citep{Alametal2015}.

\paragraph{0159+723}
There are several objects in our sample where only local standard of rest ($z=0$) absorption lines were detected. 0159+723 is an example of one of these. It shows no emission lines and has many $z=0$ absorption lines (Na I, Mg I, Ca II, etc.). Without the detection of any extragalactic lines, we cannot measure a redshift for this extragalactic radio source.

\paragraph{0253+033}
This object has many deep, broad absorption and/or emission lines. However, none of the typical AGN lines can explain all of the detected features. This object has a photometric redshift of $z=0.4$ in the Million Quasars (Milliquas) catalog\footnote{http://quasars.org/milliquas.htm} \citep{Flesch2015}, but we are unable to confirm this redshift with our spectrum. The line profiles are similar to those seen in broad absorption line (BAL) quasars (e.g., \citealp{Hazardetal1984}; \citealp{Foltzetal1987}; \citealp{Weymannetal1991}), suggesting that some of the lines may be blended and are confusing the identification of line centers. This object is classified as a BL Lacertae object (BL Lac) by \cite{DAbruscoetal2014} based on its WISE colors, although the strength of the detected lines in our spectrum casts this classification into doubt. Further study of this object is needed to measure a redshift.

\paragraph{0300+470}
No emission or absorption lines were detected for this object. It has a tentative redshift of $z=0.475$ in NED from \cite{Hughesetal1992}, but the original paper shows that this object is a BL Lac who was assigned the mean redshift of all known BL Lacs in lieu of an actual redshift. Without any detected lines or a measured redshift from previous studies, we cannot assign this object a redshift.

\paragraph{0302+625}
This object's spectrum only contains Galactic absorption lines. A few of the detected absorption lines are marked in the plot -- notably, the Balmer series. We are unable to measure a redshift for this object because no extragalactic lines are detected.

\paragraph{0420+417}
This object shows no significant emission or absorption lines. Additionally, the continuum is not detected on the blue CCD and has a low SNR on the red CCD. Because of these two factors, we selected this object for additional observation on Gemini North. These observations are discussed in Section \ref{gemini}.

\paragraph{0422+004}
In the literature, \cite{Shawetal2013} find $z=0.268$ based on Ca II H \& K and G-band absorption lines (no uncertainty is given on the redshift). We do not detect these lines -- or any other significant lines -- in our spectrum, but the continuum SNR is much lower. Based on the spectrum provided in \cite{Shawetal2013}, we accept their redshift of $z=0.268$ for inclusion in our catalog.

\paragraph{0426+245}
There are three emission lines detected for this object. From these lines, we find two possible redshifts, both of which only fit two of the three lines. The first option is $z=0.59$ based on the [O III] doublet. The scatter for this option is $\sim 4 $ \angstrom ($\sigma_z=0.001$). The second option is $z=3.12$ based on C IV and C III] (using the first half of the doublet at 7900 \text{\normalfont\AA}). The scatter for this option is $\sim 7 $ \angstrom ($\sigma_z = 0.003$). Because neither of these redshifts can explain all three emission lines, we cannot assign a redshift to this object. A higher SNR spectrum is needed to detect additional lines and determine the correct redshift. 

\paragraph{0459+135}
We find $z=0.35$ using four lines -- H$\beta$, He II, H$\delta$, and [Ne V]. It is possible that H$\alpha$ is also present, but lies in the noisy red end of the spectrum at $\approx 8860 $ \text{\normalfont\AA}. There is an overall scatter of $\sim 20 - 40 $ \angstrom in the line identifications at $z=0.35$. This equates to a redshift uncertainty of $\sigma_z = 0.01$.

\paragraph{0529+483}
This object is a blazar and has a redshift of $z=1.162$ from \cite{Halpernetal2003}. This redshift is based on a single line detection, which the authors identified as Mg II (2798 \text{\normalfont\AA}). No other information is given about their redshift determination, but it is likely they assumed this line is Mg II because it is one of the most common optical emission line seen in blazars. We also detect this line in our spectrum, but no other lines are significantly detected. Without any other lines, we are unable to measure a statistically viable redshift, despite the conclusions of \cite{Halpernetal2003}. Their redshift determination, although possible, should be regarded as tentative and is therefore not included in our catalog. 

\paragraph{1013+127}
This object has a redshift in NED of $z=0.463$, but no references are given. An additional literature search revealed no potential sources for the redshift identification. A spectrum is available in SDSS \citep{Alametal2015}, from which their algorithms measured a tentative redshift of $z=3.0140 \pm 0.0005$. However, examination of the SDSS spectrum shows no clear emission or absorption lines on which to base their measurement. There are also no clear lines in our APO spectrum, nor any indication of non-significant lines that support either reported redshift. Therefore, we conclude that no reliable redshift can be determined for this object from the available data.

\paragraph{1147+245}
This object has a redshift in NED of $z=0.2$, but no references are given. There is also a spectrum in SDSS with a redshift of $z=3.9 \pm 9.0$ \citep{Alametal2015}. The SDSS spectrum shows no statistically significant lines, which, combined with the large redshift uncertainty, leads us to disregard this redshift. Based on our APO spectrum, we find a redshift of $z=0.209 \pm 0.001$ using a cluster of three lines identified as H$\alpha$ flanked by the [N II] doublet. This is in good agreement with the NED redshift. The large bump at $\sim 4000 $ \angstrom is an artifact of flux calibration that can be seen to a lesser extent in other spectra (e.g., 0529+483 and 1013+127).

\paragraph{1444+313}
This object's spectrum only contains Galactic absorption lines. A few of the detected absorption lines are marked in the plot. We cannot measure a redshift for this object without any detected extragalactic lines.

\paragraph{1506+591}
We find $z=0.310$ using seven AGN emission lines, shown in Figure \ref{APOspectra}. There is a scatter of $\sim 6- 20 $ \angstrom in the expected line locations. This equates to a redshift uncertainty of $\sigma_z = 0.004$.

\paragraph{1525+610}
We find $z=0.2456$ from four AGN emission lines. Only the long wavelength halves of the [N II] and [O III] doublets are visible, but it is assumed that the other half of each is lost in the surrounding noise. There is a scatter of $3 - 5 $ \angstrom, indicating $\sigma_z=0.0007$. The emission line at 7540 \angstrom is an artifact from background subtraction.

\paragraph{1717+178}
No significant emission lines are detected for this object. OCARS gives this object a ``tentative'' redshift of $z=0.137$ from an observation by \cite{Sowards-Emmerdetal2005}. This redshift is based on the Ca II H \& K doublet and Mg I absorptions lines. We do not detect any of these lines, but examination of the higher SNR spectrum in \cite{Sowards-Emmerdetal2005} confirms their redshift. Thus, we use $z=0.137$ in our catalog. No uncertainty was given in the source paper.

\paragraph{1754+155}
We find $z=2.06$ from O I, N IV], O III], and Fe II. There is a scatter of $\sim 50$ \text{\normalfont\AA}, indicating $\sigma_z=0.04$. This spectrum is also well fit by $z=0.05 \pm 0.02$ where the lines from blue to red are [O II], H $\gamma$, H $\beta$, and H $\alpha$. However, images of this object in the National Geographic Society - Palomar Observatory Sky Atlas (POSS-I), AllWISE \citep{Cutrietal2013}, and VLA observations at 1.45 and 43.3 GHz\footnote{http://archive.nrao.edu/nvas/} all show a point-like object. A quasar at $z=0.05$ should be near enough for the host galaxy to be resolved in at least one of these wavelength regimes. Thus, we conclude that $z=2.06 \pm 0.04$ is the most likely redshift.

\paragraph{1823+689}
We find $z=2.143$ from Ly $\alpha$ and C IV emission lines. The sudden decrease in continuum flux leftward of Ly $\alpha$ may indicate the beginning of the Lyman $\alpha$ forest, but it is too close to the edge of the spectrum (where the CCD sensitivity also decreases) to make a definitive statement. There is a scatter of 2 \angstrom between the two lines, which equates to a redshift uncertainty of $\sigma_z = 0.001$.

\paragraph{2013+163}
This object's spectrum only contains Galactic absorption lines. A few of the detected absorption lines are marked in the plot. We cannot measure a redshift for this extragalactic object because no extragalactic lines are detected.

\paragraph{2021+317}
This object shows no significant emission or absorption lines. Additionally, the continuum is not detected on the blue CCD and has a low SNR on the red CCD. Because of these two factors, this object was selected for additional observation on Gemini North. These observations are discussed in Section \ref{gemini}.

\paragraph{2051+745}
We find $z=0.92$ from Fe II and [O II]. The Fe II feature is very broad (FWHM $\sim 100$ \text{\normalfont\AA}) and has no clear peak. This is a good fit for the large group of Fe II emission lines present at $\sim 2600$ \angstrom (rest wavelength) for some quasars. Because the Fe II feature is several blended lines, we are unable to use the line scatter to estimate the redshift uncertainty. Instead, we use the FWHM of the [O II] feature -- $\sim 50$ \text{\normalfont\AA} -- to estimate a redshift uncertainty of $\sigma_z = 0.01$.

\paragraph{2225+003}
This object shows no significant absorption or emission lines. It has a flagged redshift of $z=3.823 \pm 0.001$ as measured by the SDSS BOSS survey \citep{Dawsonetal2013}. This redshift is based primarily on Ly $\alpha$ in absorption and C IV, He II, and C III] in emission, none of which are strong lines in the SDSS spectrum. In the APO spectrum, the absorption feature is non-significantly detected and none of the other lines are visible. With only one low SNR absorption feature and a weak SDSS redshift, we cannot assign this source a reliable redshift. Note that the strong emission line at 5581 \angstrom is a sky line.

\paragraph{2315+032}
This object shows no significant lines and the continuum was not detected on the blue CCD. We are unable to measure a redshift.

\paragraph{2319+444}
There is only one significant emission line in this spectrum. The object has a redshift in NED of $z=1.31$ \citep{XuHan2014}, but the original paper gives no previous source or additional data to support this redshift. There are no common AGN lines that correspond with the detected line at this redshift, nor are there any other low SNR lines to support the redshift. Additionally, if we try other redshifts where the significant emission line corresponds with common AGN lines, there are no other low SNR lines to add validity to any of these redshifts. With only one reliable emission line, we cannot assign a redshift to this object. 

\subsection{Gemini North}\label{gemini}
Several of the objects observed at Apace Point Observatory showed weak continua and no significant emission or absorption lines. We chose two of these objects (0420+417 and 2021+317) for additional observations with the Gemini Multi-Object Spectrograph -- North (GMOS-N; \citealp{Hooketal2004}) at Gemini North Observatory\footnote{Program ID GN-2016B-Q-81}. 2021+317 was observed in 2016 on June 26 and June 28, while 0420+417 was observed in 2016 on November 8 and November 26. We observed both objects with a $1''$ slit and two grating setups: the B600 grating with no filter and the R400 grating with a OG515 filter to prevent order overlap (gratings centered at 4880 and 7640 \text{\normalfont\AA}, respectively). We used $2 \times 2$ binning, which, combined with the grating setup, resulted in final spectra with linear dispersions of 0.9 \angstrom pixel$^{-1}$ and 1.4 \angstrom pixel$^{-1}$ and wavelength coverage of 3500 \angstrom $-$ 6250 \angstrom and 6250 \angstrom $-$ 9000 \text{\normalfont\AA}, respectively. Sixteen exposures of 632 seconds each were taken for both objects -- eight for each grating. 

To reduce the Gemini North observations, we used the Gemini IRAF package; a package written specifically for reducing Gemini observations\footnote{http://www.gemini.edu/sciops/data-and-results/processing-software}. The reduction process is the same as for the images taken at Apache Point Observatory -- the images were overscan corrected, trimmed, bias subtracted, flat fielded, wavelength calibrated, and background subtracted. Then we extracted a one-dimensional science spectrum from each reduced image, median stacked the extracted spectra, and flux calibrated the resulting final spectra. 

\begin{figure}
\centerline{\includegraphics[width=.5\textwidth, trim=.6cm 0cm 2cm 1.3cm,clip]{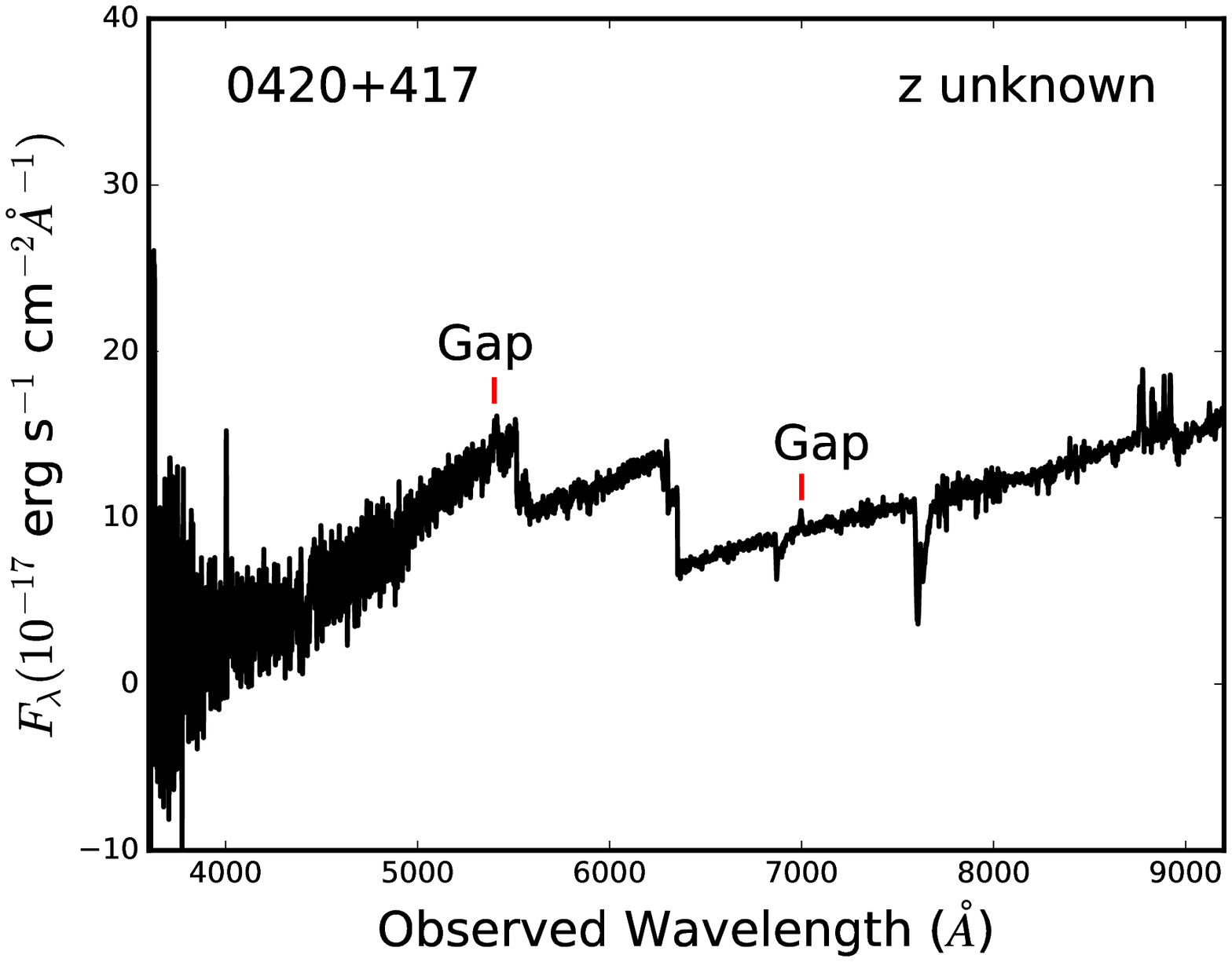}\includegraphics[width=.5\textwidth, trim=.6cm 0cm 2cm 1.3cm,clip]{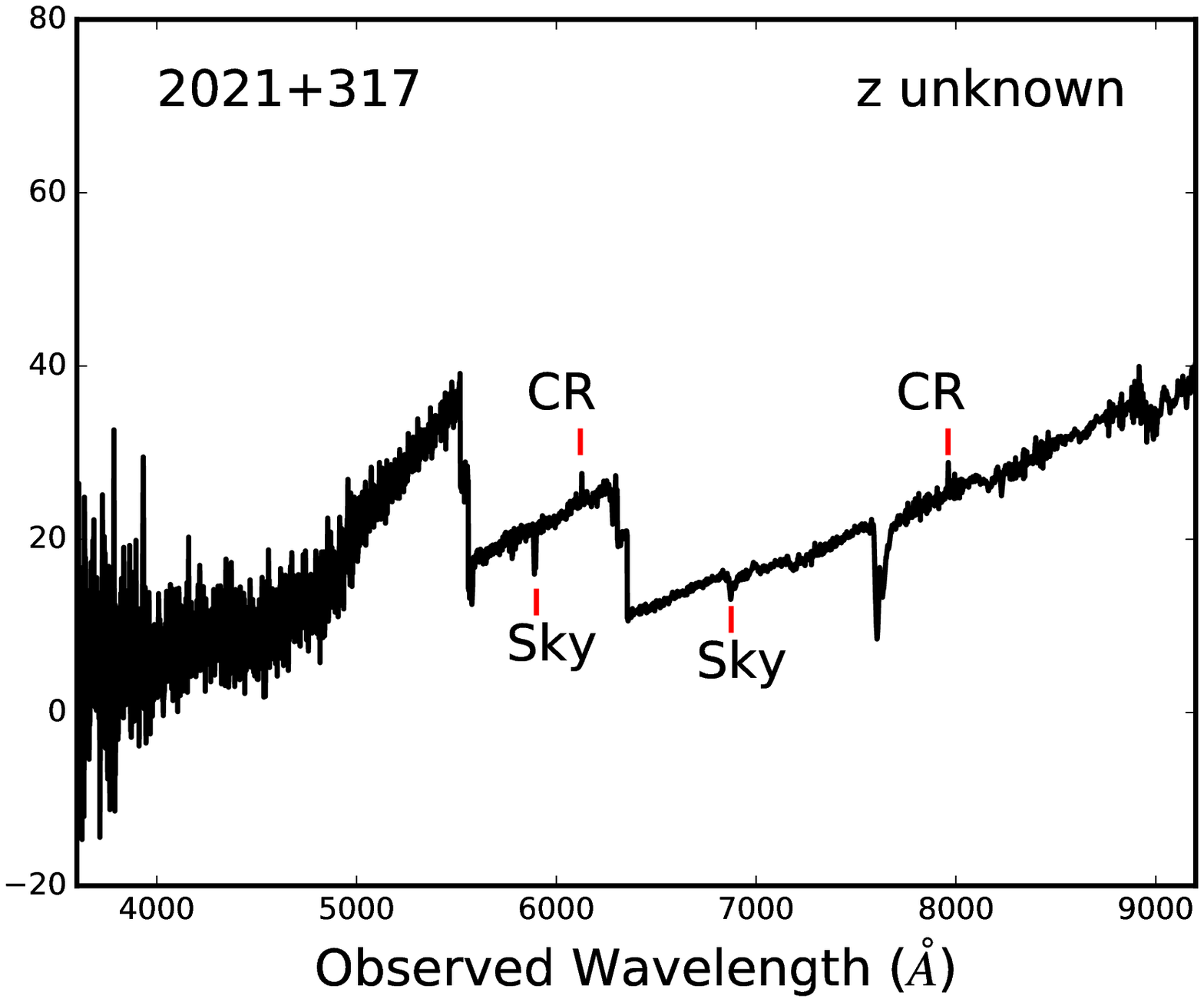}}
\caption{Flux calibrated spectra obtained from Gemini North. Jumps in the continua are not real and are due to inexact flux calibration. Several sky absorption features, cosmic rays, and artifacts from the chip gaps are marked in the plots. The night sky absorption feature at $\sim 7600$ \angstrom is also visible in both spectra. 
\\ \\
(The data used to create this figure are available in the online journal.)\label{geminispectra}}
\end{figure}

Figure \ref{geminispectra} shows the reduced spectra from the Gemini North observations. Neither object has any significant emission or absorption lines and no redshift was measured for either. The only visible lines in the spectra are due to sky absorption, cosmic rays, and artifacts from the chip gaps. Several of these lines are marked in the figure. Although the spectra are flux calibrated, the flux calibrator was observed on a different night and under different weather conditions than the targets, which were also observed on four separate nights. For precise flux calibration, the calibrator should be observed on the same night and in a similar part of the sky as the target to ensure that the atmosphere equally affects both spectra. Because this was not done, there are discontinuities in the final continua from the variations in atmospheric conditions on the four nights on which the targets were observed.

\begin{figure}
\centerline{\includegraphics[width=.5\textwidth, trim=.6cm 0cm 1.5cm 1.3cm,clip]{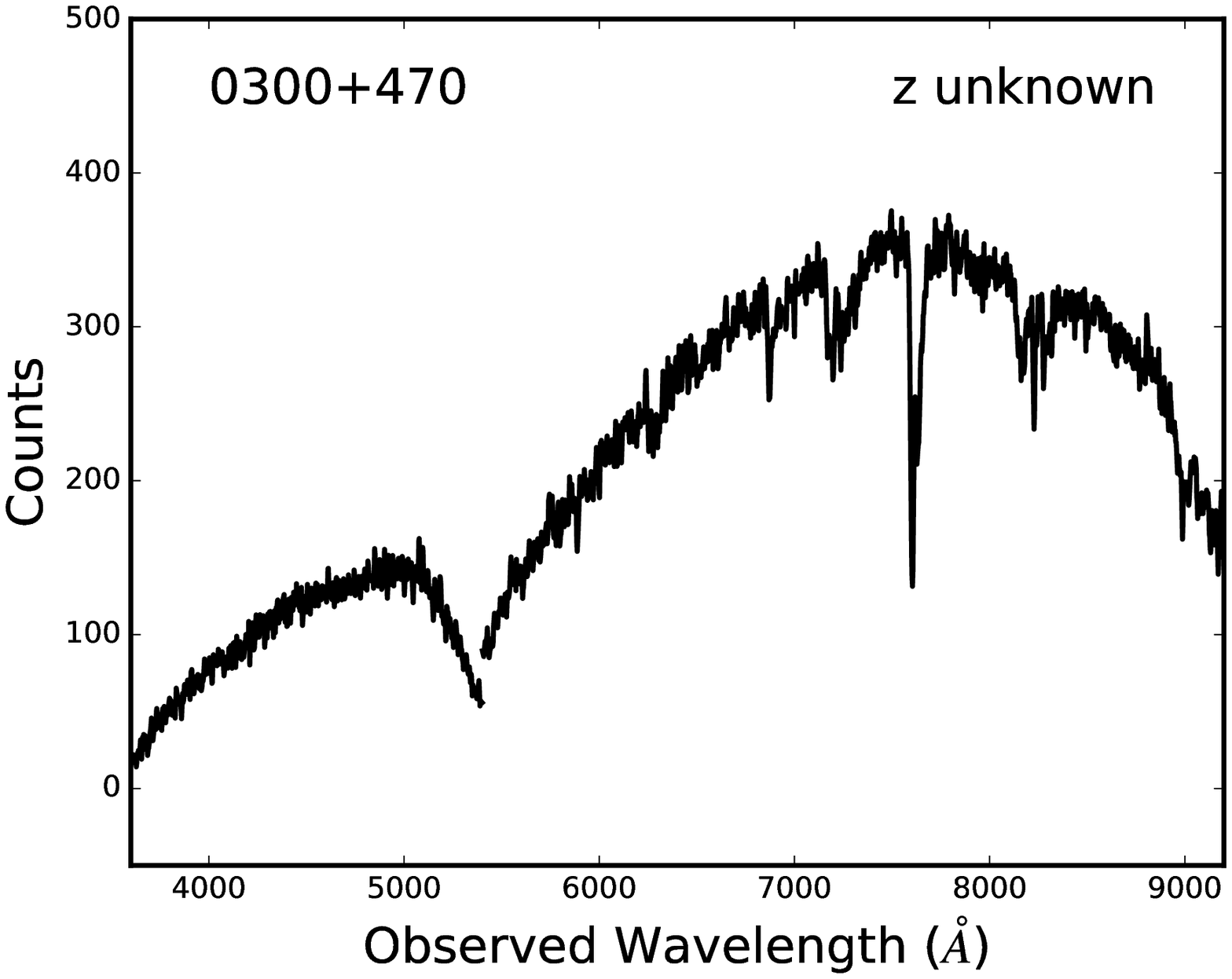}\includegraphics[width=.5\textwidth, trim=.6cm 0cm 1.5cm 1.3cm,clip]{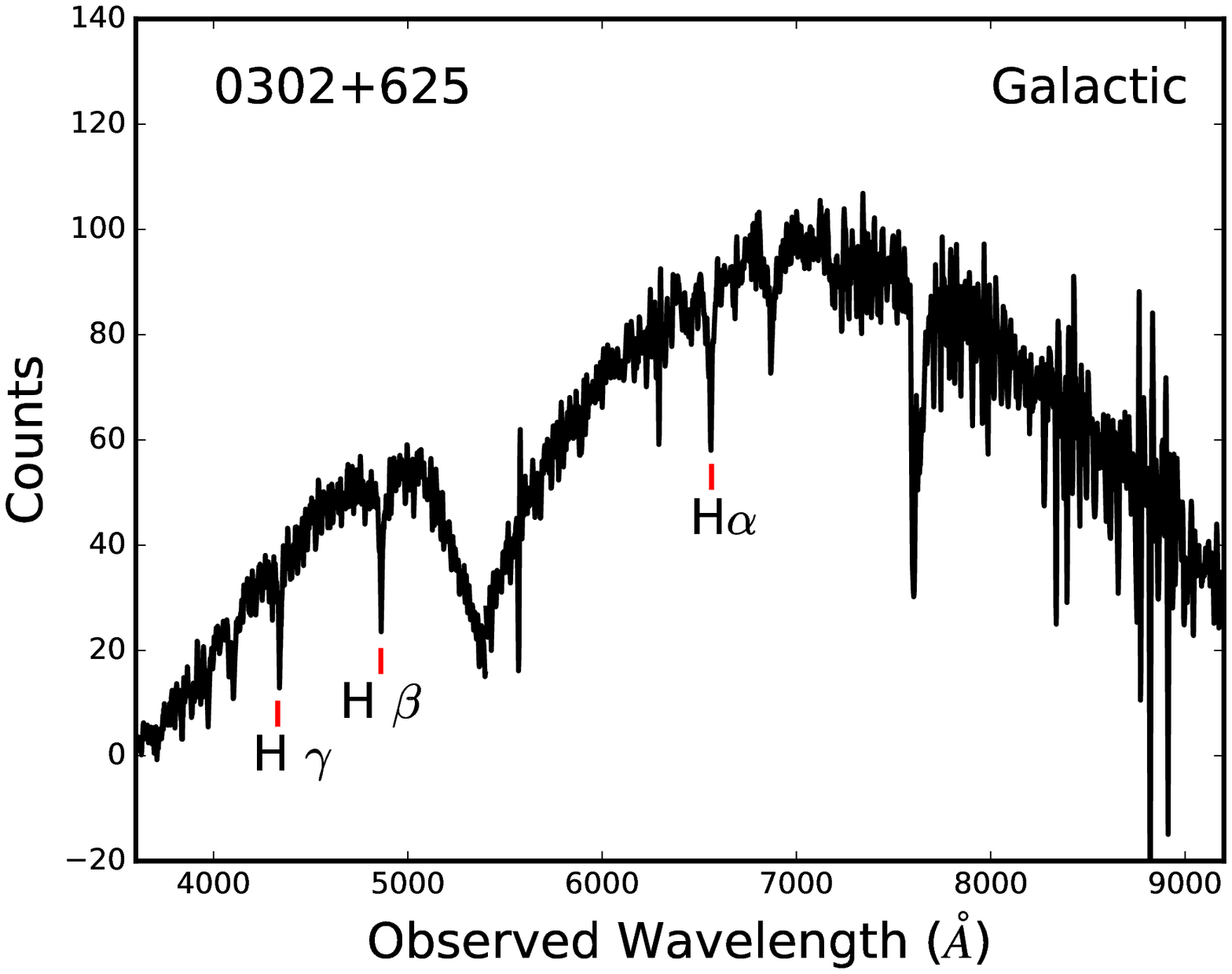}}
\vspace{0.01\textwidth}
\centerline{\includegraphics[width=.5\textwidth, trim=.6cm 0cm 1.5cm 1.3cm,clip]{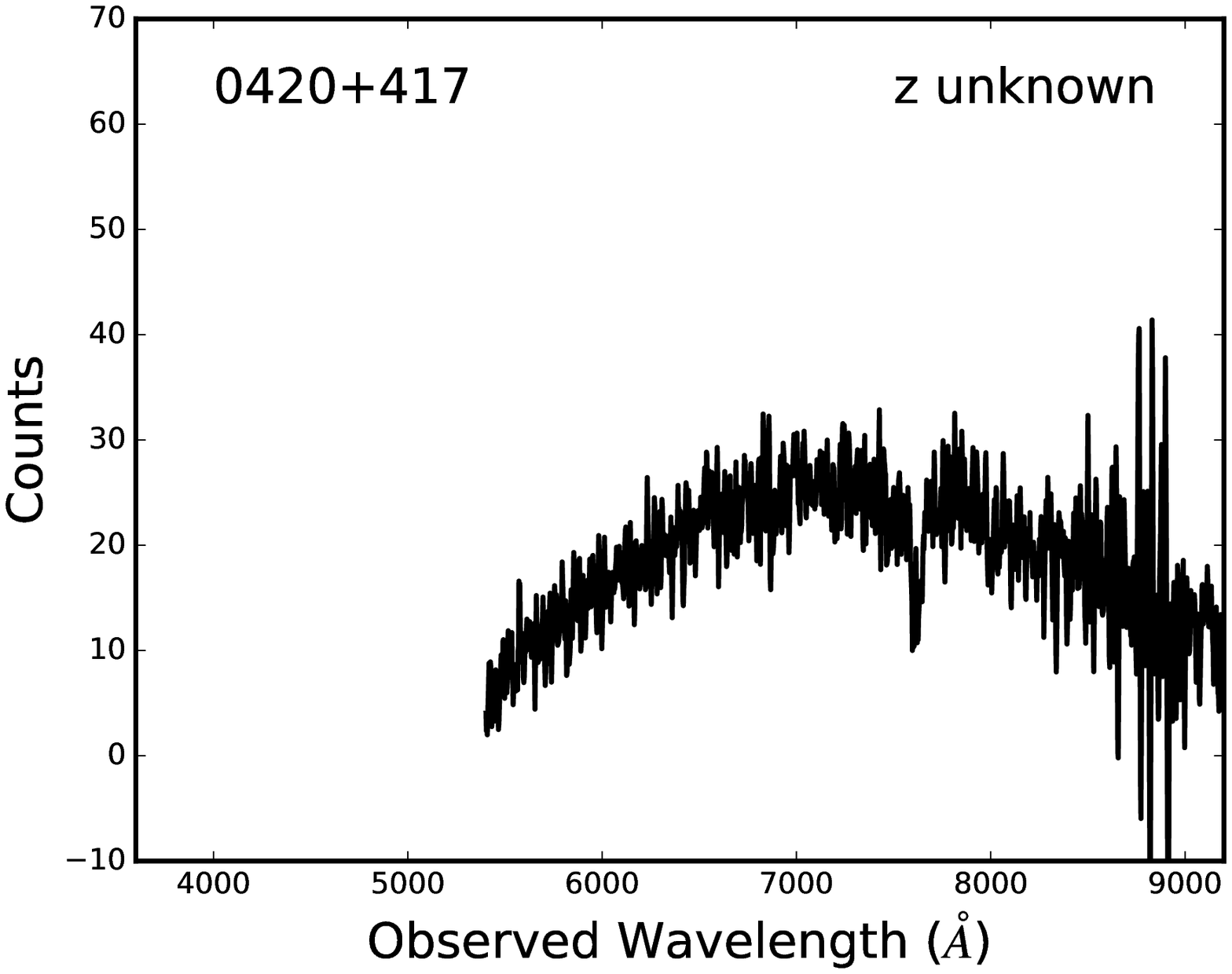}\includegraphics[width=.5\textwidth, trim=.2cm 0cm 1.5cm 1.3cm,clip]{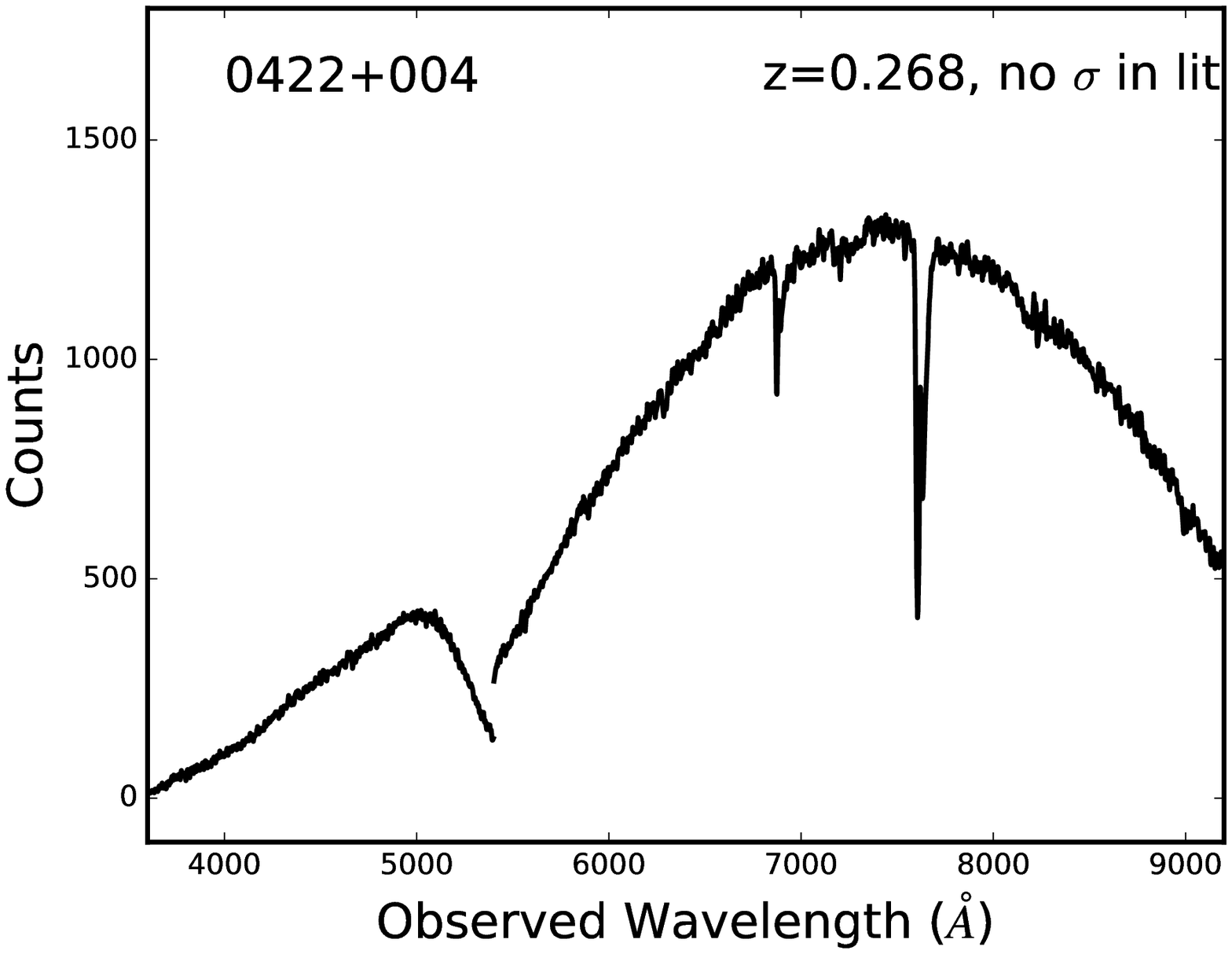}}
\vspace{0.01\textwidth}
\centerline{\includegraphics[width=.5\textwidth, trim=.6cm 0cm 1.5cm 1.3cm,clip]{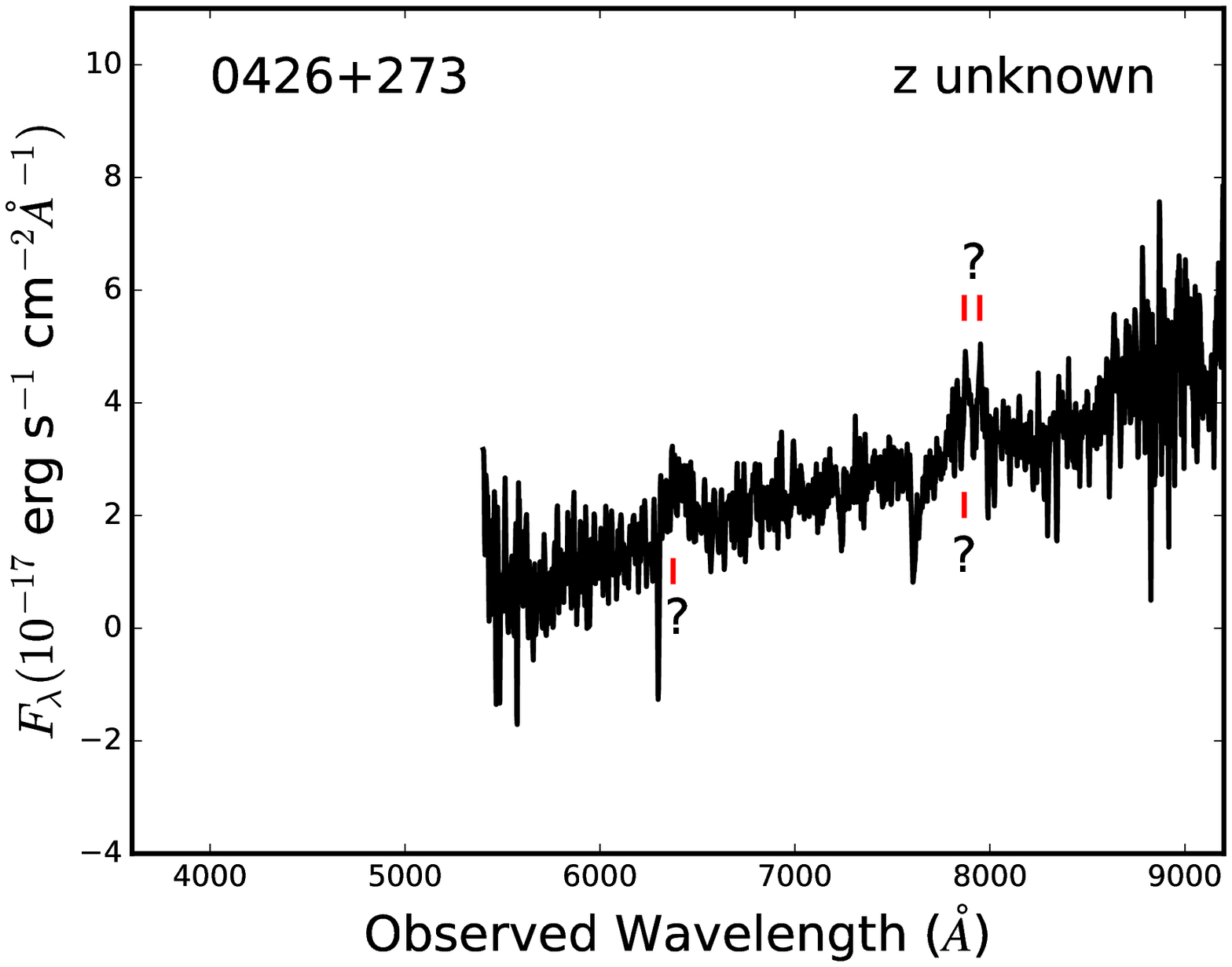}\includegraphics[width=.5\textwidth, trim=.6cm 0cm 1.5cm 1.3cm,clip]{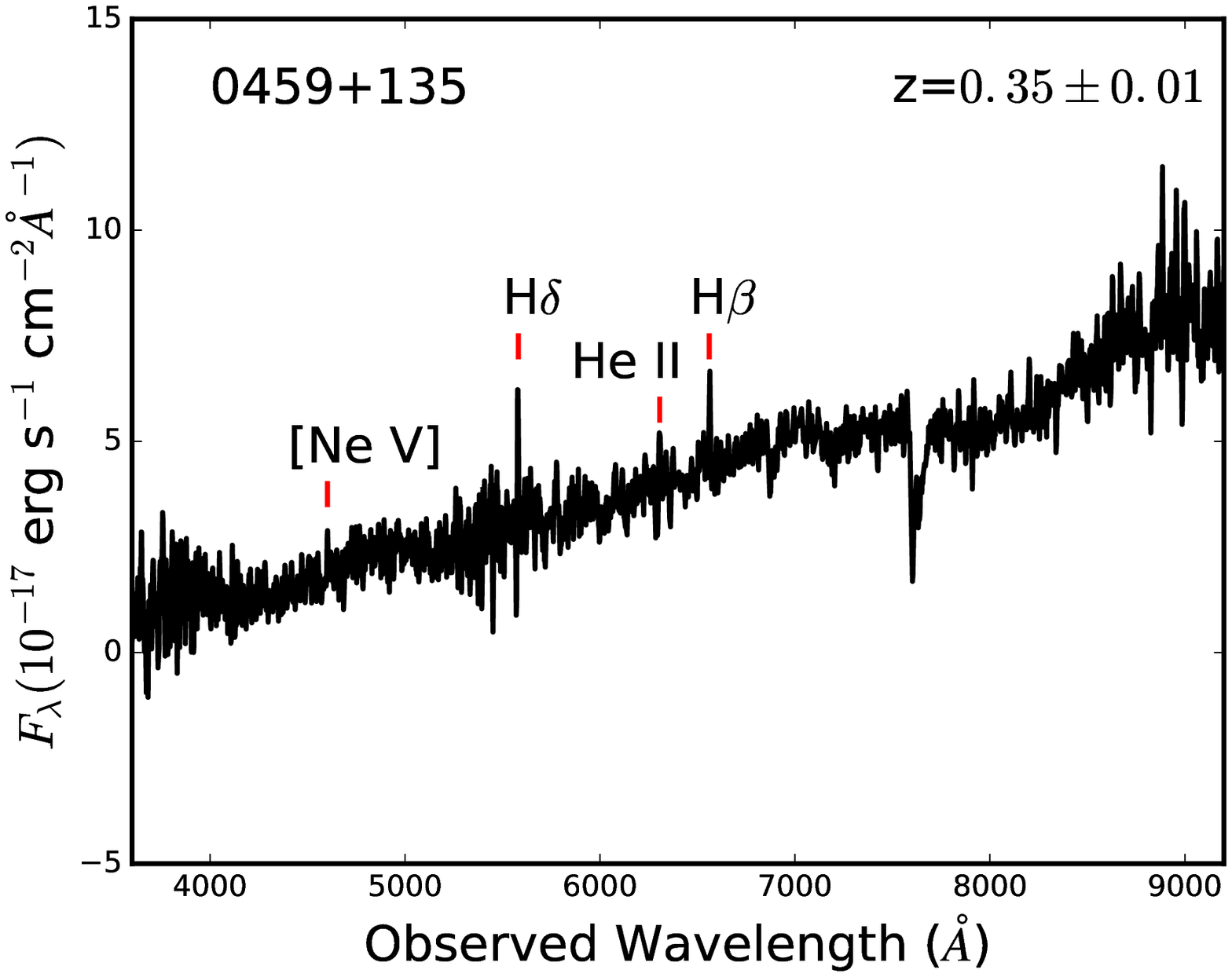}}
\end{figure}
\begin{figure}
\centerline{\includegraphics[width=.5\textwidth, trim=.6cm 0cm 1.5cm 1.3cm,clip]{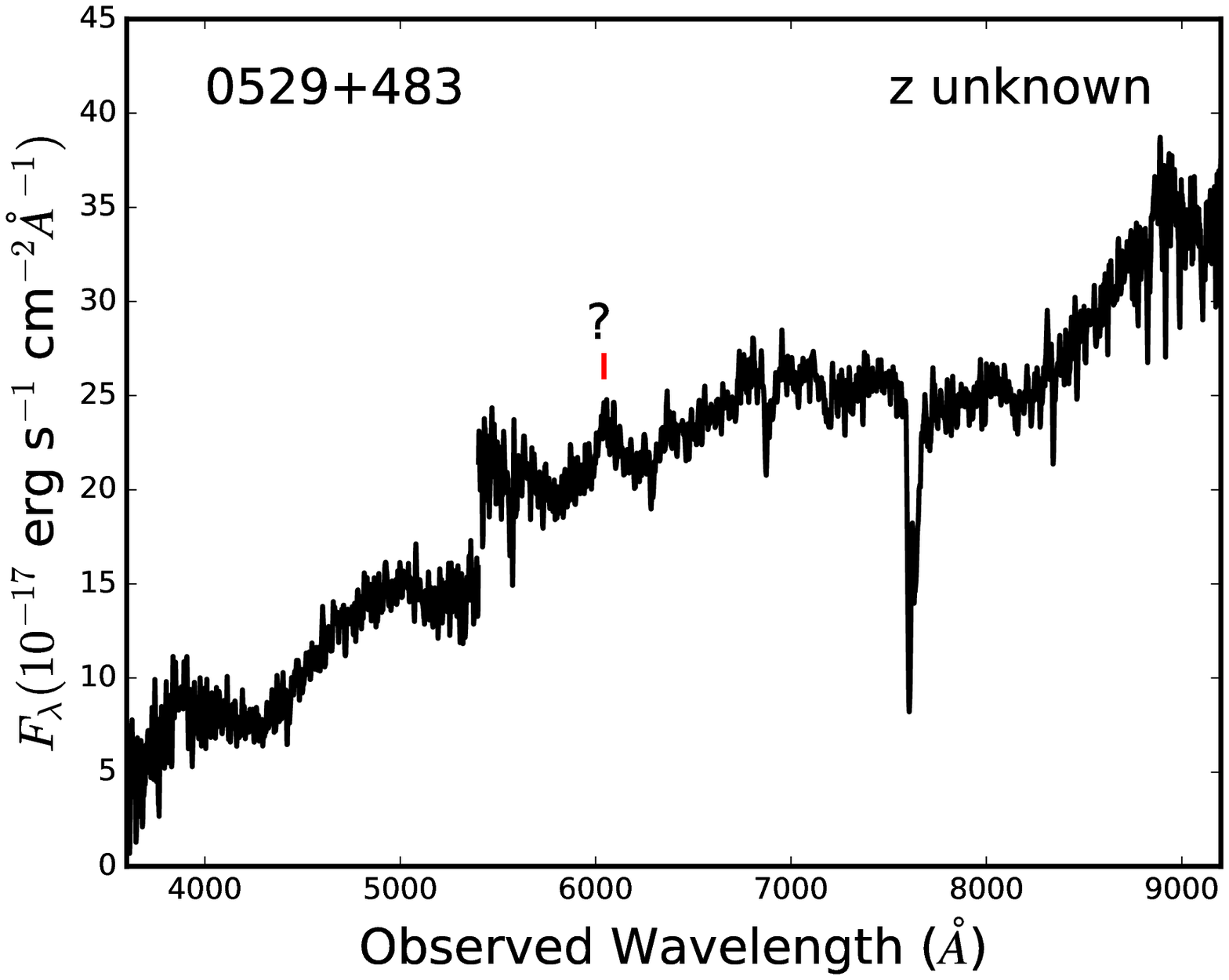}\includegraphics[width=.5\textwidth, trim=.6cm 0cm 1.5cm 1.3cm,clip]{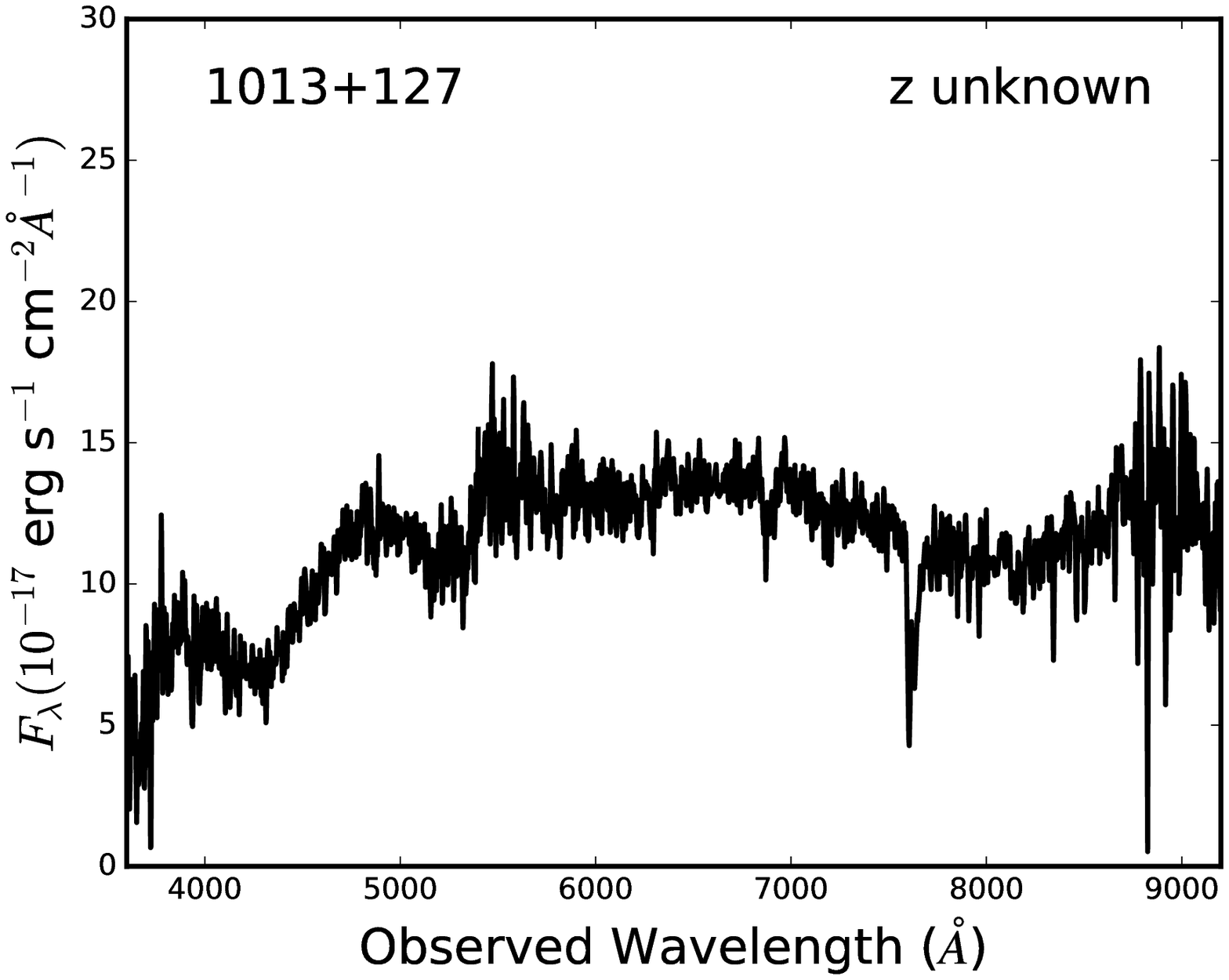}}
\vspace{0.01\textwidth}
\centerline{\includegraphics[width=.5\textwidth, trim=.6cm 0cm 1.5cm 1.3cm,clip]{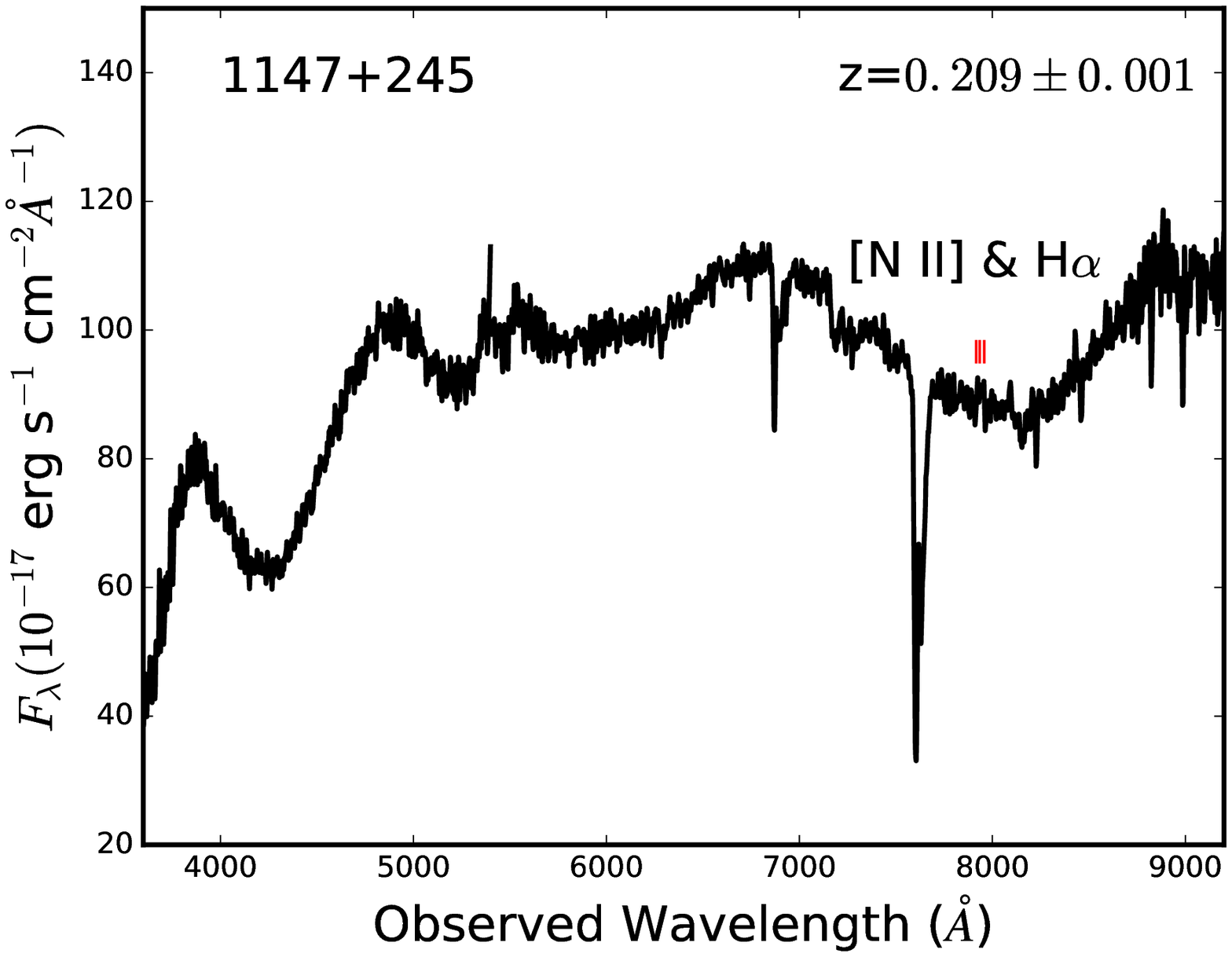}\includegraphics[width=.5\textwidth, trim=.4cm 0cm 1.5cm 1.3cm,clip]{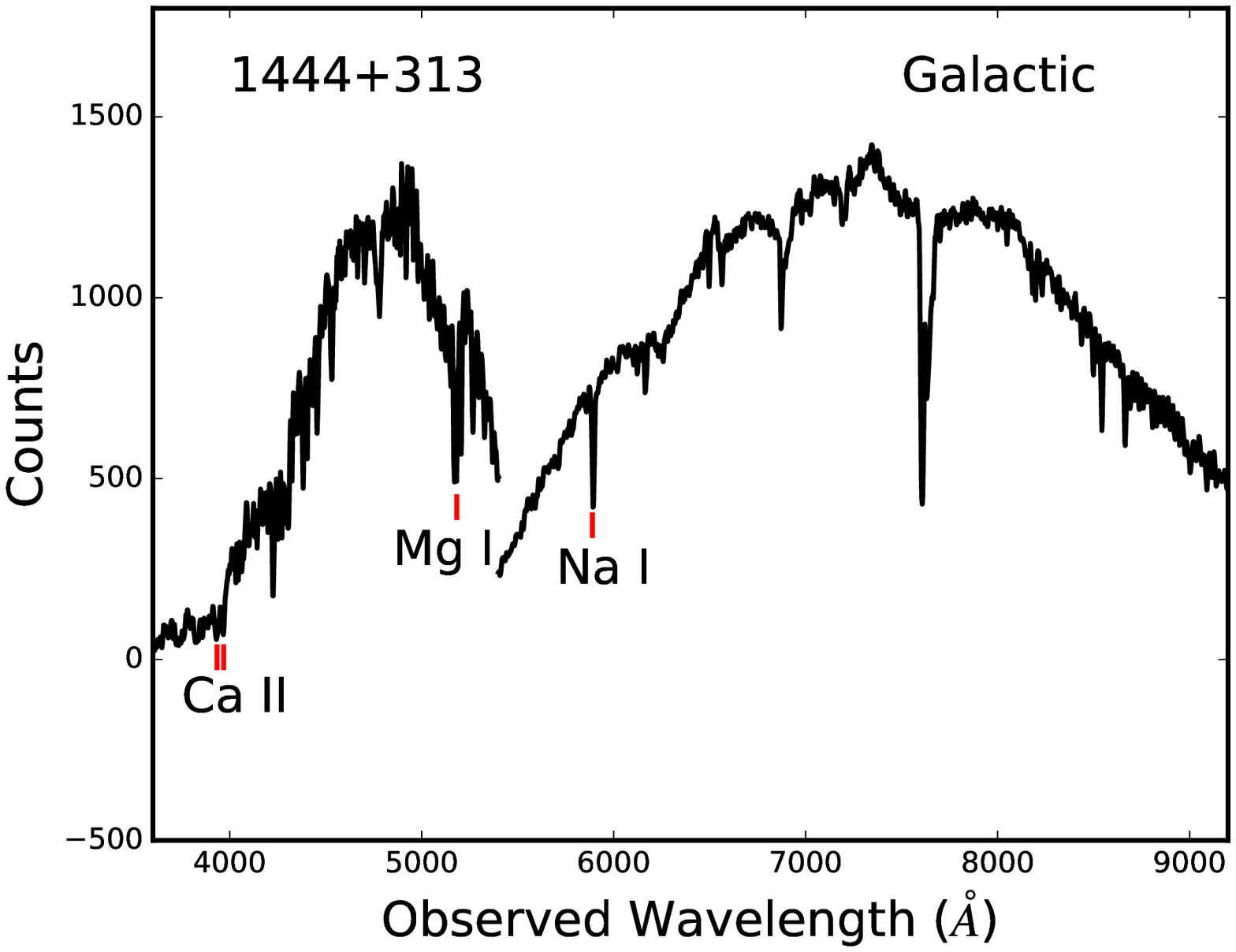}}
\vspace{0.01\textwidth}
\centerline{\includegraphics[width=.5\textwidth, trim=.6cm 0cm 1.5cm 1.3cm,clip]{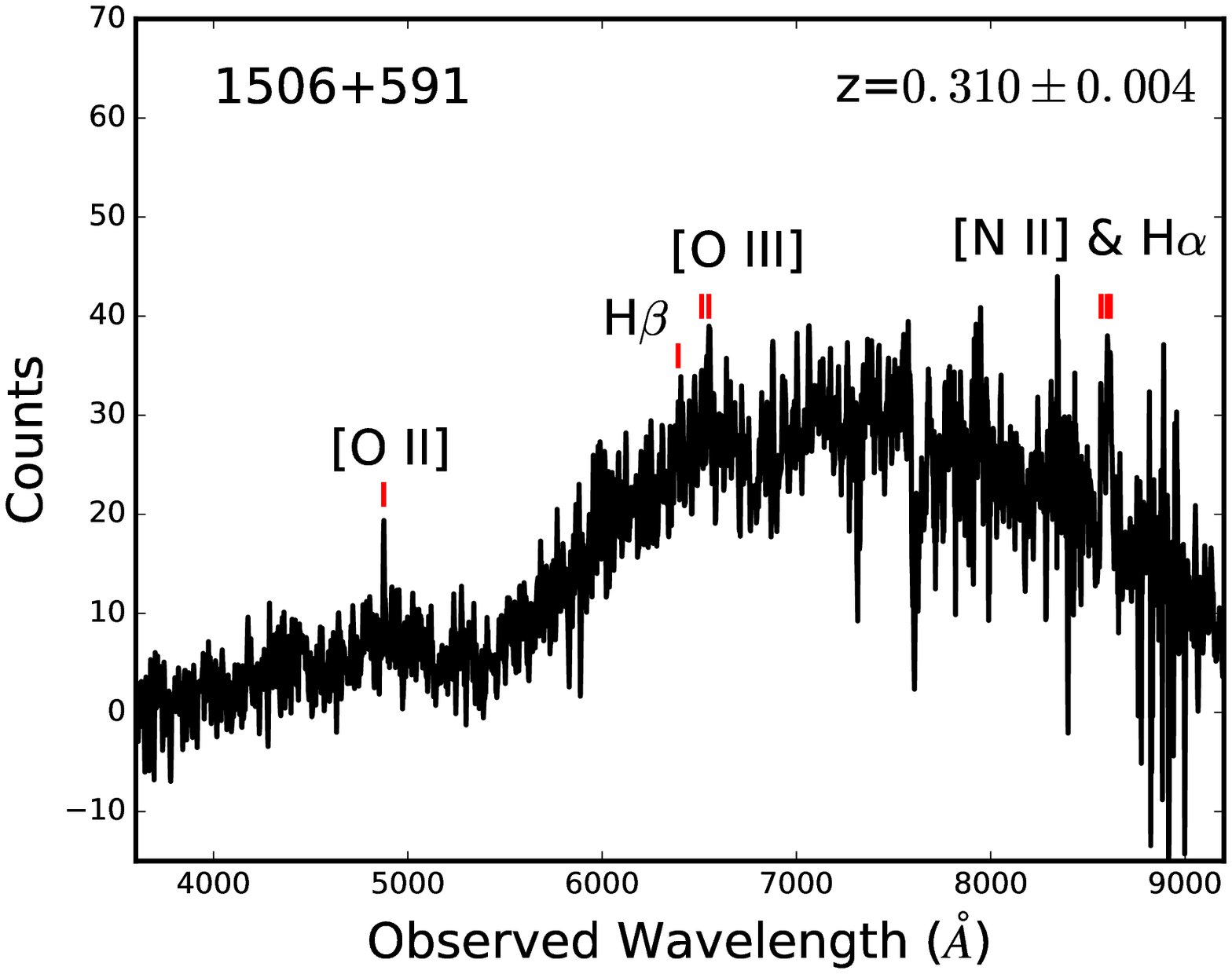}\includegraphics[width=.5\textwidth, trim=.6cm 0cm 1.5cm 1.3cm,clip]{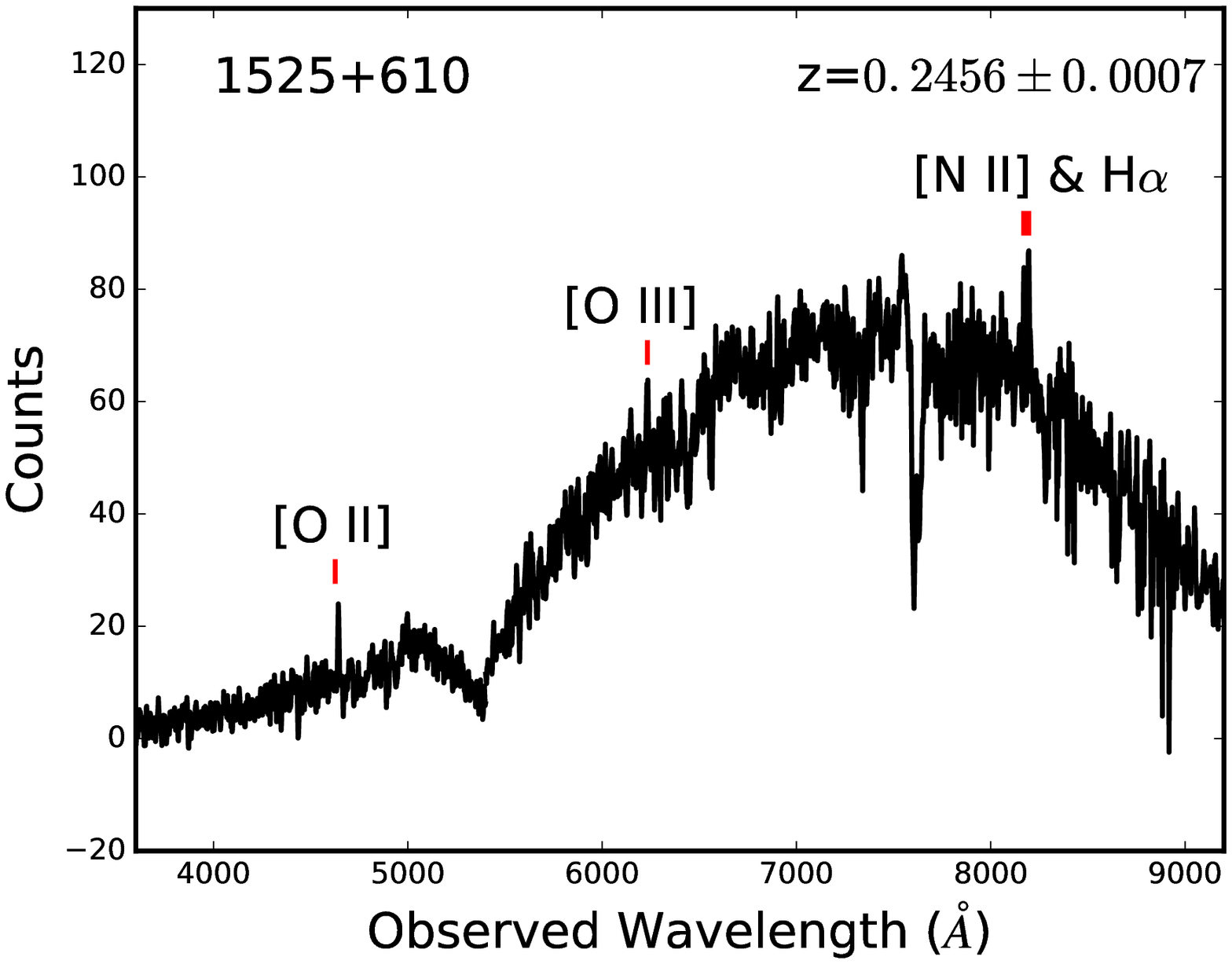}}
\end{figure}
\begin{figure} 
\centerline{\includegraphics[width=.5\textwidth, trim=.6cm 0cm 1.5cm 1.3cm,clip]{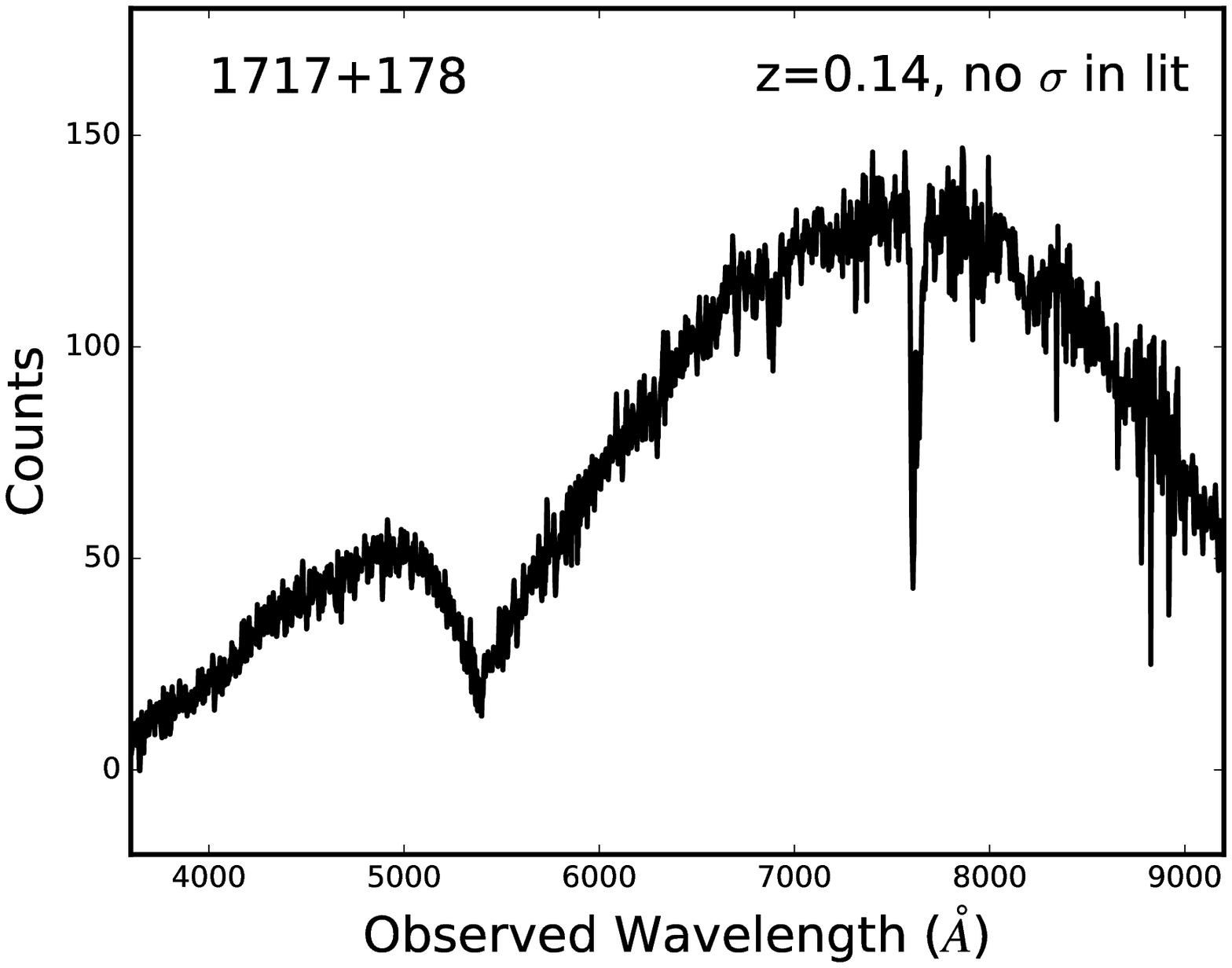}\includegraphics[width=.5\textwidth, trim=.6cm 0cm 1.5cm 1.3cm,clip]{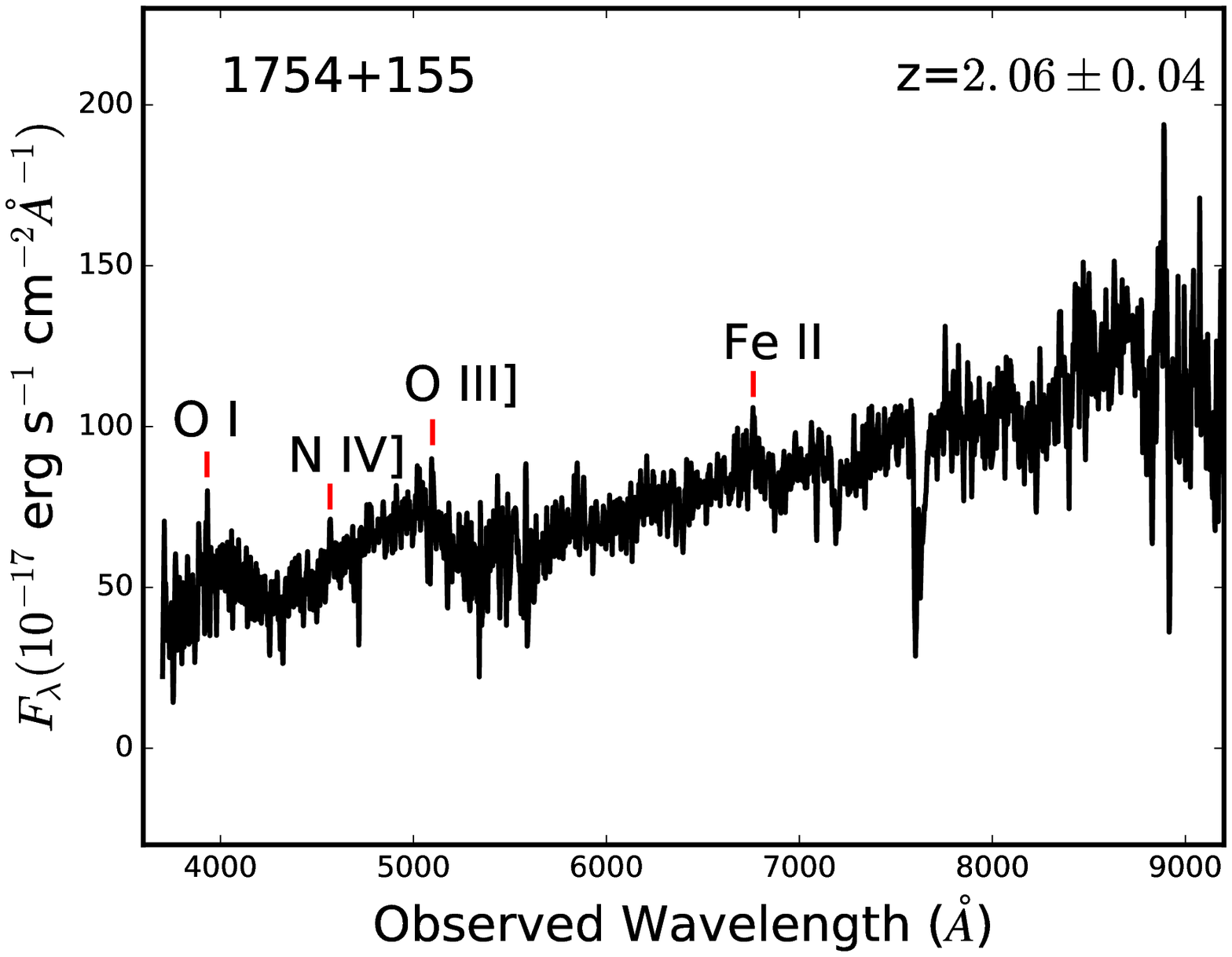}}
\vspace{0.01\textwidth}
\centerline{\includegraphics[width=.5\textwidth, trim=.6cm 0cm 1.5cm 1.3cm,clip]{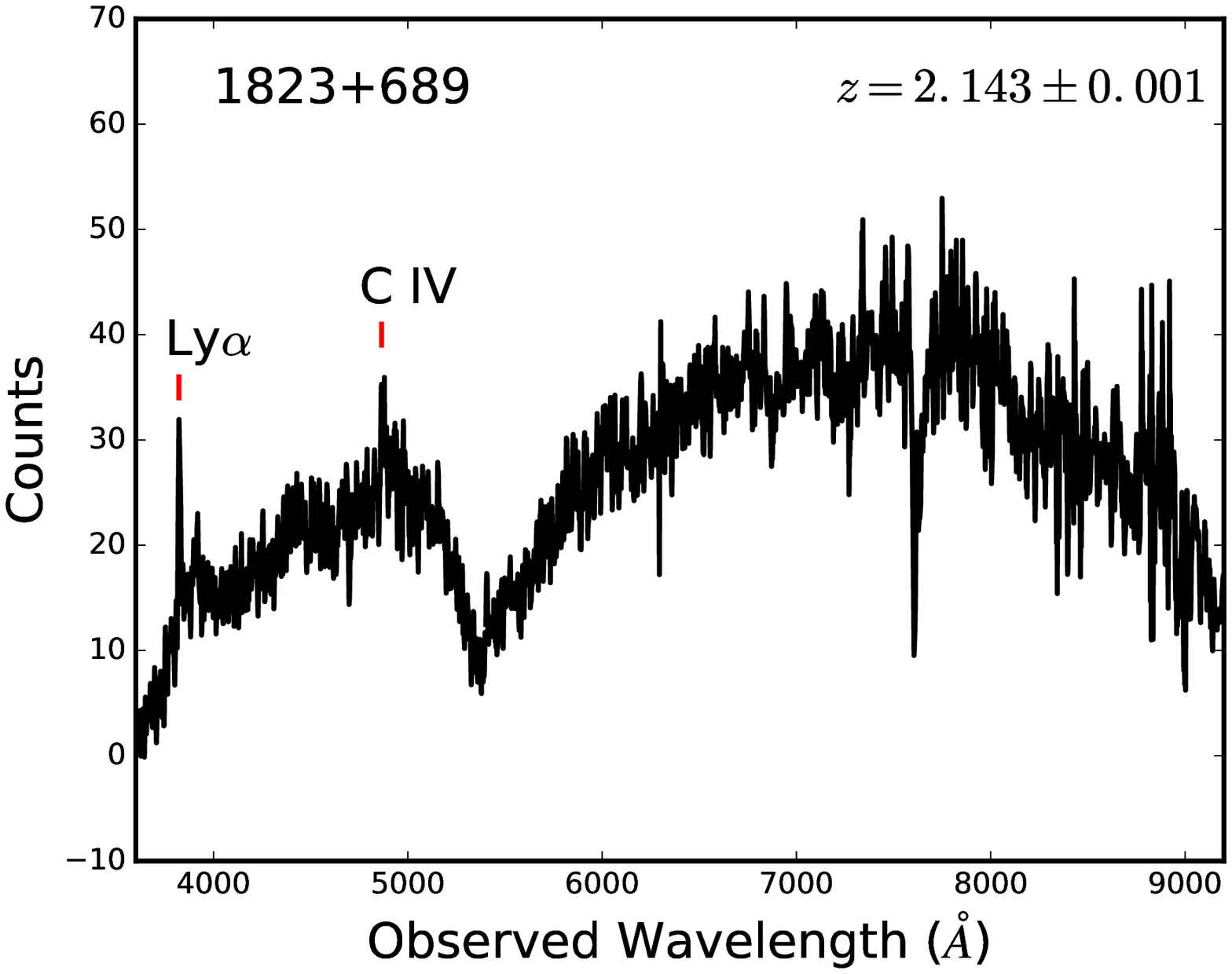}\includegraphics[width=.5\textwidth, trim=.6cm 0cm 1.5cm 1.3cm,clip]{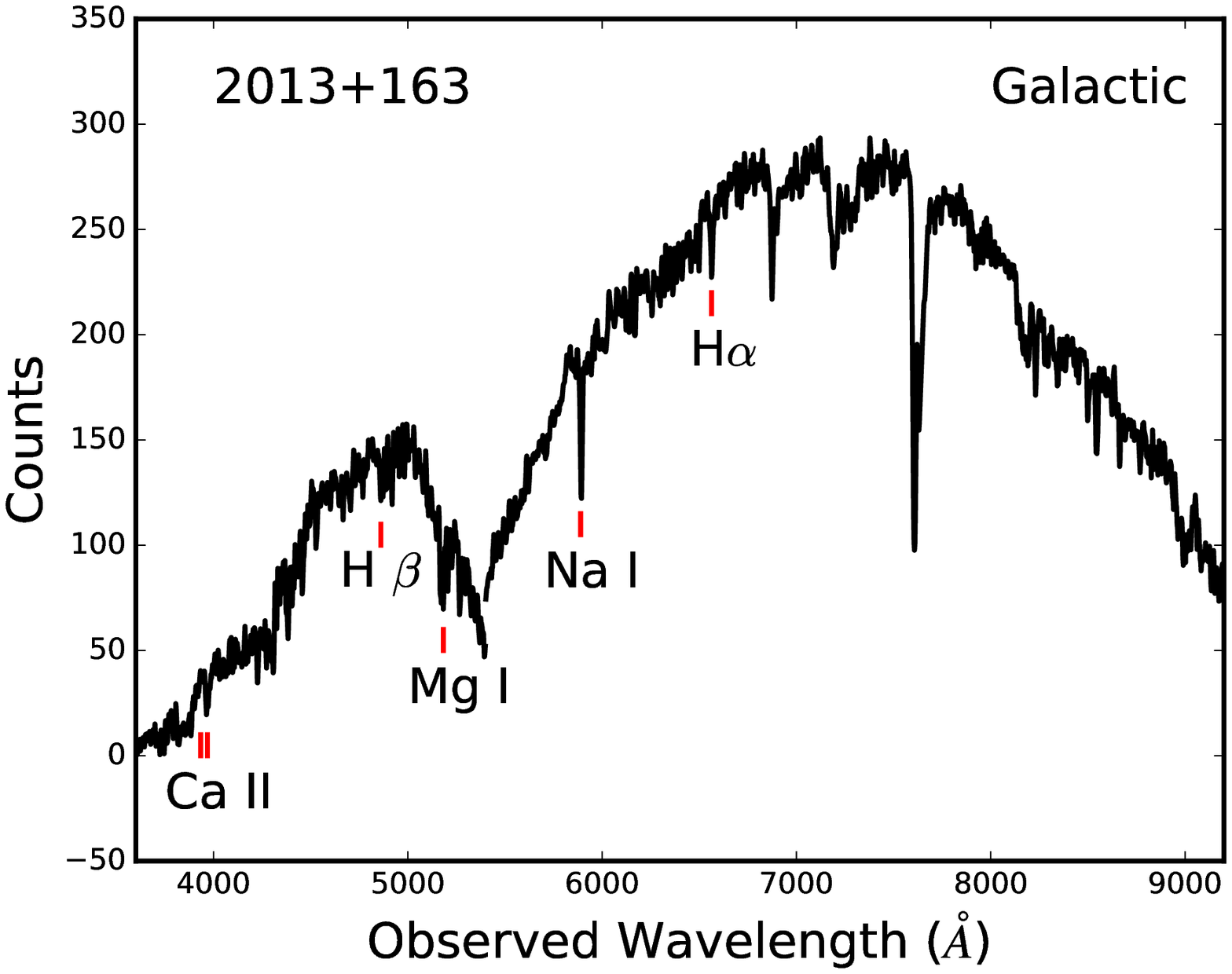}}
\vspace{0.01\textwidth}
\centerline{\includegraphics[width=.5\textwidth, trim=.6cm 0cm 1.5cm 1.3cm,clip]{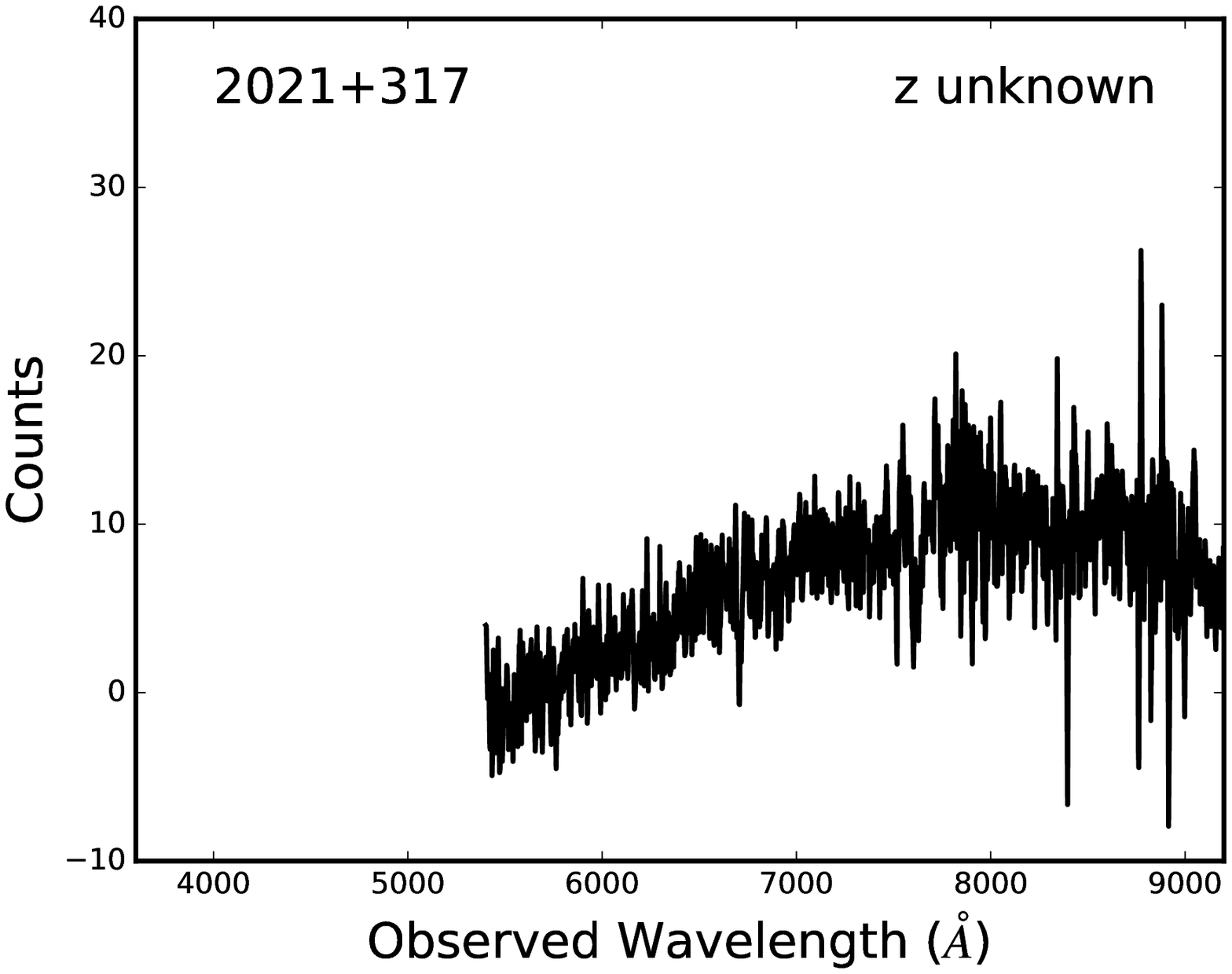}\includegraphics[width=.5\textwidth, trim=.6cm 0cm 1.5cm 1.3cm,clip]{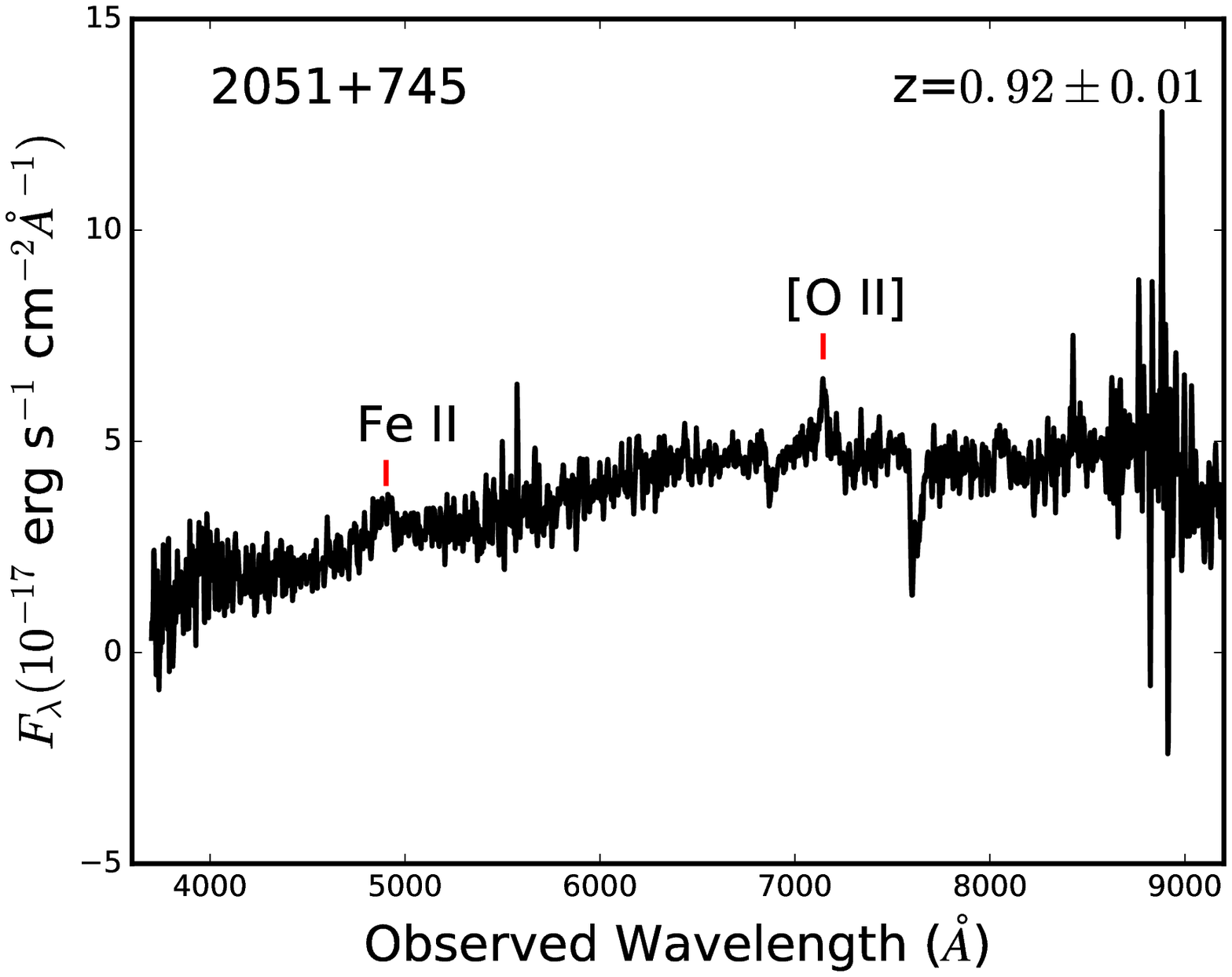}}
\end{figure}
\begin{figure}
\centerline{\includegraphics[width=.5\textwidth, trim=.6cm 0cm 1.5cm 1.3cm,clip]{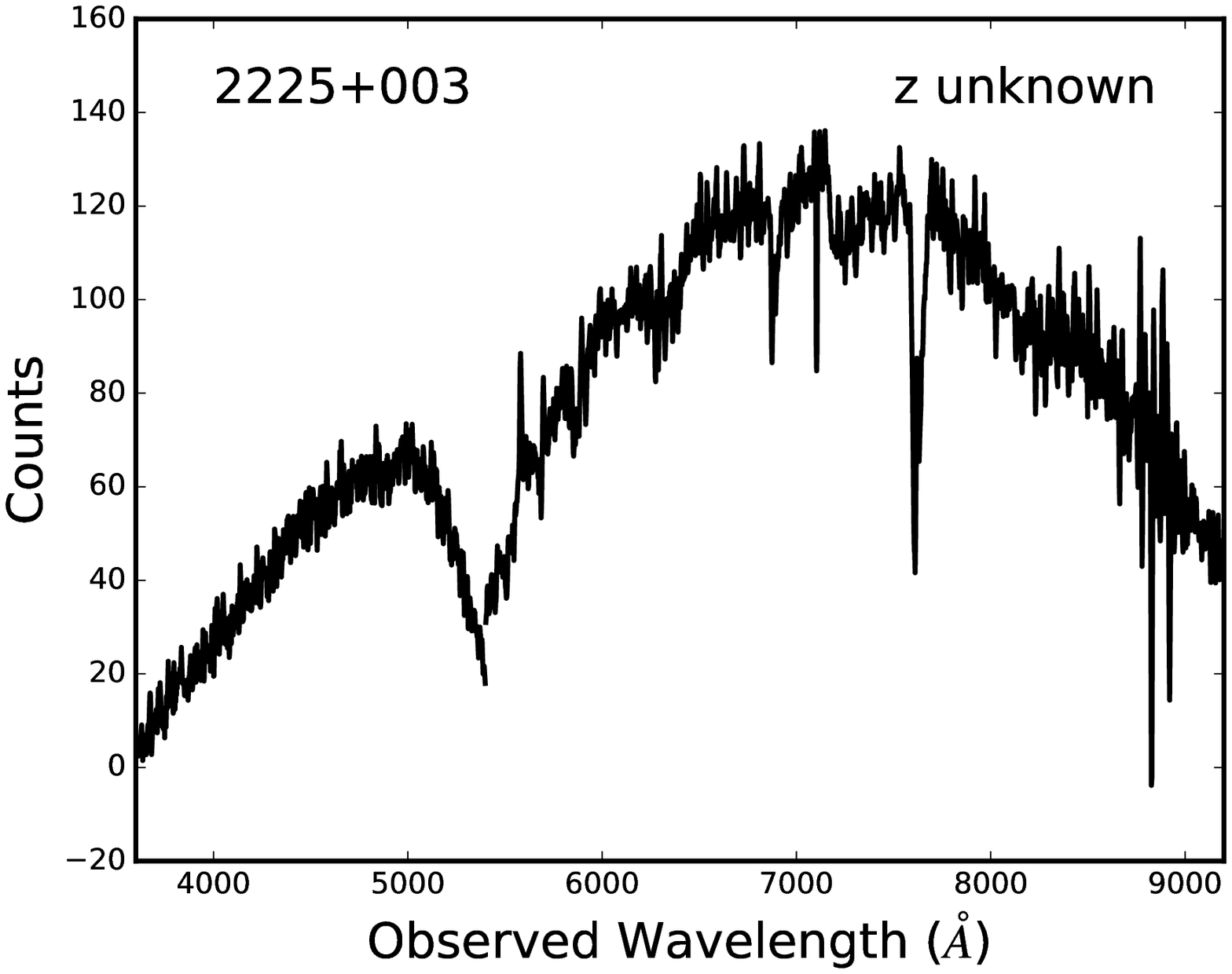}\includegraphics[width=.5\textwidth, trim=.6cm 0cm 1.5cm 1.3cm,clip]{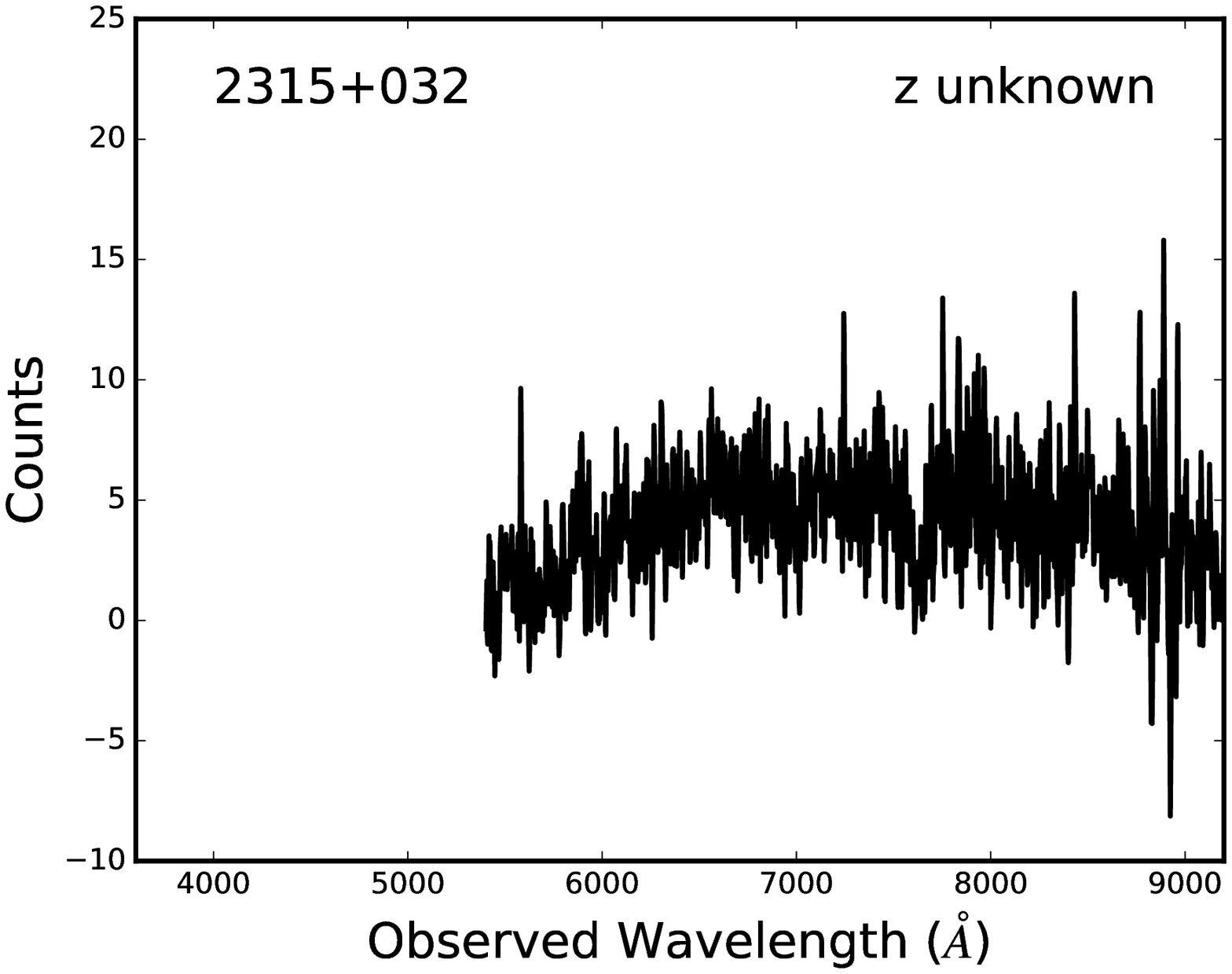}}
\vspace{0.01\textwidth}
\centerline{\includegraphics[width=.5\textwidth, trim=.2cm 0cm 1.5cm 1.3cm,clip]{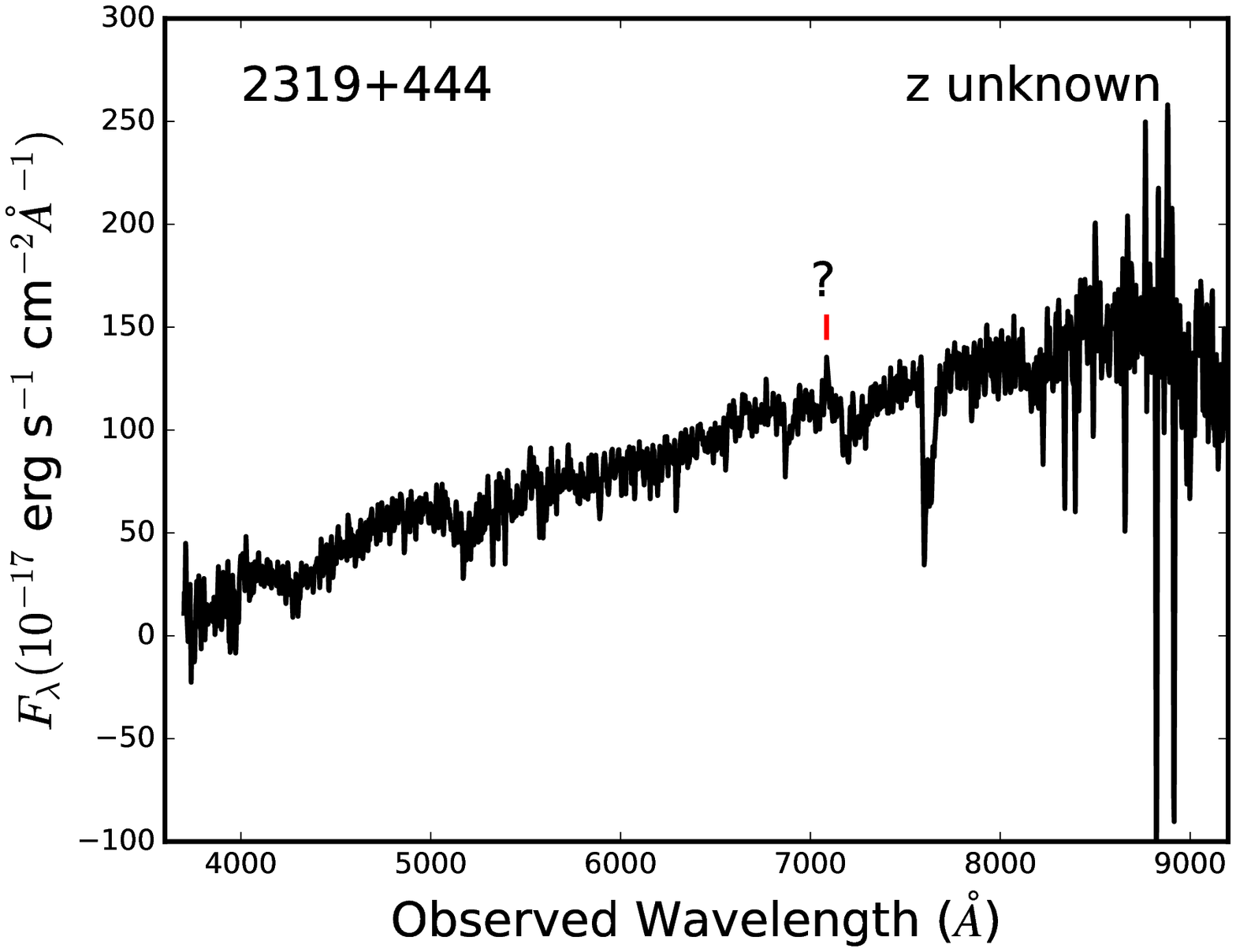}}
\caption{Calibrated, one-dimensional spectra obtained from APO. All spectra are smoothed by a 3 \angstrom wide boxcar. Some of the spectra lack flux calibration, as indicated by the y-axis label. In the spectra without flux calibration, there is a large dip at $\sim 5500$ \angstrom due to the sensitivity falloff at the edges of the red and blue CCDs. For objects where only the red half of the spectrum is plotted, the blue continuum was not significantly detected and could therefore not be extracted from the final image. Detected lines and the best-fit identifications are shown in each plot, along with their redshift. There are also two night sky absorption features visible in all plots at 6866 \angstrom and $\sim 7600$ \angstrom. For all of our redshifts, we determine the uncertainty based on the scatter in the line identifications. The two redshifts listed without uncertainties were obtained from the literature, where no uncertainties were provided. Objects labeled ``Galactic'' have only Galactic ($z=0$) lines detected. Without any detected extragalactic lines, we cannot measure redshifts for these objects. 
\\ \\
(The data used to create this figure are available in the online journal.)}
\end{figure}

\section{VLBA Relative Astrometry}\label{VLBAastro}

In addition to the coordinate time series we obtained from the Goddard VLBI global solution, we also conducted our own astrometric observations to add position epochs to existing object time series -- thereby either improving existing proper motion measurements or enabling the measurement of new proper motions. We observed 42 radio sources with the Very Long Baseline Array (VLBA) in X-band (3.6 cm / 8.3 GHz). Our targets had all been previously observed by VLBI but lacked a recent observing epoch with high astrometric precision with which to derive proper motions. Seven of our targets are well known galaxies that have been observed many times by VLBI. However, they had not been observed in at least a year (often $>$ 4 years) and never with an astrometric precision of $\sim 1 \ \mu$as. We restricted our sample to objects with declinations $> \ -10^\circ$ in order to ensure sufficient uv-coverage. We also selected targets in close separation ($< 150$ Mpc, comoving) with radio sources with well-measured proper motions. Close separation pairs of extragalactic objects are particularly useful for measuring the collapse of large-scale structure and are discussed in \cite{Darling2013} and an upcoming paper. We used the close-separation radio source with known proper motion as a phase reference and used its position relative to the target to calculate the absolute position of the target. For targets with large angular separations from their phase reference ($\theta > \ 5^\circ$), we also included another bright radio source as a second phase reference. During data reduction, we reduced some of the target observations twice; once with their close-separation companion source as the phase reference and once with the third source as the phase reference. We used the results from this second reduction to ensure that tropospheric delay errors between the target and its companion did not significantly affect our final astrometric measurements. In all cases, the resulting positions were in statistical agreement.
 
 Our VLBA observations were conducted in two campaigns from 2015 September to 2016 January and 2016 October to November. We observed with a total bandwidth of 288 MHz including 8 evenly spaced baseband channels of 32 MHz each, all with left circular polarization. Each channel contained 64 spectral points. Each target was observed in 1 minute intervals with a total on-target integration time of 3 -- 10 minutes per session, depending on the source flux density. We used 2 bit sampling at the Nyquist rate and an aggregate bit rate of 1024 megabits sec$^{-1}$. To maximize the SNR, which directly correlates with the resulting astrometric uncertainty, all of our observations used at least nine out of ten VLBA antennas. The longest baseline, Mauna Kea to St. Croix, was always used in our observations to maximize angular resolution.
 
 We included 45 minute geodetic blocks before and after each observing session to improve measurement of phase errors induced by the troposphere. For the geodetic blocks, we used a similar setup to our main observations, but with more widely spaced baseband channels in order to better model the tropospheric delay as a function of frequency (total bandwidth of 480 MHz). Additionally, we repeated each observing session within 3 -- 24 days so that each target is observed twice within a month and the final astrometric measurements are based on two independent observations. Previous VLBA astrometric campaigns (e.g., \citealp{ReidBrunthaler2004}) have found that a second set of observations can increase astrometric precision by $\sqrt{2}$ and provide an important verification of the astrometry. The period between repeated observations must be long enough that weather conditions are not correlated ($> 3$ days) but short enough that the proper motions of the extragalactic objects are negligible ($< 24$ days). For all targets, the astrometry obtained from the two separate observations were in statistical agreement.
 
\begin{figure}
\centerline{\includegraphics[width=.5\textwidth, trim= 0cm .2cm 1.3cm .5cm, clip]{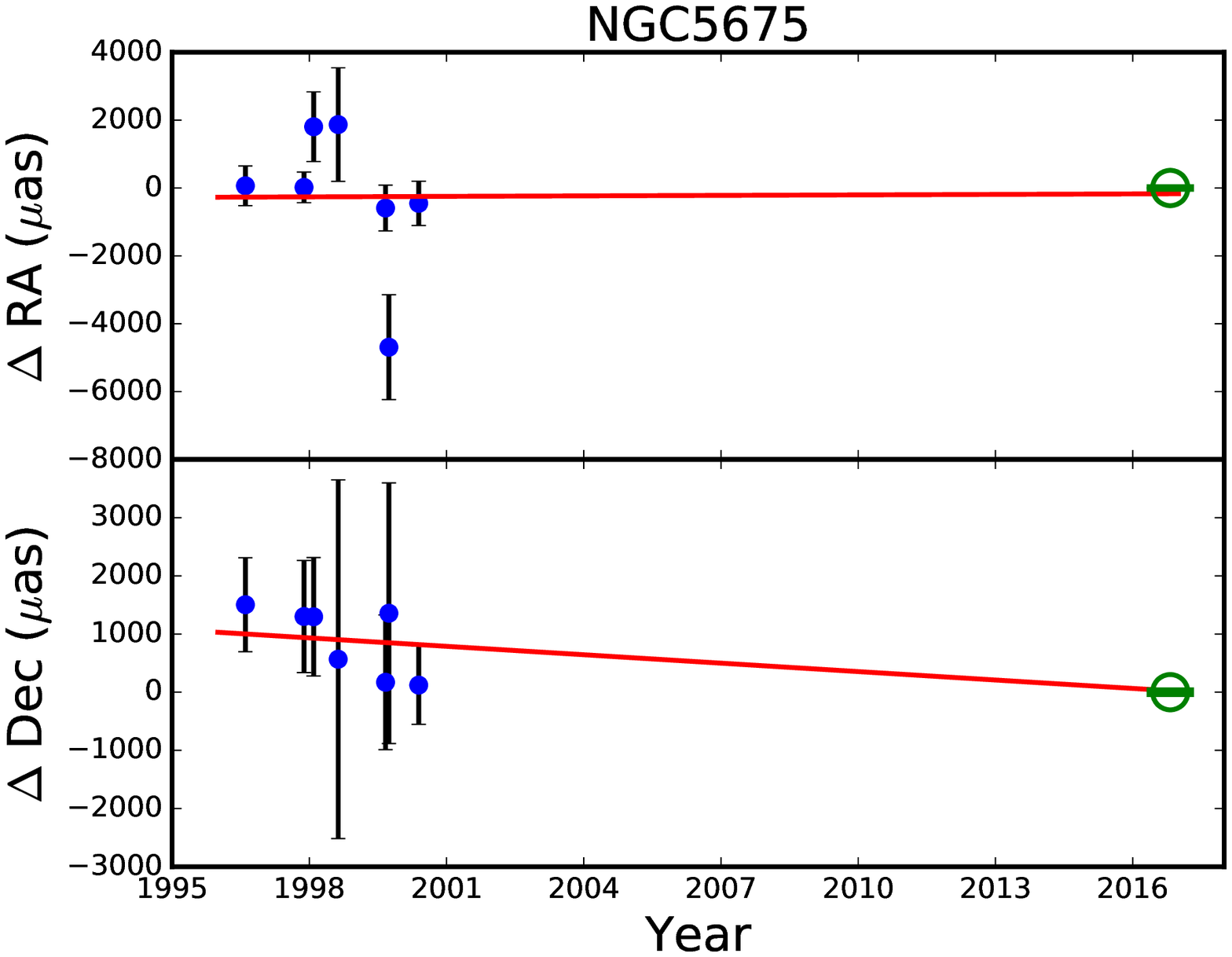}}
\caption{Time series for NGC5675, one of the radio sources observed with the VLBA. The y-axis shows the change in right ascension or declination compared to the position measured from our VLBA observations. The blue points and associated error bars are from the GSFC 2017a time series. The green circle and associated error bar is our measurement of the object's position based on our VLBA observations. The red line shows the maximum likelihood proper motion determined from the plotted points. \label{examplets}}
\end{figure}

 The data were reduced using standard VLBA reduction procedures in the Astronomical Image Processing System (AIPS; \citealp{aips1999}), including self-calibration. For each target and phase reference, we created CLEANed images with a resolution of 0.9 mas pixel$^{-1}$, using uniform weighting. The images have an average sky RMS of 0.5 mJy beam$^{-1}$ and a SNR of $\sim 2200$ (based on the integrated flux of the object and the sky RMS). We found the positions of each target and its phase reference through least-squares fitting of a two-dimensional Gaussian in the image plane. The resulting position accuracies are, on average, $4 \ \mu$as. 
 We used the relative angular separation between the target and its phase reference with known proper motion to calculate the target's absolute position, then added the new position to the target's time series. We then fit the time series using the analytic maximum likelihood parameter estimation described in Section \ref{creation}. For these new proper motions, we did not use a bootstrap re-sampling to calculate the proper motions because the majority have too few epochs to correctly implement a bootstrap. The new proper motions have a mean amplitude precision of $49.7 \ \mu$as yr$^{-1}$. Table \ref{newpm} lists the measured proper motions. Figure \ref{examplets} shows an example time series for one of the new proper motion measurements.

\begin{deluxetable}{llrlrrrrrll}
\tablecaption{Proper Motions Measured from VLBA Astrometry}
\tablewidth{0pc}
\setlength{\tabcolsep}{3pt}
\tabletypesize{\scriptsize}
\tablehead{\colhead{IVS} & \colhead{RA$^{a}$} & \colhead{$\sigma_\alpha$ }& \colhead{Dec} & \colhead{$\sigma_\delta$} & \colhead{$\mu_\alpha$} & \colhead{$\sigma_{\mu, \alpha}$} & \colhead{$\mu_\delta$} & \colhead{$\sigma_{\mu, \delta}$} & \colhead{Phase} & \colhead{MJD$^b$} \\
\colhead{Name}& \colhead{(J2000 h:m:s)} & \colhead{($\mu$s)} & \colhead{(J2000 d:m:s)} & \colhead{($\mu$as)} & \colhead{($\mu$as yr$^{-1}$)} & \colhead{($\mu$as yr$^{-1}$)} & \colhead{($\mu$as yr$^{-1}$)} & \colhead{($\mu$as yr$^{-1}$)} & \colhead{Reference} & \colhead{}
}
\startdata
NGC0315 & 00:57:48.8833591 & 6.2 & +30:21:08.8113827 & 94.8 & 2.1 & 5.1 & -29.0 & 5.9 &  NGC0262 & 57700.0 \\ 
0111+021 & 01:13:43.1449347 & 13.0 & +02:22:17.3166440 & 272.5 & 3.0 & 2.4 & -3.1 & 3.7 &  0056-001 & 57700.0 \\ 
0116+319 & 01:19:35.0033968 & 8.0 & +32:10:50.0580430 & 170.7 & -77.7 & 189.4 & 257.0 & 169.4 &  0134+311 & 57700.0 \\ 
0216+011 & 02:19:07.0245145 & 0.6 & +01:20:59.8659918 & 10.0 & 3.0 & 6.7 & 3.2 & 15.1 &  0215+015 & 57282.0 \\ 
0241+622 & 02:44:57.6966670 & 3.5 & +62:28:06.5156963 & 29.8 & -35.0 & 17.3 & 9.2 & 14.8 &  0302+625 & 57700.0 \\ 
NGC1218 & 03:08:26.2238192 & 3.3 & +04:06:39.3009211 & 102.3 & 15.9 & 4.2 & 6.2 & 8.2 &  0256+075 & 57700.0 \\ 
0415+379 & 04:18:21.2772429 & 4.1 & +38:01:35.8014697 & 44.6 & -13.6 & 4.9 & 58.5 & 5.8 &  0420+417 & 57700.0 \\ 
0651+410 & 06:55:10.0247270 & 0.3 & +41:00:10.1598403 & 2.6 & -3.4 & 3.8 & -6.5 & 6.4 &  0642+449 & 57688.5 \\ 
0836+182 & 08:39:30.7214128 & 4.6 & +18:02:47.1430822 & 179.0 & 39.4 & 9.6 & 24.9 & 18.3 &  0839+187 & 57688.5 \\ 
M81 & 09:55:33.1730490 & 1.3 & +69:03:55.0606048 & 10.0 & -5.4 & 2.3 & -19.2 & 2.0 &  0954+658 & 57688.5 \\ 
1013+127 & 10:15:44.0233882 & 44.5 & +12:27:07.0703636 & 174.7 & -10.2 & 7.3 & 11.8 & 13.2 &  1023+131 & 57688.5 \\ 
MRK180 & 11:36:26.4084426 & 6.2 & +70:09:27.3074967 & 29.8 & 11.1 & 8.0 & 28.1 & 7.2 &  1104+728 & 57688.5 \\ 
NGC3862 & 11:45:05.0090524 & 1.7 & +19:36:22.7412565 & 19.7 & 5.2 & 5.4 & -2.0 & 10.8 &  1147+245 & 57688.5 \\ 
1144+352 & 11:47:22.1305586 & 0.4 & +35:01:07.5224176 & 10.0 & 14.9 & 12.5 & -12.1 & 14.8 &  1128+385 & 57688.5 \\ 
1145+268 & 11:47:59.7639062 & 1.7 & +26:35:42.3324168 & 19.7 & 3.5 & 3.2 & -7.4 & 4.8 &  1147+245 & 57688.5 \\ 
NGC4261 & 12:19:23.2160634 & 0.6 & +05:49:29.7000238 & 15.8 & -8.8 & 18.8 & 31.5 & 22.0 &  1219+044 & 57688.5 \\ 
M84 & 12:25:03.7432602 & 0.3 & +12:53:13.1385898 & 5.6 & -82.9 & 3.5 & -51.2 & 9.5 &  3C274 & 57688.5 \\ 
1232+366 & 12:35:05.8064618 & 0.8 & +36:21:19.3212954 & 11.1 & 1.1 & 14.4 & -7.4 & 18.1 &  1226+373 & 57286.0 \\ 
1304-318 & 13:07:15.1788254 & 1.2 & -32:07:58.64190010 & 27.4 & -37.3 & 36.2 & 186.3 & 99.4 &  1313-333 & 57286.0 \\ 
1306+360 & 13:08:23.7091270 & 3.6 & +35:46:37.1639510 & 158.7 & -7.6 & 3.6 & 9.6 & 5.9 &  OP326 & 57688.5 \\ 
1424+240 & 14:27:00.3917388 & 6.1 & +23:48:00.0371945 & 115.9 & -15.7 & 6.3 & 13.4 & 10.4 &  1417+273 & 57688.5 \\ 
NGC5675 & 14:32:39.8296178 & 2.7 & +36:18:07.9320926 & 40.8 & 3.9 & 11.8 & -48.3 & 21.2 &  1424+366 & 57688.5 \\ 
1441+522 & 14:43:02.7606725 & 0.3 & +52:01:37.2986544 & 3.1 & 16.2 & 31.2 & 30.8 & 55.0 &  1418+546 & 57286.0 \\ 
1456+044 & 14:58:59.3562109 & 0.6 & +04:16:13.8209655 & 8.3 & -1.5 & 3.5 & 61.9 & 5.5 &  1502+036 & 57688.5 \\ 
1623+578 & 16:24:24.8075814 & 1.2 & +57:41:16.2810578 & 11.2 & 18.0 & 3.4 & 20.3 & 3.6 &  1637+574 & 57688.5 \\ 
1625+582 & 16:26:37.2365501 & 1.1 & +58:09:17.6680805 & 13.1 & -14.8 & 36.2 & 110.6 & 69.3 &  1637+574 & 57286.0 \\ 
1716+686 & 17:16:13.9380456 & 1.9 & +68:36:38.7449705 & 10.1 & 11.5 & 6.8 & 4.4 & 11.7 &  1642+690 & 57406.5 \\ 
1726+769 & 17:23:59.4451145 & 0.6 & +76:53:11.5518107 & 2.0 & 3.7 & 8.3 & 7.4 & 12.1 &  1803+784 & 57406.5 \\ 
1721-146 & 17:24:46.9665661 & 0.4 & -14:43:59.76127900 & 11.4 & -10.8 & 13.0 & 11.2 & 27.9 &  NRAO530 & 57406.5 \\ 
1727+502 & 17:28:18.6240785 & 0.5 & +50:13:10.4710557 & 5.8 & 70.8 & 10.9 & 64.8 & 10.4 &  1726+455 & 57688.5 \\ 
1754+159 & 17:56:33.7256016 & 1.9 & +15:53:43.8331315 & 37.4 & -8.3 & 5.6 & -5.6 & 12.2 &  1754+155 & 57688.5 \\ 
1755+055 & 17:57:58.8251366 & 0.2 & +05:31:48.0238501 & 3.6 & 44.9 & 101.5 & 132.2 & 275.0 &  1749+096 & 57406.5 \\ 
1839+548 & 18:40:57.3766918 & 1.0 & +54:52:15.9099683 & 10.9 & -3.2 & 28.0 & -26.8 & 38.1 &  1823+568 & 57406.5 \\ 
3C390.3 & 18:42:08.9901326 & 0.8 & +79:46:17.1283321 & 2.3 & 37.2 & 3.2 & 1.8 & 3.2 &  1803+784 & 57688.5 \\ 
1843+356 & 18:45:35.1088536 & 0.4 & +35:41:16.7261975 & 5.3 & -7.2 & 9.9 & -6.4 & 15.7 &  1846+322 & 57406.5 \\ 
1915+657 & 19:15:23.8190341 & 1.8 & +65:48:46.3878479 & 9.5 & 0.3 & 32.2 & 103.9 & 48.6 &  1842+681 & 57406.5 \\ 
2208-373 & 22:11:50.5258764 & 2.5 & -37:07:04.97856310 & 97.4 & -490.8 & 206.5 & 1328.6 & 961.2 &  2220-351 & 57282.0 \\ 
2305+198 & 23:08:11.6364805 & 1.0 & +20:08:42.1954031 & 25.0 & -9.6 & 27.5 & 60.7 & 35.8 &  2250+194 & 57282.0 \\ 
2319+444 & 23:22:20.3580790 & 0.7 & +44:45:42.3545965 & 9.8 & -5.9 & 3.4 & 174.3 & 5.1 &  2309+454 & 57700.0 \\ 
2324+151 & 23:27:21.9660414 & 7.0 & +15:24:37.3104954 & 87.0 & -7.7 & 22.7 & -47.0 & 29.8 &  2328+107 & 57700.0 \\ 
\enddata
\tablenotetext{a}{J2000 average positions derived from the two epochs of CLEANed VLBA images.}
\tablenotetext{b}{Average Modified Julian Date of the two observation epochs.}
\label{newpm}
\end{deluxetable}

\section{The VLBA Extragalactic Proper Motion Catalog}\label{catalog}
\begin{figure}
\begin{centering}
\includegraphics[width=0.4\columnwidth]{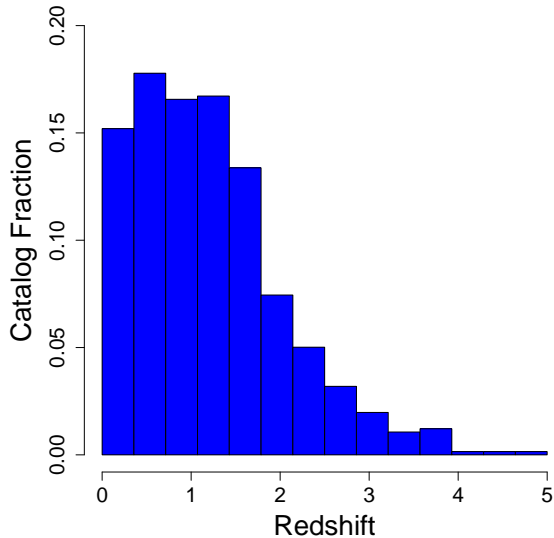}
\caption{Redshift distribution of the VLBA Extragalactic Proper Motion Catalog. M81, which is blue-shifted, is excluded from the plot.} \label{redshifthist}
\end{centering}
\end{figure}

The final extragalactic proper motion catalog contains 713 proper motions. The first ten entries of the catalog are shown in Appendix \ref{fullcatalog}. Figure \ref{skydist} shows the sky distribution of our catalog, along with the proper motion of each object. Proper motions range from $0.01 - 1359.25 \ \mu$as yr$^{-1}$. 
On average, objects were observed for 21.9 years ($\sigma = 4.4$ years) and the proper motion measurements were made based on an average of 249 group delays ($\sigma= 513$ delays). Figure \ref{redshifthist} shows the catalog redshift distribution. The mean redshift is $z=1.20$ and the standard deviation is $\sigma_z=0.84$.

\begin{figure}
\begin{centering}
\includegraphics[width=0.8\columnwidth]{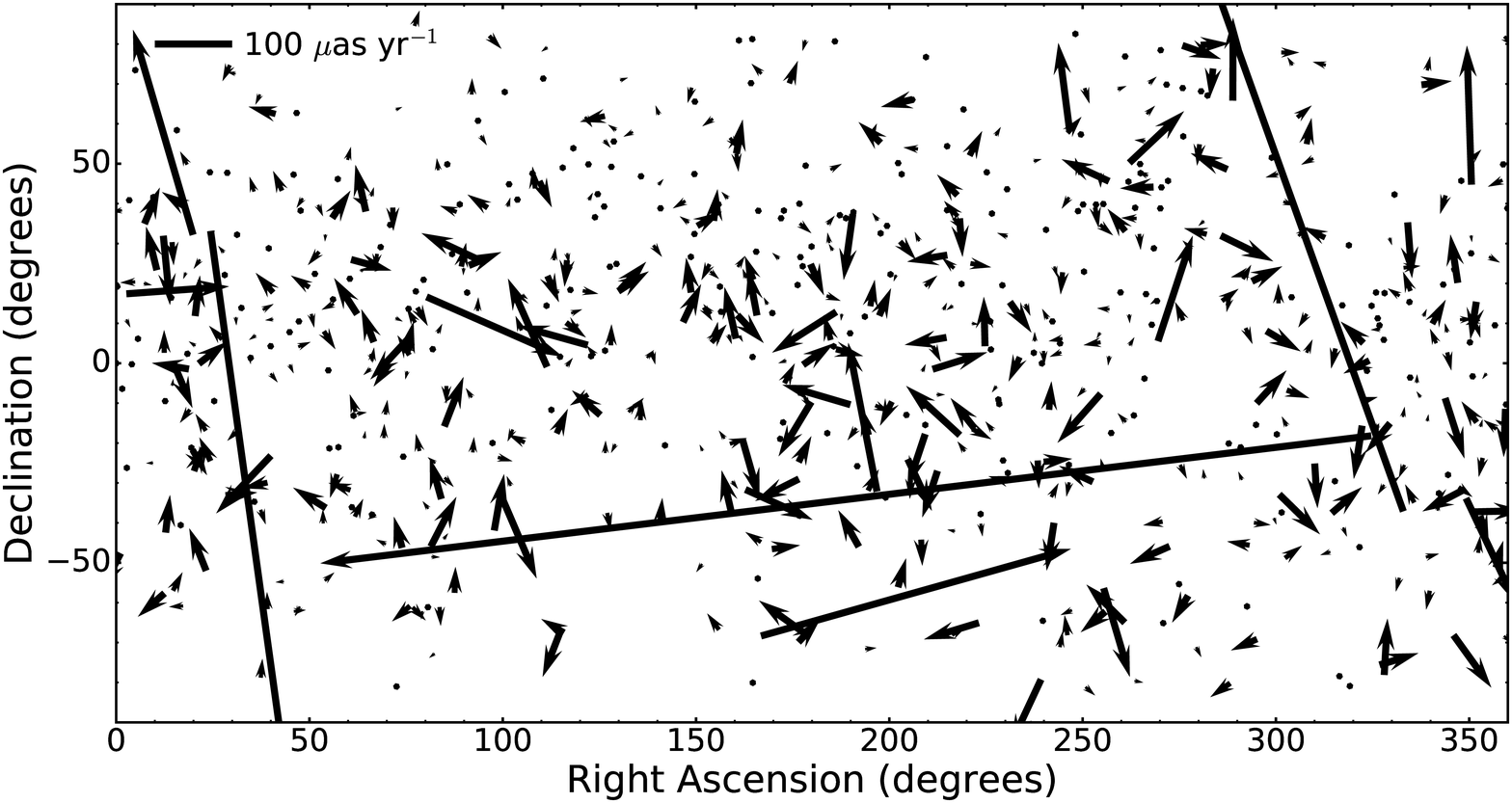}
\caption{Proper motions and sky positions of all catalog objects. The arrow tails correspond to the object positions. A bar showing the size of a proper motion with amplitude 100 $\mu$as yr$^{-1}$ is plotted at the top left.} \label{skydist}
\end{centering}
\end{figure}

\begin{figure}
\begin{centering}
\includegraphics[width=0.6\columnwidth]{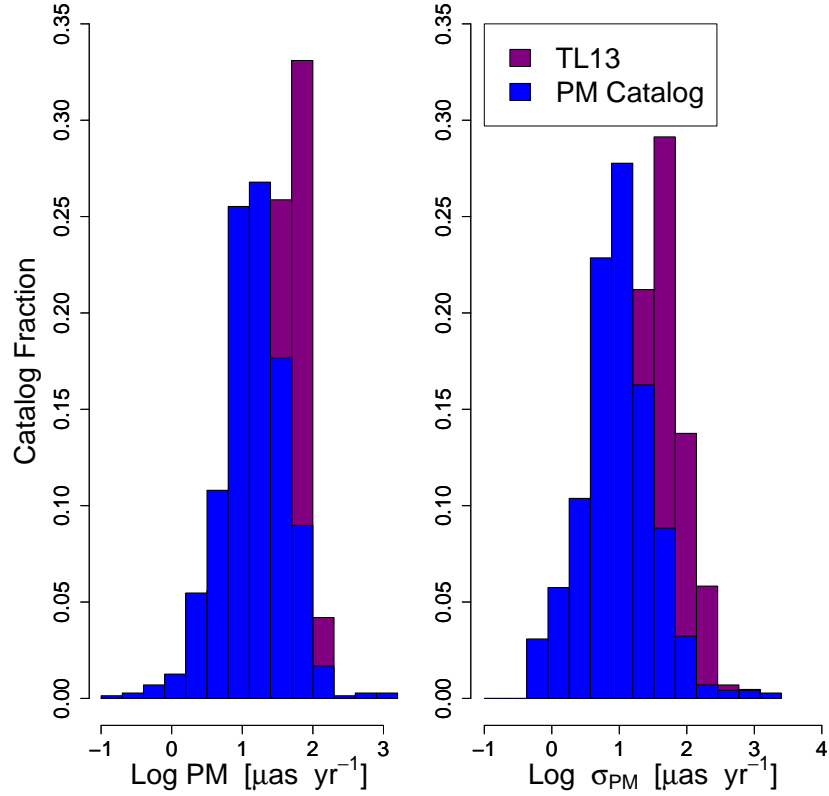} 
\caption{Proper motion amplitudes and associated uncertainties for our catalog (blue) and TL13 (purple). In both catalogs, the proper motions in right ascension and declination are calculated separately for each object, but here we show the total proper motion amplitude for illustrative purposes.} \label{PMhist}
\end{centering}
\end{figure}

Figure \ref{PMhist} shows histograms of the catalog proper motions and errors, along with the equivalent histograms for the TL13 proper motion catalog. The median proper motion amplitude of our objects is $14.7 \ \mu$as yr$^{-1}$ and the standard deviation is $82.5 \ \mu$as yr$^{-1}$. For the TL13 catalog, the median proper motion amplitude is $36.0 \ \mu$as yr$^{-1}$ with a standard deviation of $29.8 \ \mu$as yr$^{-1}$. The large difference in standard deviations is primarily because TL13 remove large proper motions that deviate from their iteratively fit dipole by more than $7 \sigma$. For completeness, we have published all of our calculated proper motions without any clipping.

 There is a large difference in the proper motion uncertainty distributions of the two catalogs -- the mean proper motion uncertainty of our catalog is 26.8 $\mu$as yr$^{-1}$, while it is 51.8 $\mu$as yr$^{-1}$ for the TL13 catalog. The lower uncertainty in our proper motions is due in part to the additional epochs of observations that have been added to the time series in the intervening years. It is also due to our new method of calculating the proper motions from the time series. Most curve-fitting programs use a non-linear parameter estimation, which arrives at the best-fit parameters through a series of guesses. Then the uncertainties of the best-fit parameters are estimated using a similar iterative approach. Instead of using this approximate technique, we derived the analytic equations for the maximum likelihood estimators. By directly calculating the best-fit line to the time series using an analytic calculation, we are also able to directly calculate the uncertainties of the best-fit parameters, rather than using an iterative method to approximate the uncertainties. However, due to the large variation in the uncertainties of individual positions within each time series, we report the standard deviation of a bootstrap distribution as the uncertainty of each proper motion (Section \ref{creation}). These standard deviations are larger than the calculated uncertainties of the maximum likelihood estimators but are a better reflection of the true parameter uncertainties. Despite this increase to the catalog proper motion uncertainties, our mean proper motion uncertainties remain significantly lower than those in TL13, demonstrating the significant effect our analytic calculations and additional epochs have on the proper motion uncertainties.

There are several systematic effects present in our catalog. First, there is a bias toward objects with positive declinations: 61\% of the catalog objects lie above the celestial equator. This is due to the higher concentration of VLBI telescopes in the Northern Hemisphere than in the Southern Hemisphere. We find that this systematic does not have a significant effect on measuring the secular aberration drift (see Section \ref{drift}). If this bias affected the measurement, we would expect to see a significant deviation in declination of the dipole apex from its expected location at the Galactic center. Section \ref{results} shows that this is not the case and that our fitted apex is within 1$\sigma$ of the Galactic center. 

\begin{figure}
\centerline{\includegraphics[width=.4\textwidth]{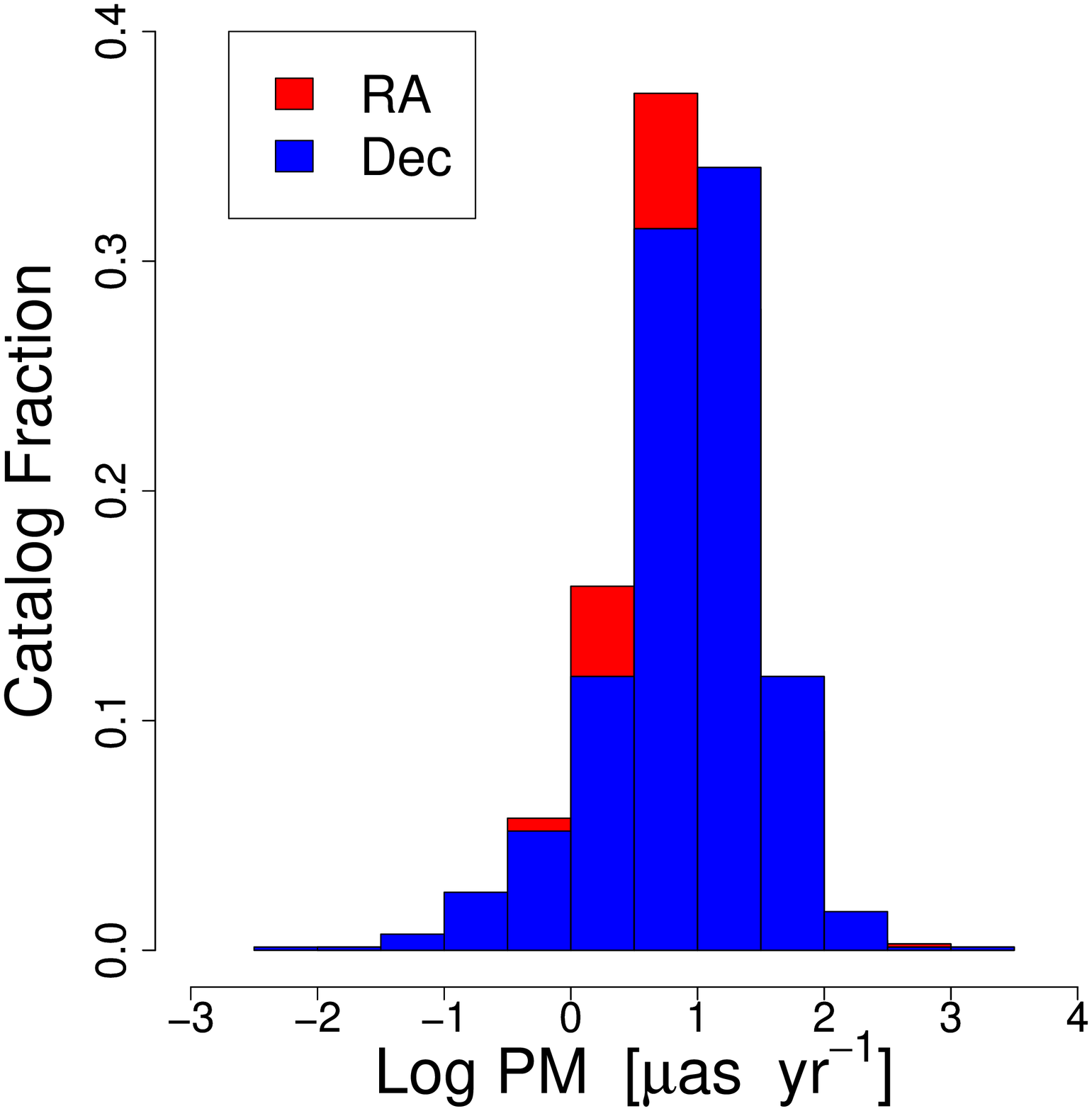} \includegraphics[width=.4\textwidth]{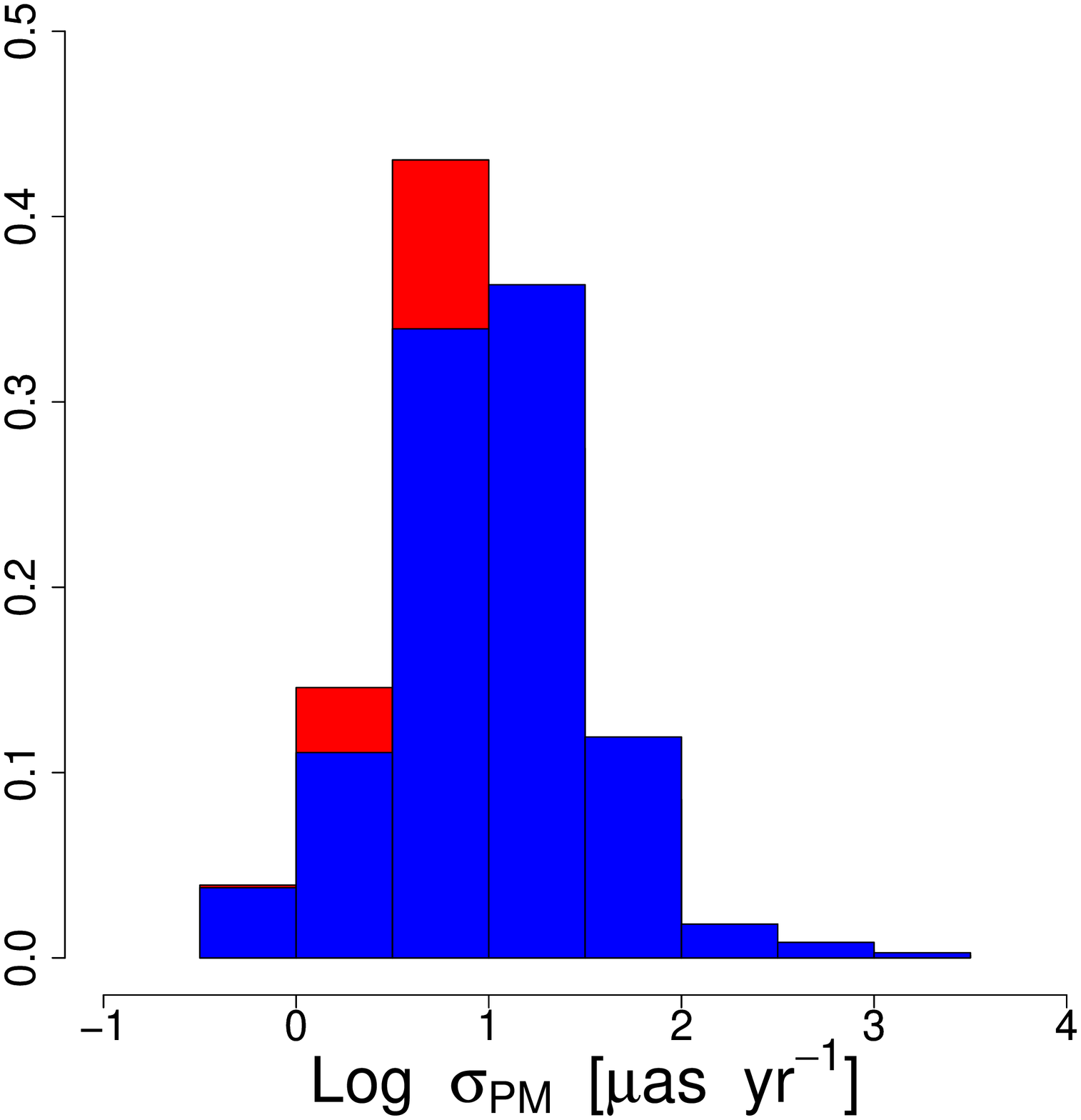} }
\caption{Proper motions and associated uncertainties for our catalog in right ascension (red) and declination (blue). The declination proper motions and uncertainties are systematically larger than those in right ascension.} \label{PMbias}
\end{figure}

Second, the declination proper motions and associated uncertainties are systematically larger than those in right ascension. Figure \ref{PMbias} shows histograms illustrating this bias. The majority of our catalog was created using geodetic experiments conducted with the VLBA, which is more than twice as long in longitude than it is in latitude (East to West the VLBA is $\sim 8600$ km and North to South it is $\sim 3400$ km), causing the primary telescope beam to be an ellipse that is, on average, longer in declination than in right ascension. Thus, the ellipticity of the primary beam allows right ascensions to be measured more precisely than declinations for most point-like objects.

\section{Secular Aberration Drift}\label{drift}
Following the formalism in \cite{Mignard2012}, the secular aberration drift can be expressed as the curl-free portion of a first-order vector spherical harmonic, resembling an electric (E) field dipole. The extragalactic proper motions due to the drift can then be expressed as an E-mode dipole as a function of sky position ($\alpha$,$\delta$) by
\begin{equation} 
\mu_\alpha=\frac{1}{2}\sqrt{\frac{3}{\pi}} \ \left ( s_{11}^{Re} \sin \alpha + s_{11}^{Im} \cos \alpha \right )
\end{equation}
and
\begin{equation}
\mu_\delta=\frac{1}{2}\sqrt{\frac{3}{\pi}} \ \left ( s_{10} \sqrt{\frac{1}{2}} \cos \delta + s_{11}^{Re} \cos \alpha \sin \delta - s_{11}^{Im} \sin \alpha \sin \delta \right ),
\end{equation}
where $s_{lm}$ are the amplitudes of the ``spheroidal'' E-mode vector spherical harmonics of degree $l$ and order $m$. These can be converted to the $d_1$, $d_2$, and $d_3$ amplitudes used in TL13 by
\begin{equation}\label{dconversion}
\begin{pmatrix}
d_1 \\
d_2 \\
d_3 
\end{pmatrix} 
= 2 \sqrt{\frac{\pi}{3}} \
\begin{pmatrix}
\phantom{-} - s_{11}^{Re} \\
\phantom{--}s_{11}^{Im} \\
\sqrt{2} \ s_{10}
\end{pmatrix}.
\end{equation}

In addition, a divergence-free first-order vector spherical harmonic, resembling a magnetic (B) field dipole can also be fit to the extragalactic proper motions using the equations
\begin{equation}
\mu_\alpha = \frac{1}{2}\sqrt{\frac{3}{\pi}} \ \left ( t_{10} \sqrt{\frac{1}{2}} \cos \delta + t_{11}^{Re} \cos \alpha \sin \delta - t_{11}^{Im} \sin \alpha \sin \delta \right ) 
\end{equation}
and
\begin{equation}
\mu_\delta = \frac{1}{2}\sqrt{\frac{3}{\pi}} \ \left( - t_{11}^{Re} \sin \alpha - t_{11}^{Im} \cos \alpha \right ),
\end{equation}
where $t_{lm}$ are the ``toroidal'' B-mode vector spherical harmonic amplitudes and the conversion between these and the toroidal amplitudes used in TL13 is
\begin{equation}\label{rconversion}
\begin{pmatrix}
r_1 \\
r_2 \\
r_3 
\end{pmatrix} 
= 2 \sqrt{\frac{\pi}{3}} \
\begin{pmatrix}
\phantom{---} t_{11}^{Re} \\
 \phantom{--} - t_{11}^{Im} \\
- \sqrt{2} \ t_{10}
\end{pmatrix}.
\end{equation}

\subsection{Data Processing and Results}\label{results}
\begin{figure}
\begin{centering}
\includegraphics[width=0.8\columnwidth]{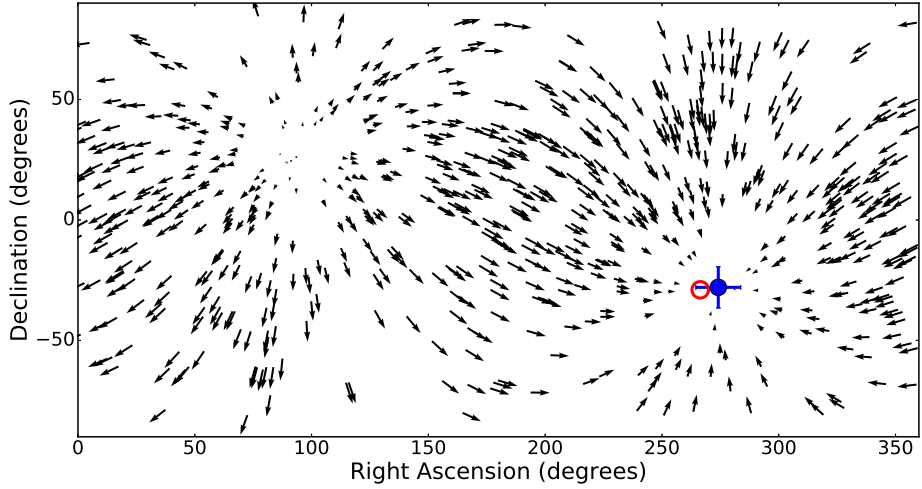} 
\caption{The maximum likelihood secular aberration drift (E-mode dipole) model estimated from the catalog proper motions. The apex of the dipole and its uncertainties are shown in blue, while the true location of the Galactic center is shown in red. The arrow tails correspond to the catalog object positions and the arrow amplitudes represent the modeled portion of the catalog proper motions due to the secular aberration drift. } \label{dipolefig}
\end{centering}
\end{figure}

\begin{figure}
\begin{centering}
\includegraphics[width=\columnwidth]{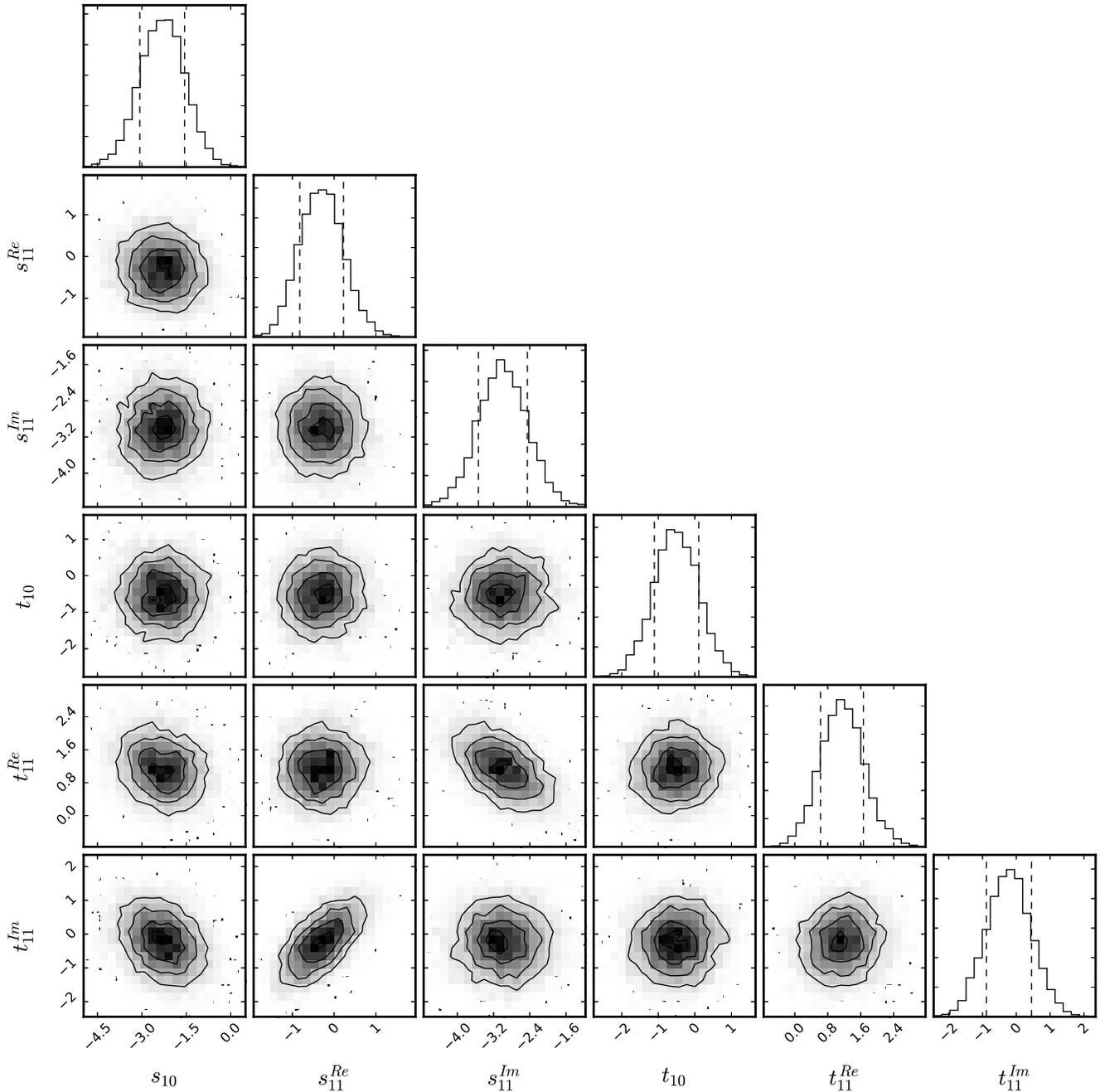} 
\caption{The two- and one-dimensional posterior probability distributions for the dipole coefficients plotted using the Python module {\tt Corner} \citep{corner}. All coefficients are in units of $\mu$as yr$^{-1}$. The dashed lines on the one-dimensional histograms show the 68\% confidence intervals for each parameter.} \label{corner}
\end{centering}
\end{figure}

We simultaneously fit the E- and B-mode dipoles to our extragalactic proper motions using a Markov Chain Monte Carlo (MCMC) Bayesian sampling of the posterior probability distribution function with the Python package {\tt lmfit}\footnote{https://lmfit.github.io/lmfit-py/} \citep{Newvilleetal2014}. The probability distributions for each dipole parameter are estimated through sampling of the log-likelihood functions where we assume the coefficients are drawn from a Gaussian distribution. Table \ref{dipoletable} lists the maximum likelihood solution for each dipole E- and B-mode vector spherical harmonic coefficient and the 68\% confidence interval spread of the probability distributions in both the spherical harmonic formalism and the TL13 formalism. Figure \ref{dipolefig} shows the maximum likelihood secular aberration drift (E-mode dipole) model and Figure \ref{corner} shows the estimated posterior probability distributions. We detect the secular aberration drift at a $6.3\sigma$ significance level, with the square root of the power equal to $4.89 \pm 0.77$ $\mu$as yr$^{-1}$, an amplitude of $1.69 \pm 0.27$ $\mu$as yr$^{-1}$, and an apex of ($275.2 \pm 10.0^\circ$, $-29.4 \pm 8.8^\circ$). The apex of our E-mode dipole is within 1$\sigma$ of the Galactic center (266.4$^\circ$, -28.9$^\circ$) and the amplitude is within 5$\sigma$ of the predicted amplitude of $5.40 \pm 0.78$ $\mu$as yr$^{-1}$. 

We do not detect the B-mode dipole at a 2.2$\sigma$ significance level. We find that including the B-mode dipole increases the significance of the E-mode dipole by $\sim 0.63\sigma$ when compared to fitting just the E-mode dipole without a rotational component. The presence of a non-significant toroidal dipole indicates that we are detecting some residual Earth rotation that is not completely removed during data processing. Table \ref{correlation} shows the correlations between the coefficients of our fit. There is a strong correlation between several coefficients of the E- and B-modes, which accounts for the necessary inclusion of the B-mode dipole.

\begin{deluxetable}{ll|ll}
\tablecaption{Secular Aberration Drift Model}
\label{dipoletable}
\tablehead{
\multicolumn{2}{c}{$\vec{Y}_{lm}$ Formalism$^{a}$} & \multicolumn{2}{c}{TL13 Formalism$^{b}$} \\
\hline
\colhead{Order} & \colhead{Amplitude} & \colhead{Term} & \colhead{Amplitude} \\
\colhead{} & \colhead{($\mu$as yr$^{-1}$)} & \colhead{} & \colhead{($\mu$as yr$^{-1}$)} }
\startdata
\sidehead{E-mode Dipole}
$s_{10}$ & $-2.40 \pm 0.75 $ & $\phantom{-}d_3 $ & $-6.94 \pm 2.17 $ \\
$s_{11}^{Re}$ & $-0.27 \pm 0.52$ & $-d_1$ & $\phantom{-}0.55 \pm 1.06$ \\
$s_{11}^{Im}$ & $-3.00 \pm 0.55$ & $\phantom{-}d_2$ & $-6.13 \pm 1.13$ \\
$\sqrt{P_1^s}$ & $\phantom{-}4.89 \pm 0.77$ & Amplitude & $\phantom{-}1.69 \pm 0.27 $ \\
\hline
\sidehead{B-mode Dipole}
$t_{10}$ & $-0.55\pm 0.59$ & $-r_3 $ & $1.59 \pm 1.71 $ \\
$t_{11}^{Re}$ & $\phantom{-}1.12 \pm 0.54 $ & $\phantom{-}r_1$ & $2.29 \pm 1.11$ \\
$t_{11}^{Im}$ & $-0.06 \pm 0.65$ & $-r_2$ & $0.12 \pm 1.33$ \\
$\sqrt{P_1^t}$ & $\phantom{-}1.68 \pm 0.75$ & Amplitude & $0.58 \pm 0.26 $ \\
\hline
E-mode Apex & \multicolumn{2}{l}{($275.2 \pm 10.0^\circ$,$-29.4 \pm 8.8^\circ$)} & \\
\enddata
\tablenotetext{a}{Vector spherical harmonic coefficients using the formalism presented in \cite{Mignard2012}.}
\tablenotetext{b}{The same dipole model as presented in the left column, but converted into the formalism used in TL13 for comparison. See Sec. \ref{drift} for conversions between the two formalisms.}
\tablecomments{$\sqrt{P_1^s}$ and $\sqrt{P_1^t}$ are the square-root of the 1st order vector spherical harmonic powers. They can be converted to the TL13 amplitudes (right column) by dividing by $2 \sqrt{2\pi / 3}$.}
\end{deluxetable}

\begin{deluxetable}{r|lll}
\tablecaption{Correlations between Spherical Harmonic Coefficients}
\label{correlation}
\tablehead{
\colhead{} & \multicolumn{3}{c}{B-mode Coefficients} \\
\colhead{E-mode Coefficients} & \colhead{$t_{10}$} & \colhead{$t_{11}^{Re}$} & \colhead{$t_{11}^{Im}$}}
\startdata
$s_{10}$ & \nodata & -0.216 & -0.282 \\ 
$s_{11}^{Re}$ & \nodata & \nodata & 0.614 \\
$s_{11}^{Im}$ & \nodata & -0.502 & \nodata \\
\enddata
\tablecomments{Unreported correlations are $<0.150$.}
\end{deluxetable}

\begin{figure}
\begin{centering}
\centerline{\includegraphics[width=.5\textwidth, trim= .3cm 0cm 1.5cm 1cm, clip]{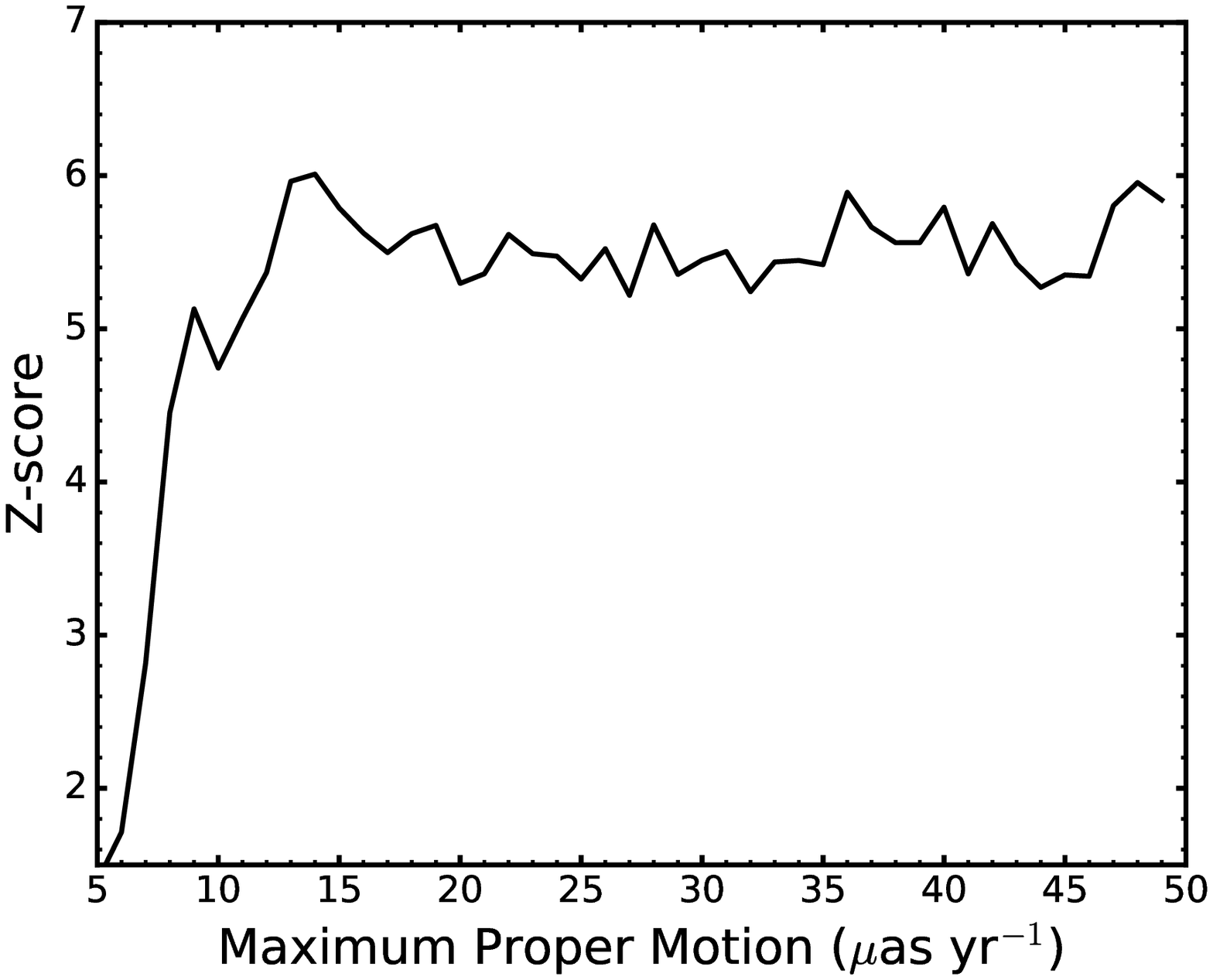} \includegraphics[width=.5\textwidth, trim= .3cm 0cm 1.5cm 1cm, clip]{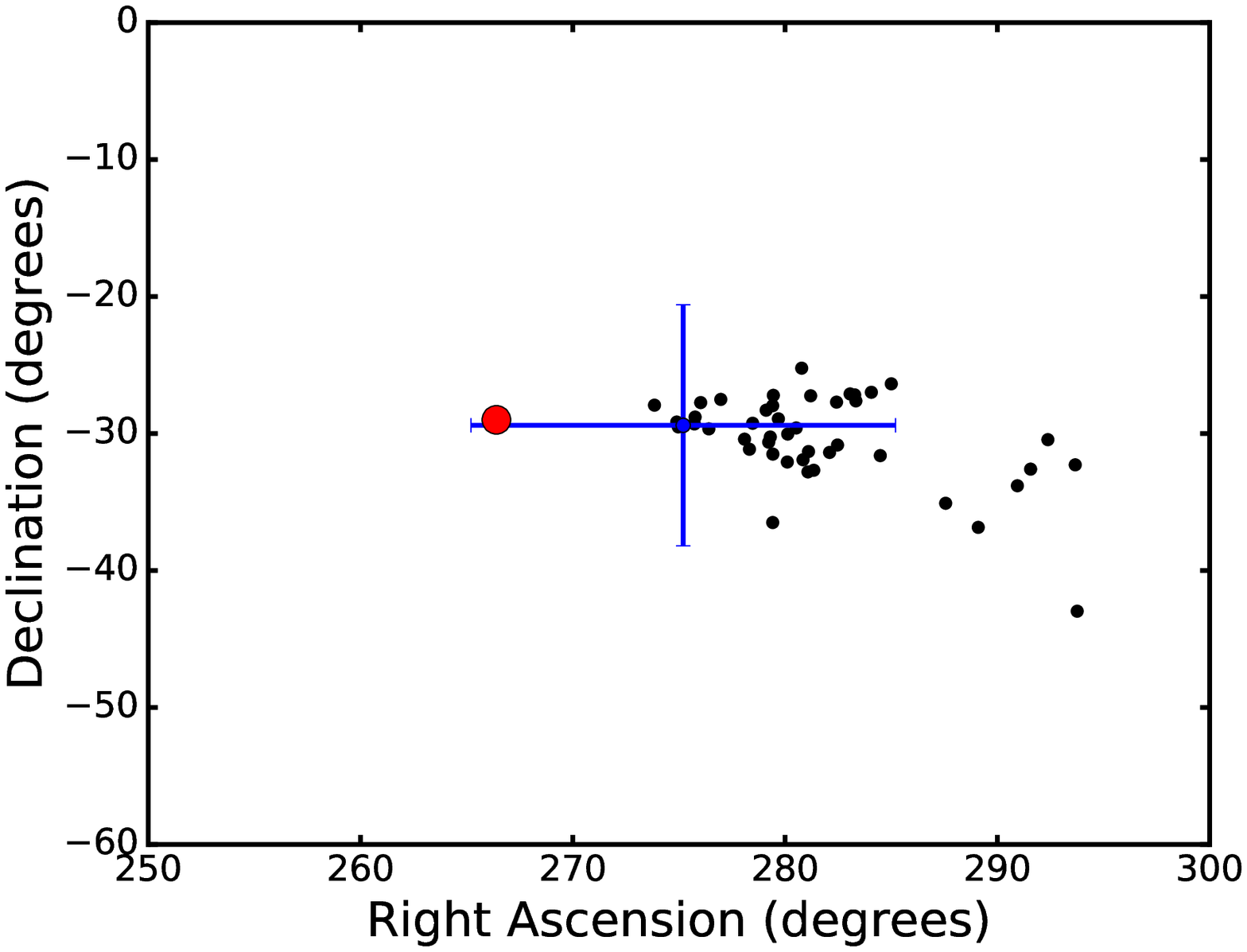}}
\caption{Left: The Z-score of our E-mode dipole model as a function of the maximum proper motion amplitude used in the fit. Z-score is calculated following \cite{Mignard2012}, Eqn. 85. The fit remains significant ($> 5 \sigma$) for all cut-offs greater than $9 \ \mu$as yr$^{-1}$. Right: The location of the E-mode dipole apex of our model as a function of maximum proper motion amplitude for proper motion cut-offs between 5 and 50 $\mu$as yr$^{-1}$ (in steps of 1 $\mu$as yr$^{-1}$). All apex locations are within 2$\sigma$ of each other. The apex for our chosen cut-off (500 $\mu$as yr$^{-1}$) and its associated uncertainty is plotted in blue. The red dot marks the location of the Galactic center. Both of these plots demonstrate the robustness of our secular aberration drift model -- both the location of the dipole apex and the significance of the fit are insensitive to our choice of maximum proper motion amplitude.} \label{maxdrift}
\end{centering}
\end{figure}

Before fitting, we removed all objects with proper motion amplitudes greater than 500 $\mu$as yr$^{-1}$ (three objects). The proper motions of these outliers are likely dominated by intrinsic radio jet motions and obscure the small signal from the secular aberration drift. To ensure that our choice of proper motion cut-off did not significantly affect the resulting dipole model, we fit the E- and B-mode dipoles for a wide range of cut-off values. Figure \ref{maxdrift} shows the results of this experiment for proper motion cut-offs between 5 and 50 $\mu$as yr$^{-1}$ (in steps of 1 $\mu$as yr$^{-1}$). We also performed the same experiment for cut-offs between 50 $\mu$as yr$^{-1}$ and 1000 $\mu$as yr$^{-1}$ (in steps of 50 $\mu$as yr$^{-1}$). We found that the Z-score (calculated following \cite{Mignard2012}, Eqn. 85) is relatively constant and the fit remains statistically significant for all cut-offs greater than $9 \ \mu$as yr$^{-1}$. Fits with cut-offs below 9 $\mu$as yr$^{-1}$ include less than 226 proper motions and the decrease in significance can be attributed to the small number of data points. Additionally, we found that the apex remains within $2 \sigma$ of the Galactic center for all cut-offs -- all apex locations more than $ 1 \sigma$ from the Galactic center were again for low cut-offs ($< 15 \ \mu$as yr$^{-1}$) with a small number of data points. Combined, these factors indicate that our model of the dipole is robust and insensitive to our choice in maximum proper motion amplitude. Therefore, we chose to use a cut-off of 500 $\mu$as yr$^{-1}$ to include the majority of the catalog in our fit.

The secular aberration drift has been measured by several previous studies, most notably TL13 and \cite{Xuetal2013}. TL13 measured a spheroidal dipole with amplitude $6.4 \pm 1.1 \ \mu$as yr$^{-1}$ pointed towards ($266 \pm 7^\circ$,$-26\pm 7^\circ$) and a toroidal component with amplitude $1.9 \pm 0.8 \ \mu$as yr$^{-1}$. Our model has a lower spheroidal dipole amplitude ($1.7 \pm 0.3 \ \mu$as yr$^{-1}$) than that measured in TL13, making their amplitude a better match for the predicted amplitude ($5.40 \pm 0.78$ $\mu$as yr$^{-1}$; see Section \ref{intro}). However, our uncertainties are also much smaller, yielding a more significant detection. The apex of both fits are statistically consistent with the Galactic center. 

One possible cause of our lower spheroidal dipole amplitude is the inclusion of a no-net-rotation (NNR) constraint in our catalog (see Section \ref{creation}). To allow for some rotation of the radio source positions with respect to the ICRS axes, the NNR constraint was incrementally relaxed for subsets of the objects when creating the Goddard 2017a coordinate time series. On the other hand, TL13 used a more relaxed constraint that allowed sources to rotate by $< 2$ arcseconds with respect to the ICRS axes. The difference in our E-mode amplitudes may be due, in part, to the different techniques used to handle the NNR constraint. The NNR constraint affects the toroidal dipole, which affects the spheroidal dipole because of the correlation between several of the E- and B-mode coefficients (Table \ref{correlation}). To this effect, \cite{Titovetal2011} found that their spheroidal amplitude decreased and the E-mode apex shifted away from the Galactic center when sources were more closely fixed to the ICRS axes and only allowed to rotate by $<2$ milliarcseconds. Our E-mode amplitude is lower than expected, but the associated apex is still in close alignment with the Galactic center, indicating that our model is only slightly affected by the NNR constraint and that the aberration drift is still detectable.

\cite{Xuetal2013} measured the secular aberration drift using a similar data set to TL13 but with a different estimation method. Instead of solving for individual proper motions and then fitting a dipole to these motions, \cite{Xuetal2013} added a three-dimensional solar acceleration vector to the global parameters in CALC/SOLVE and directly solved for this acceleration. They found a solar acceleration vector of ($7.47 \pm 0.46$, $0.17 \pm 0.57$, $3.95 \pm 0.47$) mm s$^{-1}$ yr$^{-1}$, which is equivalent to a dipole with root-power of $16.8 \pm 0.8$ $\mu$as yr$^{-1}$ ($5.8 \pm 0.3$ $\mu$as yr$^{-1}$ using the convention in TL13) pointed towards ($243.0^\circ$, $-11.5^\circ$). Their dipole is offset from the Galactic center by $\sim 18^\circ$ north and $\sim 23^\circ$ west. The offset equates to a significant acceleration component perpendicular to the Galactic plane. Our dipole model contains no significant out of plane acceleration ($0.27 \pm 0.51$ mm s$^{-1}$ yr$^{-1}$). Because our dipole apex is statistically consistent with the Galactic center, our solar acceleration vector does not contain a significant perpendicular component. 

\begin{figure}
 \begin{centering}
 \includegraphics[width=0.5\columnwidth]{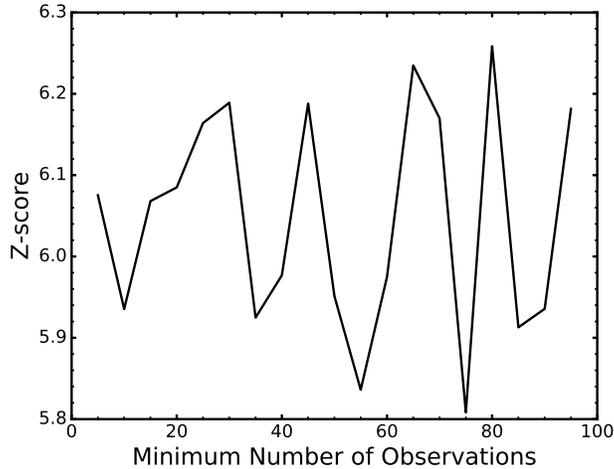} 
 \caption{The Z-score of our E-mode dipole model as a function of the minimum number of observing sessions required for a proper motion to be included in the fit. Our model remains significant regardless of the number of observing sessions.} \label{nobs}
 \end{centering}
 \end{figure}

Since the secular aberration drift is a small component of the overall proper motion of many sources, there is a concern that individual large intrinsic proper motions with small uncertainties could have a significant effect on the dipole fit. \cite{Titovetal2011} used several subsets of their data, including only ICRF2 defining sources, sources with low structure indices, and sources with more than 1000 sessions, to assess the robustness of their fit. They found that the resulting dipole did not vary significantly for any of the data subsets. We performed a similar analysis to ensure that individual outliers in proper motion do not significantly influence our model. Figure \ref{nobs} shows the results of a test where we clipped our data to only include objects whose proper motions were determined from a minimum number of observing sessions. We find that our model remains significant regardless of the number of observing sessions per object.

\begin{figure}
\begin{centering}
\centerline{\includegraphics[width=.4\textwidth]{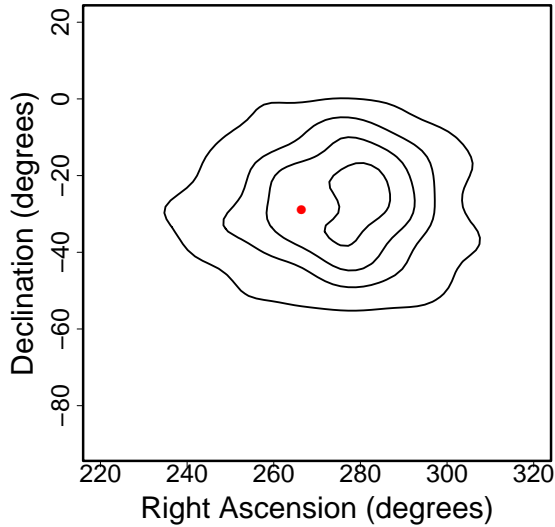}}
\caption{The location of the E-mode apex for each bootstrap re-sampling. The contours show the density of the bootstrap distribution. From the inner contour outward, the contours contain 10\%, 30\%, 60\%, and 80\% of the data. This shows that the apex of the E-mode dipole is insensitive to exclusion of subsets of the data. The Galactic center is marked in red.} \label{bootstrapplots}
\end{centering}
\end{figure}

To further assess the effect of large intrinsic proper motions on our secular aberration drift model, we performed 1,000 bootstrap re-samplings of the fit. Table \ref{bootstrap} lists the median values and 68\% confidence intervals from the bootstrap distribution. We find a median root-power of 5.11 $\mu$as yr$^{-1}$ with a 68\% confidence interval of $3.61 - 6.71 \ \mu$as yr$^{-1}$. Our dipole model has a root-power of $4.89 \pm 0.77 \ \mu$as yr$^{-1}$, which is in good agreement with these values, indicating that our model is not significantly influenced by individual data points. Figure \ref{bootstrapplots} shows the location of the E-mode apex for all bootstrap iterations. This plot demonstrates that the location of the E-mode apex is not significantly influenced by individual proper motions.

\begin{deluxetable}{lcc}
\tablecaption{Median Values of 1,000 Bootstrap Dipole Fits}
\label{bootstrap}
\tablehead{
\colhead{} & \colhead{Median$^{a}$} & \colhead{68\% Confidence Interval}} 
\startdata
$\sqrt{P_1^s}^b$ & 5.11 & $3.61 - 6.71$ \\
$\sqrt{P_1^t}$ & 2.85 & $1.65 - 4.45$ \\
E-mode Apex RA & 284.6$^\circ$ & $274.1^\circ - 302.9^\circ$ \\
E-mode Apex Dec & $-29.2^\circ$ & $-48.2^\circ - -9.1^\circ$ \\
\enddata
\tablenotetext{a}{All values are in units of $\mu$as yr$^{-1}$, except for the E-mode apex, which is in units of degrees.}
\tablenotetext{b}{Square-root of the 1st order vector spherical harmonic power. Can be converted to the TL13 amplitude by dividing by $2\sqrt{2\pi/3}$.}
\end{deluxetable}

\section{Conclusions}\label{Conclusions}
In this paper we presented the VLBA Extragalactic Proper Motion Catalog containing 713 proper motions with average uncertainties of 24 $\mu$as yr$^{-1}$. The catalog is created primarily from archival Goddard VLBI data, with redshifts obtained from OCARS. In addition, we added or updated 40 extragalactic proper motions and 10 redshifts through our own VLBA and APO observations, respectively. 

We used the resulting catalog to measure the secular aberration drift at a $6.3 \sigma$ significance. An accurate measurement of the aberration drift is important so that it can be fully removed from extragalactic proper motions before using those proper motions to study cosmological effects. We detect a spherical dipole with a root-power of $4.89 \pm 0.77 \ \mu$as yr$^{-1}$ and an apex at ($275.2 \pm 10.0^\circ$, $-29.4 \pm 8.8^\circ$). We simultaneously fit a toroidal dipole with low significance ($2.2 \sigma$) to increase the significance of the E-mode dipole. Overall, our model of the E-mode dipole proves robust to a number of tests and remains statistically significant for many subsets of the data.

Although the E-mode dipole is significant, its amplitude is much lower than expected. This difference may be caused by the NNR constraint used when the majority of our radio source positions were calculated. Even if the NNR constraint has prevented us from recovering the full dipole, our catalog can still be used for cosmological studies by removing the residual dipole. 

In an upcoming paper, we subtract this dipole and use the remaining proper motions to search for relative proper motions between close separation extragalactic objects following \cite{Darling2013}. The relative proper motions of extragalactic objects contain signatures of both the Hubble expansion and the collapse of large-scale structure, enabling a detection or measurement of both effects. In a second paper, we fit a quadrupolar pattern to the catalog proper motions in order to obtain limits on the stochastic gravitational wave background \citep{Darlingetal2017}. 

With continued geodetic observations and new Goddard VLBI global solutions, the number of available extragalactic proper motions is expected to continue to increase in the coming years. Additional observing epochs of existing proper motions will also increase the overall accuracy of future catalogs. Both of these will contribute to more accurate measurements of the secular aberration drift and enable more detailed cosmological studies. In the near future, with the release of {\it Gaia} \citep{GaiaCollaboration2016} proper motions, we expect to significantly expand our catalog. The {\it Gaia} proper motions will be less precise -- astrometry of $\sim 1.69$ mas for objects in the secondary data set of DR1 \citep{Lindegrenetal2016} -- than those produced by VLBI, but there will be $\sim 10^6$ new extragalactic proper motions \citep{Robinetal2012}. The dramatic increase in sample size should enable statistically significant detections of the secular aberration drift with relative precision of 10\% \citep{Mignard2002} and cosmological effects despite the decrease in overall catalog proper motion accuracy. The {\it Gaia} proper motions will also be less affected by intrinsic proper motion, which will enable a higher significance detection of correlated motions like the secular aberration drift and cosmological effects (see \citealp{Darlingetal2017} for further discussion). Additionally, the Next Generation Very Large Array (ngVLA) with VLBA baselines is expected to obtain proper motions of $\sim 10 \ \mu$as yr$^{-1}$ \citep{Boweretal2015}. The addition of a central array with a large collecting area to be used in conjunction with the VLBA will enable rapid, high SNR detections of many radio sources within a single observing epoch and is expected to increase the number of epochs for many catalog objects, thereby increasing the overall proper motion precision of the catalog.
 
\section*{Acknowledgments}
The authors thank the referee, Hana Kr\'{a}sn\'{a} (Technische Universit\"{a}t Wien), for many useful suggestions, and David Gordon (NASA Goddard Space Flight Center) and Amy Mioduszewski (NRAO) for helpful discussions. The authors acknowledge support from the NSF grant AST-1411605. All catalogue comparisons were performed using the tabular manipulation tool
STILTS \citep{Taylor2006}. This research has made use of NASA Goddard Space Flight
Center's VLBI source time series 2017a solution, prepared by
David Gordon. This research has made use of the NASA/IPAC Extragalactic Database (NED) which is operated by the Jet Propulsion Laboratory, California Institute of Technology, under contract with the National Aeronautics and Space Administration. This research has made use of the SIMBAD database,
operated at CDS, Strasbourg, France. The National Radio Astronomy Observatory is a facility of the National Science Foundation operated under cooperative agreement by Associated Universities, Inc. IRAF is distributed by the National Optical Astronomy Observatories, which are operated by the Association of Universities for Research in Astronomy, Inc., under cooperative agreement with the National Science Foundation.
 
 Based on observations obtained at the Gemini Observatory (processed using the Gemini IRAF package), which is operated by the Association of Universities for Research in Astronomy, Inc., under a cooperative agreement with the NSF on behalf of the Gemini partnership: the National Science Foundation (United States), the National Research Council (Canada), CONICYT (Chile), Ministerio de Ciencia, Tecnolog\'{i}a e Innovaci\'{o}n Productiva (Argentina), and Minist\'{e}rio da Ci\^{e}ncia, Tecnologia e Inova\c{c}\~{a}o (Brazil).

\facilities{NMSU:1m, Gemini:Gillett, VLBA}

\software{AIPS \citep{aips1999}, Corner \citep{corner}, IRAF (including Gemini IRAF), LMFIT \citep{Newvilleetal2014}, STILTS \citep{Taylor2006}}


\newpage
\appendix

\section{The VLBA Extragalactic Proper Motion Catalog Sample}\label{fullcatalog}

Table \ref{catalog} shows the first 10 objects from the VLBA Extragalactic Proper Motion Catalog. The entire catalog contains 713 radio sources with the same
columns and is available online and at http://vizier.u-strasbg.fr/. From left to right, the columns are the IVS name of the source, the source's right ascension and associated uncertainty (in milliseconds) from the most recent VLBA observing epoch, the declination and associated uncertainty (in milliarcseconds) from the most recent VLBA observing epoch, the proper motion in right ascension and associated uncertainty (both in $\mu$as yr$^{-1}$), the number of sessions used to determine the source's proper motion in right ascension, the reduced $\chi^2$ of the derived proper motion in right ascension, the same quantities for the proper motion in declination, the number of years of VLBI observations used in determining the proper motion, the Modified Julian Date of the most recent VLBI session used to measure the object's position, a flag to indicate proper motions added or updated by this paper, the redshift of the source, a flag indicating the quality of the reported redshift, and the source from which the redshift data were obtained. The redshift flag is obtained from the OCARS catalog \citep{Malkin2016} and indicates the following potential limitations to the reported redshift: ``(a) photometric, (b) unreliable or doubtful identification, (c) substantially different estimates in the literature, (d) lower limit, and (e) imaging.'' See the OCARS catalog for more information about the redshift of a particular source.   
 
 \newpage

 \movetabledown=1.8in
\begin{rotatetable}

\begin{deluxetable*}{lcccrrrrcrrrccccccc}
\tablecaption{The VLBA Extragalactic Proper Motion Catalog Sample}
\tabletypesize{\scriptsize}
\setlength{\tabcolsep}{3pt}
\tablewidth{0pt}
\tablehead{\colhead{IVS} & \colhead{RA} & \colhead{$\sigma_\alpha$ }& \colhead{Dec} & \colhead{$\sigma_\delta$} & \colhead{$\mu_\alpha$} & \colhead{$\sigma_{\mu, \alpha}$} & \colhead{$N_\alpha$} & \colhead{$\chi^2_\alpha$} & \colhead{$\mu_\delta$} & \colhead{$\sigma_{\mu, \delta}$} & \colhead{$N_\delta$} & \colhead{$\chi^2_\delta$} & \colhead{Length} & \colhead{Last Obs.} & \colhead{New} & \colhead{z} &\colhead{z$_{\mbox{flag}}$} & \colhead{z} \\
\colhead{}& \colhead{(J2000 h:m:s)} & \colhead{(ms)} & \colhead{(J2000 d:$'$:$''$)} & \colhead{(mas)} & \colhead{($\mu$as yr$^{-1}$)} & \colhead{($\mu$as yr$^{-1}$)} & \colhead{} & \colhead{} & \colhead{($\mu$as yr$^{-1}$)} & \colhead{($\mu$as yr$^{-1}$)} & \colhead{} & \colhead{} & \colhead{(years)} & \colhead{(MJD)} & \colhead{PM} & \colhead{} & \colhead{} & \colhead{Source}}
\decimals
\startdata
2358+189 & 00:01:08.621513 & 0.608 & +19:14:33.800894 & 1.93 & 5.59 & 8.49 & 97 & 1.5 & -1.12 & 9.45 & 97 & 1.1 & 21.2 & 57819.9 &    & 3.10 &  &  OCARS \\ 
0002-478 & 00:04:35.655473 & 0.015 & -47:36:19.604493 & 0.26 & -24.75 & 13.45 & 45 & 1.4 & -47.34 & 20.18 & 45 & 1.1 & 22.8 & 57659.8 &    & 0.88 &  &  OCARS \\ 
0003+380 & 00:05:57.175296 & 0.172 & +38:20:15.157041 & 17.96 & -13.99 & 9.58 & 29 & 2.4 & -2.21 & 11.12 & 29 & 1.7 & 22.9 & 57102.7 &    & 0.23 &  &  OCARS \\ 
0003-066 & 00:06:13.892882 & 0.003 & -06:23:35.335489 & 0.05 & 0.27 & 1.29 & 1437 & 2.0 & 3.64 & 1.80 & 1437 & 2.2 & 23.5 & 57840.9 &    & 0.35 &  &  OCARS \\ 
IIIZW2 & 00:10:31.005918 & 0.029 & +10:58:29.504257 & 1.35 & 5.17 & 12.17 & 62 & 2.8 & -10.08 & 12.58 & 62 & 1.7 & 26.7 & 57776.2 &    & 0.09 &  &  OCARS \\ 
0007+171 & 00:10:33.991756 & 0.522 & +17:24:18.762349 & 0.49 & 132.71 & 26.67 & 26 & 3.8 & 9.21 & 14.77 & 26 & 1.3 & 20.0 & 55244.7 &    & 1.60 &  &  OCARS \\ 
0008-264 & 00:11:01.246780 & 0.017 & -26:12:33.377442 & 0.23 & 0.43 & 7.31 & 180 & 1.9 & 0.26 & 9.41 & 180 & 2.0 & 24.6 & 57840.9 &    & 1.10 &  &  OCARS \\ 
0010+405 & 00:13:31.130193 & 0.008 & +40:51:37.144086 & 0.12 & -3.90 & 3.06 & 67 & 1.2 & 7.63 & 4.64 & 67 & 1.6 & 25.6 & 57784.9 &    & 0.26 &  &  OCARS \\ 
0013-005 & 00:16:11.088362 & 0.095 & -00:15:12.449356 & 2.57 & 1.61 & 5.51 & 59 & 1.4 & 3.95 & 7.81 & 59 & 1.3 & 23.7 & 57751.0 &    & 1.58 &  &  OCARS \\ 
0017+200 & 00:19:37.854481 & 0.003 & +20:21:45.644651 & 0.03 & -19.24 & 9.99 & 214 & 2.7 & -0.74 & 4.25 & 214 & 2.2 & 21.2 & 57840.9 &    &  &  &  OCARS \\ 
\enddata
\tablenotetext{a}{The first 10 objects from the VLBA Extragalactic Proper Motion Catalog. The complete catalog is available online. See the Appendix for a description of the columns.}\label{catalog_sample}
\end{deluxetable*}
\end{rotatetable}

\newpage



\bibliographystyle{aasjournal}
\bibliography{PMCatalog}


\end{document}